\PassOptionsToPackage{dvipsnames}{xcolor}
\documentclass[twocolumn,tighten,twocolappendix]{aastex631}
\usepackage{gensymb}
\usepackage{natbib}      
\usepackage{graphicx}
\usepackage{amsmath}
\usepackage{times}
\usepackage{amssymb}
\usepackage{fancyhdr}
\usepackage{epsfig}
\usepackage{float}
\usepackage{url}
\usepackage{array}
\usepackage{verbatim}     
\usepackage{calc}
\usepackage{boldline}
\usepackage{tabu}
\usepackage{rotating}     
\usepackage{xcolor,colortbl}

\usepackage{multirow}

\usepackage{newtxtext,newtxmath}
\usepackage{aecompl}
\usepackage{ae,aecompl}
\usepackage{hyperref}
\usepackage{cleveref}
\usepackage{tabularx}
\usepackage{fontawesome}

\hypersetup{linkcolor=red,citecolor=MidnightBlue,filecolor=cyan,urlcolor=blue}
\newcommand{\p}[1]{{\color{magenta}{#1}}}
\newcommand{\q}[1]{{\color{red}{#1}}}

\begin{document}
%

\title{Enhancing Morphological Measurements of Cosmic Web with Delaunay Tessellation Field Estimation}

\correspondingauthor{Yu Liu; Yu Yu; Pengjie Zhang}
\email{liuyu9@tsinghua.edu.cn; yuyu22@sjtu.edu.cn; zhangpj@sjtu.edu.cn}

\shorttitle{MFs with DTFE}
\shortauthors{Yu Liu et al.}

\author[0000-0002-9734-906X]{Yu Liu}
\affiliation{Department of Astronomy, Tsinghua University, Beijing, 100084, P.R. China}
\affiliation{Department of Astronomy, School of Physics and Astronomy, Shanghai Jiao Tong University, Shanghai, 200240, P.R. China}
\affiliation{Key Laboratory for Particle Astrophysics and Cosmology (MOE)/Shanghai Key Laboratory for Particle Physics and Cosmology, P.R. China}
\author[0000-0002-9359-7170]{Yu Yu}
\affiliation{Department of Astronomy, School of Physics and Astronomy, Shanghai Jiao Tong University, Shanghai, 200240, P.R. China}
\affiliation{Key Laboratory for Particle Astrophysics and Cosmology (MOE)/Shanghai Key Laboratory for Particle Physics and Cosmology, P.R. China}
\author[0000-0003-2632-9915]{Pengjie Zhang}
\affiliation{Department of Astronomy, School of Physics and Astronomy, Shanghai Jiao Tong University, Shanghai, 200240, P.R. China}
\affiliation{Key Laboratory for Particle Astrophysics and Cosmology (MOE)/Shanghai Key Laboratory for Particle Physics and Cosmology, P.R. China}
\affiliation{Tsung-Dao Lee Institute, Shanghai Jiao Tong University, Shanghai, 200240, P.R. China}
\author[0000-0001-5277-4882]{Hao-Ran Yu}
\affiliation{Department of Astronomy, Xiamen University, Xiamen, Fujian, 361005, P. R. China}
\textit{}

\begin{abstract}
The density fields constructed by traditional mass assignment methods are susceptible to irritating discreteness, which hinders morphological measurements of cosmic large-scale structure (LSS) through Minkowski functionals (MFs). For alleviating this issue, fixed-kernel smoothing methods are commonly used in literatures, at the expense of losing substantial structural information. In this work, we propose to measure MFs with Delaunay tessellation field estimation (DTFE) technique, with the goal to maximize extractions of morphological information from sparse tracers. We perform our analyses starting from matter fields and progressively extending to halo fields. At matter field level, we elucidate how discreteness affects the morphological measurements of LSS. Then, by comparing with traditional Gaussian smoothing scheme, we preliminarily showcase the advantages of DTFE for enhancing measurements of MFs from sparse tracers. At halo field level, we first numerically investigate various systematic effects on MFs of DTFE fields, which are induced by finite voxel sizes, halo number densities, halo weightings, and redshift space distortions (RSDs), respectively. Then, we explore the statistical power of MFs measured with DTFE for extracting cosmological information encoded in RSDs. We find that MFs measured with DTFE exhibit improvements by $\sim$ $2$ orders of magnitude in discriminative power for RSD effects and by a factor of $\sim$ $3$-$5$ in constraining power on structure growth rate over the MFs measured with Gaussian smoothing. These findings demonstrate the remarkable enhancements in statistical power of MFs achieved by DTFE, showing enormous application potentials of our method in extracting various key cosmological information from galaxy surveys.
\end{abstract}


\section{Introduction} \label{sec:intro}
Ambitious on-going and up-coming cosmological surveys [e.g.,  HETDEX (\citealt{2008ASPC..399..115H}),  LSST (\citealt{2009arXiv0912.0201L}), Euclid (\citealt{2011arXiv1110.3193L}), 4MOST (\citealt{2012SPIE.8446E..0TD}), PFS (\citealt{2014PASJ...66R...1T}), SPHEREX (\citealt{2014arXiv1412.4872D}), DESI (\citealt{2016arXiv161100036D}), WFIRST (\citealt{2018arXiv180403628D}), and CSST (\citealt{2019ApJ...883..203G})], particularly the fifth-generation surveys [e.g.,  WST (\citealt{2019BAAS...51g..45E}), MSE (\citealt{2019arXiv190303158P}),  MegaMapper (\citealt{2019BAAS...51g.229S}),  and MUST\footnote{MUltiplexed
Survey Telescope (MUST): \url{https://must.astro.tsinghua.edu.cn/must/en/index.html}}], will provide high-precision map of cosmic large-scale structure (LSS), which encodes a wealth of valuable cosmological information about our Universe. Efficient extractions of these critical information will help us greatly deepen our understanding of many key fundamental questions in cosmology (e.g., dark energy properties, neutrino masses, gravity and inflation models, etc.). This necessitates the development of powerful statistical tools to characterize or quantify LSS properties from various angles.

Two-point statistics (i.e., two-point correlation function and power spectrum) have played crucial roles in the analyses of LSS data, especially for studies on baryonic acoustic oscillations (e.g., \citealt{2005ApJ...633..560E}; \citealt{2017MNRAS.470.2617A}; \citealt{2021MNRAS.500..736B}; \citealt{2023MNRAS.525.5406M}). These statistics characterize the clustering properties of LSS, but can only give a complete description for Gaussian random field. In reality, LSS has evolved into a highly non-Gaussian field in the present-day Universe, which makes them can not capture appreciable non-Gaussian information on small scales, thus requiring measurements of an infinite hierarchy of $N$-point statistics (i.e., $N$-point correlation functions and polyspectra). At present, accurate measuring and theoretical modelling higher-order $N$-point statistics are challenging\footnote{Still, some specific progresses have been made recently in this direction (see \citealt{2021arXiv210610278P}; \citealt{2021arXiv210801714H}; \citealt{2022MNRAS.509.2457P} and Refs. therein).}, due to the complexities in all possible combinations of multiplets. Moreover, even if the first $N$-order information is extracted, other interesting information may still remain in higher-order terms.

Consequently, various summary statistics for non-Gaussian information have been proposed as potential supplements to two-point statistics, e.g., count in cells (\citealt{1991ApJ...369..273D}; \citealt{2020MNRAS.495.4006U}; \citealt{2020MNRAS.498L.125R}), void statistics (\citealt{2014PhRvD..90j3521C}; \citealt{2016MNRAS.459.2670Z}; \citealt{2019MNRAS.488.4413K}), peak statistics (\citealt{2012PhRvD..85b3011G}; \citealt{2014MNRAS.442.2534S}; \citealt{2015PhRvD..91f3507L}; \citealt{2015MNRAS.450.2888L}), Voronoi statistics (\citealt{2020MNRAS.495.3233P}; \citealt{2021PhRvD.103j3522J}), and even scattering transform (\citealt{2020MNRAS.499.5902C}; \citealt{2021MNRAS.507.1012C}), etc. In particular, morphological statistical methods [e.g., $\ensuremath{\beta}$-Skeleton (\citealt{2019MNRAS.485.5276F}; \citealt{2021ApJ...922..204S}), Betti numbers (\citealt{2017MNRAS.465.4281P}; \citealt{2019MNRAS.485.4167P}; \citealt{2021MNRAS.505.1863G}), persistent topology (\citealt{2019MNRAS.486.1523E, 2023MNRAS.520.2709E}; \citealt{2021MNRAS.507.2968W}; \citealt{2023arXiv231113520J}; \citealt{2024MNRAS.529.4325B}; \citealt{2024arXiv240313985Y}), minimal spanning tree (\citealt{1985MNRAS.216...17B}; \citealt{2021arXiv211112088N}), cosmic web skeleton (\citealt{2006MNRAS.366.1201N}; \citealt{2008MNRAS.383.1655S}), wavelet analyses (\citealt{1993MNRAS.260..365M}; \citealt{2012A&A...542A..34A}), shape statistics (\citealt{1998ApJ...495L...5S}; \citealt{2003MNRAS.344..602B}), genus statistics (\citealt{1986ApJ...306..341G}; \citealt{1986ApJ...309....1H}; \citealt{2021ApJ...907...75A}), Minkowski tensors (\citealt{2018ApJ...858...87A}; \citealt{2018ApJ...863..200A}), Minkowski functionals (\citealt{1994A&A...288..697M}; \citealt{2020PhRvD.101f3515L}; \citealt{2022PhRvD.105b3527M}), etc.] are actively employed to characterize the geometrical and topological properties of LSS, providing alternative ways to capture higher-order information directly, in complementary to traditional $N$-point formalism.  

As a conceptual generalization of genus (\citealt{1994A&A...288..697M}; \citealt{1997ApJ...482L...1S}), Minkowski functionals (MFs) can elegantly describe the global characterizations of morphological properties of LSS, and have been employed in various cosmological studies [e.g., detecting primordial non-Gaussianities (\citealt{2006ApJ...653...11H}; \citealt{2008MNRAS.385.1613H}; \citealt{2016A&A...594A..17P}; \citealt{2013MNRAS.429.2104D}), serving as standard ruler (\citealt{2010ApJ...715L.185P}; \citealt{2011MNRAS.412.1401Z}; \citealt{2014MNRAS.437.2488B}), constraining cosmological parameters (\citealt{2018ApJ...853...17A}; \citealt{2021ApJ...907...75A}; \citealt{2022ApJ...928..108A}), probing neutrino masses (\citealt{2019JCAP...06..019M}; \citealt{2020PhRvD.101f3515L}; \citealt{2022arXiv220402945L}),  analyzing effects of redshift space distortions (\citealt{2013MNRAS.435..531C}; \citealt{2021arXiv210803851J}), testing cosmologies and gravities (\citealt{2003PASJ...55..911H}; \citealt{2012ApJ...747...48W}; \citealt{2017PhRvL.118r1301F}), etc.]. One popular method [other method can also be found in literatures, i.e., the germ-grain approach (cf.\ \citealt{1994A&A...288..697M}; \citealt{2014MNRAS.443..241W})] for measuring MFs is the iso-density approach, which we will focus on in our study (see Section \ref{sec:MFs}). The MFs measured in this way can be well theoretically modelled for Gaussian (\citealt{9789814368223_0003}; \citealt{1997ApJ...482L...1S}; \citealt{2003ApJ...584....1M}) and weakly non-Gaussian (\citealt{1994ApJ...434L..43M}; \citealt{2003ApJ...584....1M}; \citealt{2009PhRvD..80h1301P}; \citealt{2010PhRvD..81h3505M};  \citealt{2012PhRvD..85b3011G}; \citealt{2021PhRvD.104j3522M}; \citealt{2022PhRvD.105b3527M}) fields [even for the fields with RSDs (\citealt{1996ApJ...457...13M}; \citealt{2013MNRAS.435..531C})], showing distinct advantages over the germ-grain approach.

The iso-density approach estimates four MFs with a series of excursion sets, which are specified by a series of iso-density contours of LSS. Therefore, this method relies heavily on the reconstruction of underlying continuous density field from discrete point tracers. In realistic applications, the number densities of tracers (i.e., halos/galaxies) are low, thereby inducing significant shot noises (i.e., discreteness effects) in the tracer fields constructed by commonly used mass assignment methods [i.e., Nearest Grid Point (NGP), Cloud-in-Cell (CIC), Triangular-Shaped Cloud (TSC), etc.]. As a result, this renders the tracer fields to be exceedingly discontinuous. To alleviate this problem, smoothing methods with fixed kernel widths, at least larger than the mean tracer spacing, are commonly employed in previous studies. These methods can help eliminate noise components and produce nicely smoothed continuous fields. However, meanwhile, the recipes of fixed smoothing will also erode the texture of underlying density distribution, i.e., discarding substantial structural information below the scales of kernel widths, consequently downgrading the statistical power of MFs in various cosmological studies (cf.\ \citealt{2010ApJ...722..812Z}; \citealt{2023arXiv230504520J}).

In this study, we opt for Delaunay Tessellation Field Estimator (DTFE)\footnote{Delaunay Tessellation Field Estimator (DTFE): \url{https://github.com/MariusCautun/DTFE}} (\citealt{2000A&A...363L..29S}; \citealt{2009LNP...665..291V}; \citealt{2009arXiv0912.3448V}; \citealt{2011arXiv1105.0370C}; \citealt{2011ascl.Soft05003c}) to address this issue\footnote{Previously, \citealt{2005ApJ...634..744M} attempted to use wavelet-denoising method to alleviate the problem induced by fixed-kernel smoothing. Additionally, \citealt{2010ApJ...722..812Z} endeavored to boost topological information of genus statistics with DTFE.}. This method can help to recover the largest number of structural elements from point sets, allowing for extracting maximum amount of morphological information with MFs. Therefore, the statistical power of MFs with DTFE (hereafter DTFE MFs) can be significantly enhanced, over the fixed smoothing methods (\citealt{2010ApJ...722..812Z}). We first demonstrate this at matter field level and further extend it to halo field level. In particular, for DTFE MFs of halo fields, we numerically explore various main systematic effects, caused by finite voxel size, halo number density, halo-weighting scheme, and RSDs, on their measurements, and showcase their strong discriminative and constraining power in RSD studies. Hopefully, the applications of DTFE will dramatically improve the performance of MFs in extracting various key cosmological information (e.g., neutrino masses (\citealt{2020PhRvD.101f3515L}), modified gravities (\citealt{2017PhRvL.118r1301F}), primordial non-Gaussianities (\citealt{2006ApJ...653...11H}; \citealt{2008MNRAS.385.1613H}), and cosmological parameters (\citealt{2020ApJ...896..145A}; \citealt{2022ApJ...928..108A}), etc.) from sparse tracers.

Historically, due to its exceptional characteristics, DTFE has been applied in various advanced pipelines for characterizing, identifying, and classifying structures of LSS. Examples include MMF (\citealt{2007A&A...474..315A}) and NEXUS (\citealt{2013MNRAS.429.1286C}), which leverage multi-scale geometry of structural components, and SpineWeb (\citealt{2010ApJ...723..364A}) and DisPerSE (\citealt{2011MNRAS.414..350S}; \citealt{2011MNRAS.414..384S}), which utilize topology of cosmic web via Morse theory (\citealt{morse1934calculus}; \citealt{milnor1963morse}; \citealt{jost2008riemannian}). Also, DTFE has been employed in WVF (\citealt{2007MNRAS.380..551P}), the first watershed-based void finder that can identify cosmic voids irrespective of sizes and shapes. Subsequently, \citealt{2008MNRAS.386.2101N} proposed a closely related technique, ZOBOV, based on Voronoi Tessellation Field Estimator (VTFE) (cf.\ \citealt{2007PhDT.......486S}). Thereafter, later watershed-based void finding pipelines, such as VIDE (\citealt{2015A&C.....9....1S}), REVOLVER (\citealt{2019PhRvD.100b3504N}), and V$^2$ (\citealt{2022JOSS....7.4033D}; \citealt{2023ApJS..265....7D}), are built upon ZOBOV. In addition, DTFE can construct not only continuous density fields but also volume-covering velocity fields and their corresponding divergences, shears, and vorticities (\citealt{1996MNRAS.279..693B}; \citealt{2007MNRAS.382....2R}; \citealt{2009LNP...665..291V}).

This paper is structured as follows. In Section \ref{sec:MFs}, we give a brief overview for MFs. In Section \ref{sec:DTFE}, we describe the basic DTFE algorithm for generating continuous fields from point sets. In Section \ref{sec:N-body}, we introduce in detail the data used in this work. In Section \ref{sec:DM MFs}, we preliminarily demonstrate the advantages of DTFE in measuring MFs from sparse point set, at matter field level. Then, at halo field level, we investigate various main systematic effects on DTFE MFs in Section \ref{sec:Systematic Effects} and show strong statistical power of DTFE MFs in extracting cosmological information in Section \ref{sec:DTFE power}. Finally, we give summaries and discussions in Section \ref{sec:Summary}. Appendix \ref{app:Smoothing Lengths} shows the details of our strategy to determine the smoothing lengths used in Gaussian smoothing method. Appendix \ref{app:RSD Signals} displays the RSD signals and the associated RSD signal-to-noise (S/N) ratios extracted from DTFE MFs. Appendix \ref{app:Fisher Forecast} presents the technical details of Fisher forecasts employed in our study.

\section{Minkowski Functionals} \label{sec:MFs} 
MFs are a family of morphological (i.e., geometrical and topological) descriptors with properties of additivity, motion invariance, and conditional continuity for any manifold in $D$-dimensional space. These descriptors are originally derived from the theory of convex bodies and integral geometry, and were first introduced into cosmology by Ref.\ \citealt{1994A&A...288..697M} to characterize the morphology (i.e., geometry and topology) of cosmic web. According to Hadwiger's theorem (\citealt{Hadwiger1957Vorlesungen}), the topo-geometrical properties of any given manifold in $D$-dimensional space can be completely characterized by $D+1$ MFs. Thus, it opens a unique way to comprehensively access all orders of N-point correlation information at once.

In LSS studies, the manifolds ($\mathbb{M}$) of typical interests are excursion sets ($E_{\nu}$) of $3D$ cosmological scalar fields (i.e., dark matter fields or halo/galaxy fields), 
\begin{equation}
E_{\nu}=\{\mathbf{x} \in \mathbb{M}: \nu(\mathbf{x}) \geq \nu\},
\end{equation}
where $E_{\nu}$ is the set of all points $\mathbf{x}$ with density $\nu(\mathbf{x})\geq \nu$ and $\nu$ is the density threshold serving as diagnostic parameter for displaying morphological features. Four MFs quantify the enclosed volume ($V_0$) of $E_{\nu}$, as well as the surface's area ($V_1$), integrated mean curvature ($V_2$), and Euler characteristic ($V_3$) of the set boundary $\partial E_{\nu}$ (i.e., the iso-density surface),
\begin{equation}
\begin{aligned}
    &V_{0}(\nu)=\frac{1}{|\mathscr{D}|}\int_{E_{\nu}}d^3x,\\
    &V_{1}(\nu)=\frac{1}{6|\mathscr{D}|}\int_{\partial E_{\nu}}dS(\mathbf{x}),\\
    &V_{2}(\nu)=\frac{1}{6\pi|\mathscr{D}|}\int_{\partial E_{\nu}}\left(\frac{1}{R_1(\mathbf{x})}+\frac{1}{R_2(\mathbf{x})}\right)dS(\mathbf{x}),\\
    &V_{3}(\nu)=\frac{1}{4\pi|\mathscr{D}|}\int_{\partial E_{\nu}}\frac{1}{R_1(\mathbf{x})R_2(\mathbf{x})}dS(\mathbf{x}),
\end{aligned}
\end{equation}
where $R_1(\mathbf{x})$ and $R_2(\mathbf{x})$ are two principal radii of curvature of the set's surface orientated toward lower-density regions. These quantifiers provide complete morphological description, including size (i.e., $V_0$ and $V_1$), shape (i.e., $V_2$), and connectivity (i.e., $V_3$) of the excursion sets. In particular, according to Gauss-Bonnet theorem, the last MF (i.e., Euler characteristic) is simply related to the number of isolated regions (balls), empty regions inside balls (bubbles), and holes in ball surfaces (tunnels) per unit volume,
\begin{equation}
V_3=\frac{1}{|\mathscr{D}|}(N_{\mathrm{ball}}+N_{\mathrm{bubble}}-N_{\mathrm{tunnel}}),
\end{equation}
having direct relation with genus ($G=1-V_3$), which is the first topological descriptor (e.g., \citealt{1986ApJ...306..341G}) widely used in cosmic web's topological analyses. 

Two standard grid-based numerical algorithms to compute MFs from regularly gridded density field\footnote{In literatures, several novel triangulation-based algorithms (cf.\ \citealt{2003MNRAS.343...22S}; \citealt{2010arXiv1006.4178A}; \citealt{2021MNRAS.508.3771L}), which estimate MFs by using triangulated iso-density surfaces, have also been proposed in succession, aiming to improve the accuracies of MFs' estimations.}, i.e., Koenderink invariant from differential geometry (\citealt{schneider_1993}) and Crofton's formula from integral geometry (\citealt{Hadwiger1957Vorlesungen}; \citealt{https://doi.org/10.1002/zamm.19790590633}), have been developed in literatures (\citealt{1997ApJ...482L...1S}). In this work, we employ Crofton's formula to quote our results, because it is more stable and is the most commonly used method in previous works. All MFs are measured as functions of density threshold $\nu\equiv 1+\delta$ with $\nu\in \left[0.003, 1000\right]$, where $\delta$ is the density contrast. The error bars are estimated via the standard errors of MFs, i.e., $s_\mathrm{e}=\sigma/\sqrt{n_\mathrm{f}}$. Here, $\sigma$ is the standard deviation of MFs, calculated from $n_\mathrm{f}$ subfields obtained by equally subdividing the original field. For DM and halo fields, the numbers of subfields are $n_\mathrm{f}=4^3=64$ and $n_\mathrm{f}=8^3=512$, respectively. Since the error bars are too tiny to be visible in most scenarios, they are omitted in all figures in this paper, except for Fig.\ \ref{fig:Gau_len} and Fig.\ \ref{fig:RSD_SNR}. Note that all MFs are visualized by logarithmic x-axis, considering that probability distribution function of LSS roughly obeys log-normal form at low redshift. 

\section{Delaunay Tessellation Field Estimation} \label{sec:DTFE}
Given the positions of a set of points $\mathbf{x}_{i}$ (generators) with weights $w_i$ ($i=1, 2, 3 \dots N$) in $D$-dimensional space, the first step of DTFE procedure is to self-adaptively tessellate the space into a union of space-filling and mutually disjoint Delaunay cells [i.e., simplexes, which are triangles (tetrahedra) in $2D$ ($3D$) space] using the Delaunay tessellation technique, which imposes the circumsphere of each tetrahedron does not contain any generators. Under the assumption of uniform sampling (i.e., the point set is an unbiased sample of the underlying density field), the estimated density at each generator $\mathbf{x}_{i}$ is determined by the normalized inverse of the volume of its contiguous Voronoi cell $V\left(\mathcal{W}_{i}\right)$ (see \citealt{2009LNP...665..291V} for more technical details),
\begin{equation} \label{eq:1}
\widehat{\rho}\left(\mathbf{x}_{i}\right)=(1+D) \frac{w_{i}}{V\left(\mathcal{W}_{i}\right)},
\end{equation}
where the contiguous Voronoi cell $\mathcal{W}_{i}$ is the union of all adjacent Delaunay cells $\mathcal{D}_{\mathbf{x}_{i}}^{\mathrm{adj}}$ with $\mathbf{x}_{i}$ as one of their $D+1$ vertices,
\begin{equation}
V\left(\mathcal{W}_{i}\right)=\sum{V\left(\mathcal{D}_{\mathbf{x}_{i}}^{\mathrm{adj}}\right)}.
\end{equation}
In real survey data (i.e, galaxy samples), the point set is actually always modulated by specified selection process (i.e., systematic non-uniform sampling), which can be quantified by a priori selection function $\psi\left(\mathbf{x}_{i}\right)$ varying with sky position and redshift. In this scenario, the equation (\ref{eq:1}) will be generalized to be
\begin{equation}
\widehat{\rho}\left(\mathbf{x}_{i}\right)=(1+D) \frac{w_{i}}{\psi\left(\mathbf{x}_{i}\right) V\left(\mathcal{W}_{i}\right)}.
\end{equation}

To obtain a continuous field, the density values inside a Delaunay cell, at position $\mathbf{x}=\left(x^1, x^2, x^3 \dots  x^D \right)$, then are estimated by multi-dimensional linear interpolation from the $D+1$ density values of cell's vertices $\widehat{\rho}\left(\mathbf{x}_{n}\right)$ ($n=0,1,2,3 \dots D$) 
\footnote{Here, the subscript $n$ seems to conflict with the previous subscript $i$. To clarify this confusion for careful readers, we'd like to note that the subscript $n$ used here is only reserved for labelling one of the $D+1$ vertices of a certain Delaunay cell.},
\begin{equation}
\widehat{\rho}(\mathbf{x})=\widehat{\rho}\left(\mathbf{x}_{0}\right)+\widehat{\nabla \rho} \cdot\left(\mathbf{x}-\mathbf{x}_{0}\right).
\end{equation}
Here, $\widehat{\nabla \rho}$ is the linear constant gradient inside the cell, which can be calculated by
\begin{equation}
\widehat{\nabla \rho}=\left(\begin{array}{c}
\frac{\partial \widehat{\rho}}{\partial x^1} \\
\frac{\partial \widehat{\rho}}{\partial x^2} \\
\frac{\partial \widehat{\rho}}{\partial x^3} \\
\vdots \\
\frac{\partial \widehat{\rho}}{\partial x^D} 
\end{array}\right)^T=\mathbf{J}^{-1}\left(\begin{array}{c}
\Delta \widehat{\rho}_{1} \\
\Delta \widehat{\rho}_{2} \\
\Delta \widehat{\rho}_{3} \\
\vdots \\
\Delta \widehat{\rho}_{D}
\end{array}\right) ;
\end{equation}
\begin{equation}
\mathbf{J}=\left(\begin{array}{ccccc}
\Delta x^1_{1} & \Delta x^2_{1} & \Delta x^3_{1}  & \cdots  & \Delta x^D_{1} \\
\Delta x^1_{2} & \Delta x^2_{2} & \Delta x^3_{2} & \cdots & \Delta x^D_{2} \\
\Delta x^1_{3} & \Delta x^2_{3} & \Delta x^3_{3} & \cdots & \Delta x^D_{3} \\
\vdots & \vdots & \vdots & \ddots  &  \vdots \\
\Delta x^1_{D} & \Delta x^2_{D} & \Delta x^3_{D} & \cdots & \Delta x^D_{D}
\end{array}\right),
\end{equation}
where $\Delta \widehat{\rho}_{n} = \widehat{\rho}\left(\mathbf{x}_n\right)-\widehat{\rho}\left(\mathbf{x}_0\right)$ and $\Delta x^j_{n} = x^j_{n} - x^j_{0}$, for $j=1, 2, 3 \dots D$ as well as $n=1, 2, 3 \dots D$. Employing this interpolation scheme for each Delaunay cell, then we can straightforwardly produce a continuous space-filling field (i.e., DTFE density field) on a regular grid, which exploits the same anisotropic and self-adaptive scaling features of Delaunay tessellation and guarantees mass conservation, i.e., its volume integral can reproduce the total mass, 
\begin{equation}
\widehat{W}=\int \widehat{\rho}(\mathbf{x}) \mathrm{d} \mathbf{x}=\sum_{i=1}^{N} w_{i}=W=\mathrm{cst.}.
\end{equation}

DTFE in essence is a first-order version of natural neighbour interpolation procedure. Nevertheless, its self-adaptive nature enables the automatic capture of subtle structural elements in high-density regions with maximum possible resolutions. Meanwhile, it can also properly smooth low-density regions to avoid irritating discreteness. Consequently, DTFE can sharply reconstruct conspicuous structures and hierarchical features of cosmic web from a spatial distribution of sparse tracers, which is of crucial importance to extract maximum amount of morphological information with MFs. In following sections, we will illustrate its compelling performance with quantitative results to highlight the virtues.

\section{Cosmological $N$-body Simulations and Data Samples}  \label{sec:N-body} 
\begin{figure*}
\centering 
\includegraphics[width=0.96\textwidth]{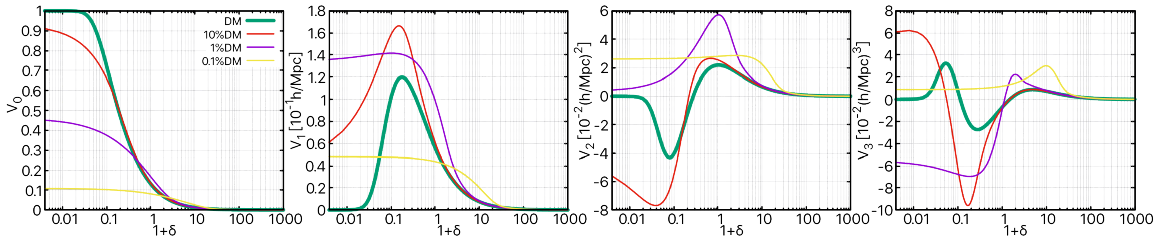}
\caption{The MFs of noise-free field and $10\%$, $1\%$, $0.1\%$ downgraded fields, which are constructed by CIC method with $N_\mathrm{g}=1024^3$ grid cells. They are represented by bold green lines and red, blue, yellow lines, respectively. Here, the bold green lines can be treated as the ideal LSS's MFs, without discreteness effects.  By comparing the lines with different colors, this plot demonstrates how discreteness affects the measurements of MFs.}
\label{fig:SN}
\end{figure*}

In this work, we adopt two high-resolution pure cold dark matter (DM) $N$-body simulations for different application requirements. One simulation is dubbed as WMAP$\_3072\_600$, which is from the CosmicGrowth simulation suite (\citealt{2019SCPMA..6219511J}), realized by running an adaptive parallel particle–particle–particle-mesh ($\mathrm{P}^3\mathrm{M}$) $N$-body code (\citealt{2002ApJ...574..538J}; \citealt{2007ApJ...657..664J}). This simulation incorporates $N_\mathrm{p}=3072^3$ DM particles with mass resolution of $5.5\times10^8\, h^{-1}M_\odot$ in a periodic cubic box with size of $L=600\,h^{-1}\mathrm{Mpc}$, adopting a WMAP cosmology, i.e.,  $[\Omega_\mathrm{c}$, $\Omega_\mathrm{b}$, $h$, $n_\mathrm{s}$, $\sigma_8]$ = $[0.2235$, $0.0445$, $0.71$, $0.968$, $0.83]$. The other one is called TianZero simulation (\citealt{2017RAA....17...85E}; \citealt{2017NatAs...1E.143Y}), realized by using a publicly-available $\mathrm{P}^3\mathrm{M}$ $N$-body code {\tt CUBEP3M} (\citealt{2013MNRAS.436..540H}). This simulation, parameterized with $[\Omega_\mathrm{c}$, $\Omega_\mathrm{b}$, $h$, $n_\mathrm{s}$, $\sigma_8]$ = $[0.27$, $0.05$, $0.67$, $0.96$, $0.83]$, evolves $N_\mathrm{p} = 6912^3$ DM particles with mass resolution of $4.6\times10^8\,h^{-1}M_\odot$ in a cubic box of width $L = 1200\,h^{-1}\mathrm{Mpc}$. Both simulations assume flat cosmology,  imposing $\Omega_{\Lambda}=1-\Omega_\mathrm{m}$, where $\Omega_\mathrm{m} = \Omega_\mathrm{b} + \Omega_\mathrm{c}$. 

The WMAP$\_3072\_600$ and TianZero are employed to perform DM and halo field level analyses, respectively. As for WMAP$\_3072\_600$, in addition to the complete sample of particles (at $z=0$) with number density of $\bar{n}_\mathrm{p}=1.34\times10^2\,(h^{-1}\mathrm{Mpc})^{-3}$, we also construct six ($10\%$, $1\%$, $0.1\%$, $0.01\%$, $0.001\%$, $0.0001\%$) downgraded particle subsamples produced by random-down-sampling processes without repetition, to study shot noise effects on MFs' measurements. In TianZero, halos are identified by using {\tt CUBEP3M}’s own on-the-fly spherical overdensity (SO) halo finder, which is set to resolve halo masses down to $2.3 \times 10^{11}\,h^{-1}M_{\odot}$ with a minimum of $500$ particles per halo. Similarly, to investigate the impacts of halo number densities on MFs' measurements, we construct three halo catalogues (at $z=0.01$) with number densities of $\bar{n}_\mathrm{h}=1.6 \times 10^{-2}\,(h^{-1} \mathrm{Mpc})^{-3}$, $\bar{n}_\mathrm{h}=1.6 \times 10^{-3}\,(h^{-1} \mathrm{Mpc})^{-3}$, and $\bar{n}_\mathrm{h}=1.6 \times 10^{-4}\,(h^{-1} \mathrm{Mpc})^{-3}$, by discarding halos with masses below mass cutoffs of $M_{\mathrm{min}} \simeq 2.3 \times 10^{11}\,h^{-1}M_\odot$, $M_{\mathrm{min}} \simeq 3.2 \times 10^{12}\,h^{-1}M_\odot$, and $M_{\mathrm{min}} \simeq 3.1 \times 10^{13}\,h^{-1}M_\odot$, respectively. The sample selection is chosen due to the consideration that galaxy samples in most observations are determined with faint flux limits (or low mass limits).  Note that throughout this paper, subhalos are excluded from analyses.

\section{The MFs of DM Fields} \label{sec:DM MFs} 
In this section, we strive to preliminarily demonstrate the superiorities of DTFE method in measuring MFs from sparse tracers. The analyses are performed at matter field level. 

\subsection{The shot noise effects on MFs} \label{sec:Shot Noise Effects}
Based on the full and $10\%$, $1\%$, $0.1\%$, $0.01\%$, $0.001\%$, $0.0001\%$ downgraded particle samples (cf.\ Section \ref{sec:N-body}), we construct seven DM fields in real space, by employing a representative mass assignment method, i.e., CIC interpolation. Each field is interpolated on $N_\mathrm{g} = 1024^3$ regular grid, with grid cell size $L_\mathrm{g} =L/\left(N_\mathrm{g}\right)^{1/3} \simeq 0.59 \, h^{-1}\mathrm{Mpc}$,
\begin{equation}
\delta(\mathbf{x}) \equiv \frac{n_{\text {p}}(\mathbf{x})}{\bar{n}_{\mathrm{p}}}-1,
\end{equation}
where $n_{\text {p}}$ is particle number density and $\bar{n}_{\text {p}}$ is its mean value. Due to the extremely high particle number density, the DM field constructed from full particle sample can be safely regarded as the underlying noise-free DM field (hereafter noise-free field). The downgraded samples are unbiased samples of the underlying density field, such that the downgraded DM fields (hereafter downgraded fields) can be used to study any pure shot noise effects on MFs. Then, we measure the MFs for each field (hereafter CIC MFs). For simplicity, the results exclusively for the noise-free field and $10\%$, $1\%$, $0.1\%$ downgraded fields are presented in Fig.\ \ref{fig:SN}. 

We find that the shapes of MFs are severely distorted by shot noises. The issue becomes increasingly conspicuous as the particle number density decreases. This is because particles successively become poorer tracers of the underlying DM field, making the downgraded fields choppier (cf.\ the top panels of Fig.\ \ref{fig:CIC_CIC+GS_DTFE}). Low-density regions are more vulnerable to the down-sampling process, since shot noises mostly affect these regions. In particular, in poorly sampled regions, with sparse or even no particles, density fields are severely discontinuous or even blank (i.e., $1+\delta = 0$) (cf.\ the top panels of Fig.\ \ref{fig:CIC_CIC+GS_DTFE}). As particle number density decreases, blank regions become larger, thereby the volume fraction of non-zero density regions becomes smaller (cf.\ the left panel of Fig.\ \ref{fig:SN} and top panels of Fig.\ \ref{fig:CIC_CIC+GS_DTFE}). Due to the existences of blank regions, the regions with $1+\delta > 0$ only account for a certain proportion of total box volume (cf.\ the left panel of Fig.\ \ref{fig:SN}). Thus, MFs have step changes\footnote{Note that the values of MFs at $1+\delta = 0$, which respectively are $V_0=1$, $V_1=0$, $V_2=0$, and $V_3=0$, cannot be plotted in Fig.\ \ref{fig:SN}, because of the logarithmic x-axis.}, when $1+\delta = 0 \Rightarrow 1+\delta > 0$. 

Moreover, shot noises give rise to various spurious structures on the excursion sets at different density thresholds, depending on down-sampling level. Specifically, for $10\%$ sampling, at low-density thresholds, more and larger isolated under-dense regions (bubbles) are produced by discrete-sampling process, such that $V_1$ and $V_3$ become larger and $V_2$ becomes smaller, relative to the case of noise-free field (the same below). At median density thresholds, we observe a lower negative minimum in $V_3$, indicating more porous structures (tunnels), produced by down sampling in the surfaces of excursion sets. As particle number density decreases, downgraded fields will be increasingly dominated by ``meatball topology'' (i.e., a preponderance of isolated high-density regions). Because of this, for the extreme case of $0.1\%$ sampling, $V_2$ and $V_3$ are always positive when $1+\delta< 1$. For the same case, we see that $V_2$ and $V_3$ have higher maximums at high-density thresholds, suggesting that the abundance of isolated regions is increased by breaking up structures into multiple objects as particles are taken out. For the intermediate case of $1\%$ sampling, the corresponding results naturally fall somewhere in between those of the former two cases. 

In conclusion, the discreteness effects pose challenges in accurately delineating iso-density contours (i.e., the excursion sets), thereby hindering the proper reflections of intrinsic morphological properties of particle-traced LSS with MFs.
Furthermore, in Appendix \ref{app:Smoothing Lengths}, interested readers can find additional results illustrating shot noise effects on MFs with Gaussian smoothing (cf.\ \citealt{2014ApJS..212...22K}; \citealt{2017ApJ...836...45A} for genus scenarios).

\subsection{The MFs with Gaussian smoothing} \label{sec:DM CIC+GS MFs}
\begin{figure*}
\centering 
\includegraphics[width=0.853\textwidth]{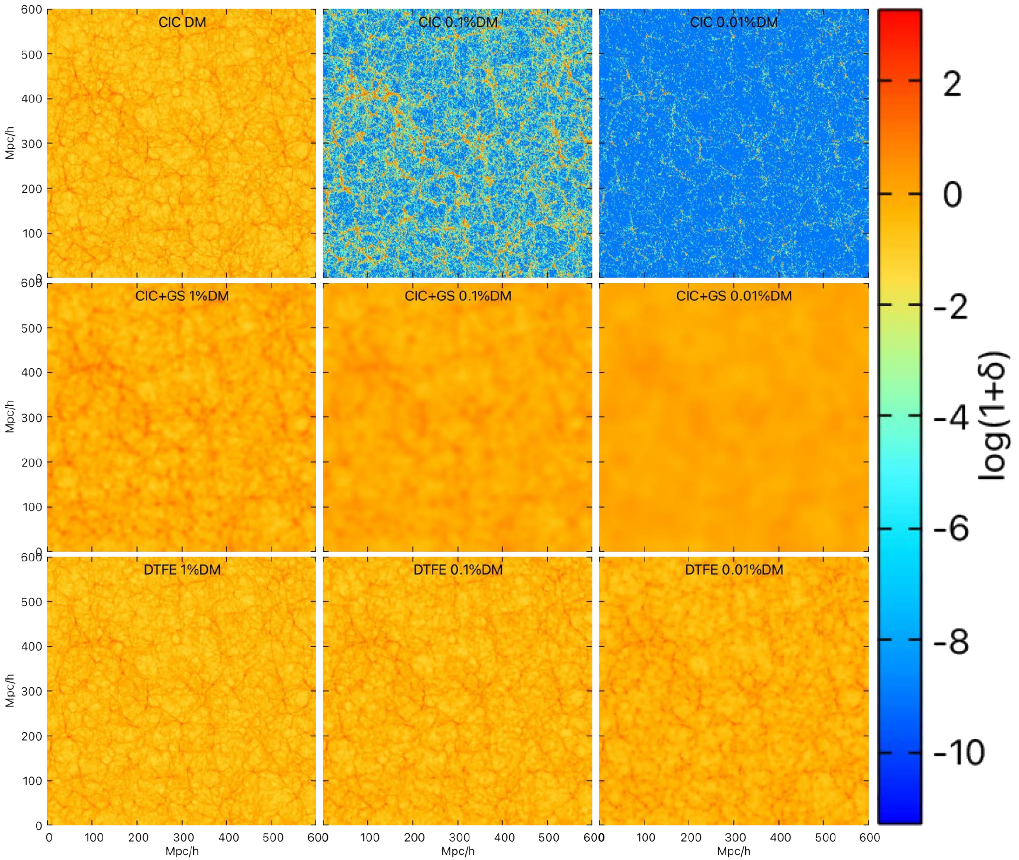}
\caption{The visualization of two-dimensional slices of downgraded fields constructed by CIC (top),  CIC+GS (middle), and DTFE (bottom) methods. As a reference, the noise-free field constructed by CIC method is also presented in the left-top panel. Except for this panel, the left, middle, and right panels represent the $1\%$, $0.1\%$, and $0.01\%$ downgraded fields, respectively. All these fields are created by using $N_\mathrm{g}=1024^3$ grid cells within a box of size $L=600\,h^{-1}\mathrm{Mpc}$.  Each two-dimensional slice displays the same region of simulation box, with the same projection depth of $5.27\,h^{-1}\mathrm{Mpc}$. For CIC+GS case, the smoothing lengths of $R_\mathrm{G}=2.93\,h^{-1}\mathrm{Mpc}$, $R_\mathrm{G}=5.86\,h^{-1}\mathrm{Mpc}$, and $R_\mathrm{G}=11.72\,h^{-1}\mathrm{Mpc}$ are adopted for the $1\%$, $0.1\%$, and $0.01\%$ downgraded fields, respectively (cf.\ Appendix \ref{app:Smoothing Lengths}). In comparison with the left-top panel, the top panels display the emergence of discreteness in CIC DM fields as particle number density is successively downgraded. These severe discreteness effects will hamper proper morphological measurements of LSS with MFs (cf.\ Fig.\ \ref{fig:SN}). The middle panels display the continuous downgraded fields obtained by using CIC+GS method. This method inevitably destroys large amounts of intrinsic structural elements of LSS by introducing excessive smoothing. Therefore, it will significantly diminish the morphological information contents captured by MFs. The bottom panels show the continuous downgraded fields constructed by using DTFE method. Benefiting from the anisotropic and spatial self-adaptive feature, DTFE can help introduce minimum smoothing and retain intrinsic structural elements of LSS to the maximum extent possible.  Compared with CIC+GS method, DTFE can greatly improve the morphological information contents captured by MFs (cf.\ Fig.\ \ref{fig:CIC+GS_DTFE_MFs}).}
\label{fig:CIC_CIC+GS_DTFE}
\end{figure*}

\begin{figure*}
\centering 
\includegraphics[width=0.69\textwidth]{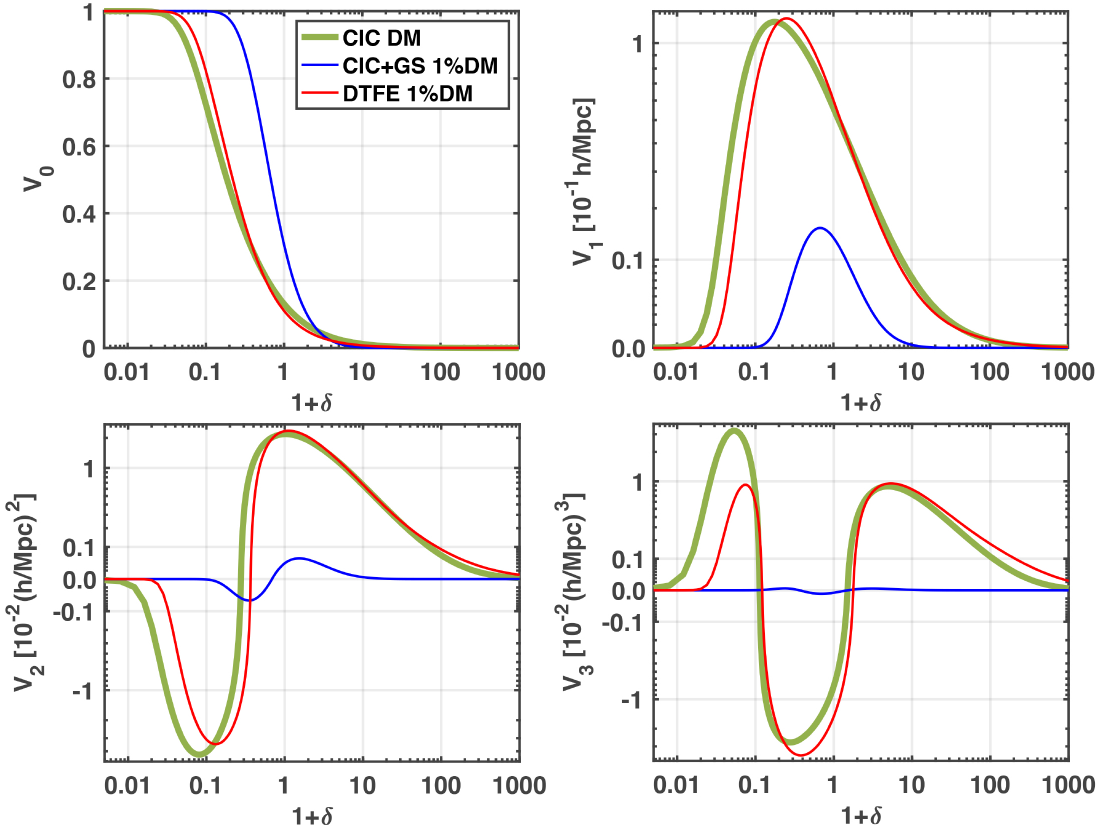}
\caption{The MFs of noise-free field and two $1\%$ downgraded fields, where the noise-free field is constructed by CIC method, and the two $1\%$ downgraded fields are constructed by CIC+GS and DTFE methods, respectively (cf.\ Fig.\ \ref{fig:CIC_CIC+GS_DTFE}). For CIC+GS case, the smoothing length adopted is $R_\mathrm{G}=2.93\,h^{-1}\mathrm{Mpc}$. As we can see, the amplitudes of DTFE MFs of $1\%$ downgraded field (i.e., the red lines) are very close to those of CIC MFs of noise-free field (i.e., the bold green lines; also cf.\ Fig.\ \ref{fig:SN}). It means that DTFE indeed can help retain intrinsic structural elements of LSS. While, the amplitudes of CIC+GS MFs (i.e., the blue lines) are much lower than those of MFs in other two cases. This is because CIC+GS scheme erases large amounts of intrinsic structural elements of LSS. Note that these results are plotted with logarithmic y-axis to better visualize the MFs in CIC+GS case.}
\label{fig:CIC+GS_DTFE_MFs}
\end{figure*}

Gaussian smoothing with a fixed-kernel size is the most commonly used method to tackle the issue of shot noises (e.g., \citealt{2005ApJ...633....1P}; \citealt{2010ApJS..190..181C}; \citealt{2014ApJ...796...86P}; \citealt{2014ApJS..212...22K}). In this subsection, to produce continuous fields, we smooth the $1\%$, $0.1\%$, $0.01\%$ downgraded fields with Gaussian window functions
\begin{equation}
W(\mathbf{r})=\frac{1}{(2\pi)^{3 / 2} R_{\mathrm{G}}^{3}} \exp \left(-\frac{|\mathbf{r}|^{2}}{2 R_{\mathrm{G}}^{2}}\right)
\end{equation} 
of smoothing lengths $R_{\mathrm{G}}=2.93\,h^{-1}\mathrm{Mpc}$, $R_{\text {G}}=5.86\,h^{-1}\mathrm{Mpc}$, $R_{\mathrm{G}}=11.72\,h^{-1}\mathrm{Mpc}$, respectively (cf.\ the middle panels of Fig.\ \ref{fig:CIC_CIC+GS_DTFE}). Then, we measure the corresponding MFs of these smoothed fields. Hereafter, we refer the MFs measured with this scheme as CIC+GS MFs. The smoothing lengths adopted here can provide enough smoothing for adequately suppressing discreteness effects without discarding too much structural information (cf.\ Appendix \ref{app:Smoothing Lengths} for the details of $R_{\mathrm{G}}$ determinations). If smoothing length is too small (e.g., $R_{\mathrm{G}} < \bar{d} \sim \bar{n}_\mathrm{p}^{-1/3}$, where $\bar{d}$ is mean tracer spacing), the algorithm may trend to pick out isolated high-density regions (\citealt{1987ApJ...319....1G}; \citealt{1989ApJ...340..625G}), leading to the so-called `meatball shift' (cf.\ \citealt{2010ApJ...722..812Z}). Actually, the free parameter of $R_{\mathrm{G}}$ is always empirically specified with the requirement of $R_{\mathrm{G}} \ge \bar{d}$. Therefore, the determination of a smoothed field is not unique.

The results for $1\%$ downgraded field are presented in Fig.\ \ref{fig:CIC+GS_DTFE_MFs}, alongside the CIC MFs of noise-free field serving as references. It seems that CIC+GS method to some degree recovers the shapes of MFs of noise-free field, but showing much lower amplitudes for $V_1$, $V_2$, and $V_3$. The amplitude suppression depends on the adopted smoothing length, i.e., larger smoothing length leading to severer amplitude suppression (cf.\ Fig.\ \ref{fig:Gau_len}). Moreover, after smoothing, MF curves are evidently squeezed towards the direction of intermediate density thresholds (i.e., $1 + \delta \sim 1$), indicating that the smoothed fields become more uniform. This is because the abundance of structural elements is substantially reduced by the smoothing process, significantly diluting intrinsic structures of LSS, especially for strongly clustered regions (cf.\ the middle panels of Fig.\ \ref{fig:CIC_CIC+GS_DTFE}). On the other hand, the MFs of smoothed fields approach those of Gaussian random fields, such that they may describe more properties of the Gaussian kernels, than the real morphology of cosmic web (cf.\ \citealt{2005ApJ...634..744M}). Predictably, the recipe of CIC+GS will downgrade the discriminative and constraining powers of MFs in various LSS studies [e.g., neutrino mass effects (\citealt{2020PhRvD.101f3515L}), semi-analytic models of galaxy formation (\citealt{2010ApJ...722..812Z}), etc.], thus limiting its applications.

\begin{figure*}
\centering 
\includegraphics[width=1.0\textwidth]{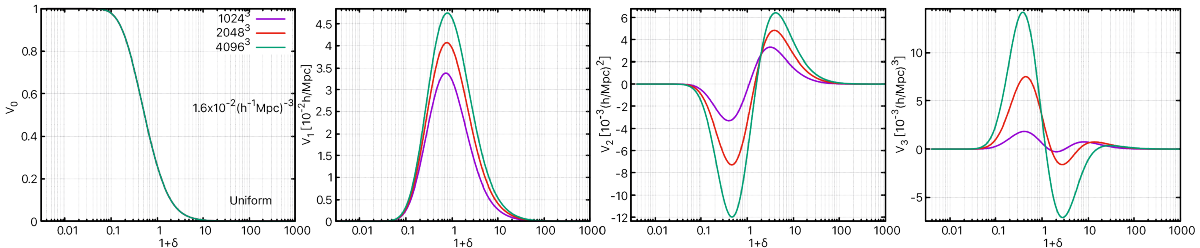}
\caption{The DTFE MFs of uniform-weighted halo fields with various grid resolutions in real space. The corresponding halo number density for these fields is $1.6 \times 10^{-2}\,(h^{-1} \mathrm{Mpc})^{-3}$. The blue,  red, and green lines correspond to halo fields with $N_\mathrm{g}=1024^3$, $N_\mathrm{g}=2048^3$, and $N_\mathrm{g}=4096^3$ grid cells, respectively. As shown, the amplitudes of $V_1$, $V_2$, and $V_3$ become greater as $N_\mathrm{g}$ increases. This means that under DTFE scheme, higher grid resolution can help resolve more subtle structural elements from point set.}
\label{fig:grid_res}
\end{figure*}

\subsection{The MFs with DTFE} \label{sec:DM DTFE MFs} 
The goal of our efforts is trying to maximize the extractions of true morphological properties with MFs from a set of points. This necessitates an optimal method for reconstructing continuous fields, such that their morphology can faithfully mirror the authentic underlying morphology of the point sets as well as possible. As LSS exhibits a multiscale nature, the smoothing scheme of this method should possess characteristics of anisotropy and spatial self-adaptivity. In fact, these characteristics align precisely with the iconic features of the DTFE method advocated in this paper. Additionally, DTFE also has the virtue of mass conservation and is not reliant on any priori parameters, ensuring the uniqueness of the constructed continuous fields (cf.\ Section \ref{sec:DTFE}). In this subsection, we explore how DTFE method contributes to recovering the morphology of noise-free field from downgraded particle samples.

We construct three DTFE fields with $N_\mathrm{g}=1024^3$ regular grid cells in real space (cf.\ the bottom panels of Fig.\ \ref{fig:CIC_CIC+GS_DTFE}), using the same downgraded samples as employed in Section \ref{sec:DM CIC+GS MFs},
\begin{equation}
\delta_{\mathrm{m}}(\mathbf{x}) \equiv \frac{\widehat{\rho}_{\mathrm{p}}(\mathbf{x})}{\bar{\widehat{\rho}}_{\mathrm{p}}}-1,
\end{equation}
and then directly measure their corresponding MFs. For comparisons, the results for $1\%$ downgraded field are also presented in Fig.\ \ref{fig:CIC+GS_DTFE_MFs}. As shown, the shapes of DTFE MFs exhibit noticeable non-Gaussian features, unfolded through the asymmetries of the MF curves. In contrast to CIC+GS MFs, they more closely resemble the MFs of the noise-free field, with much higher amplitudes for $V_1$, $V_2$, and $V_3$. This improvement is attributed to DTFE's capability to resolve a larger number of structural elements from point distribution (cf.\ the bottom panels of Fig.\ \ref{fig:CIC_CIC+GS_DTFE}). Nevertheless, these amplitudes are comparatively smaller than those of CIC MFs of the noise-free field, particularly at low-density thresholds, due to the inevitable losses of information\footnote{Note that the amplitudes of DTFE MFs will be further suppressed for the cases of downgraded fields with lower particle number densities, resulting from information losses (cf.\ the bottom panels of Fig.\ \ref{fig:CIC_CIC+GS_DTFE} and Section \ref{sec:hnd} for the scenarios of halo fields)}. On the other hand, as depicted in Fig.\ \ref{fig:CIC+GS_DTFE_MFs}, DTFE MFs instead exhibit slightly higher amplitudes at high-density thresholds. This is probably because DTFE can resolve more structural elements in high-density regions, where the impacts of shot noises are less pronounced, than the traditional CIC method. To conclude, these results suggest the great advantages of DTFE method in MFs' measurements, i.e., making morphological information more accessible and ultimately providing strong statistical power (cf.\ Section \ref{sec:DTFE power}).

\section{Systematic Effects on DTFE MFs of Halo Fields} \label{sec:Systematic Effects} 
In realistic data analyses, the measurements of MFs are inevitably affected by many systematic effects. These effects can introduce diverse modifications to the shapes and amplitudes of MFs and should be accurately taken into account to draw any sensible conclusions. Hence, it is theoretically interesting to investigate these systematic effects both analytically and numerically. Previous related works mainly focus on genus statistics (e.g., \citealt{2005ApJ...633....1P}; \citealt{2014ApJS..212...22K}; \citealt{2017ApJ...836...45A}; \citealt{1988ApJ...328...50M}; \citealt{1991ApJ...378..457P}; \citealt{2012ApJ...751...40J}; \citealt{1993ApJS...86....1M}). In particular, finite pixel (voxel) size effects were analytically studied for 2- (\citealt{1989ApJ...345..618M}) and 3-dimensional genus (\citealt{1986ApJ...309....1H}); \citealt{1996ApJ...460...51M} analytically investigated the RSD effects in linear regime; In weakly non-linear regime, \citealt{1994ApJ...434L..43M} and \citealt{2003ApJ...584....1M} provided an analytic formula for the effects of non-linear gravitational evolution, which was confirmed by \citealt{1996ApJ...460...51M}. For MFs, \citealt{2021arXiv210803851J} recently numerically studied the RSD effects and provided insights into the distinctions in CIC+GS MFs between redshift and real spaces. 

Using a state-of-the-art simulation (i.e., TianZero; cf.\ Section \ref{sec:N-body}), in this section, we investigate various dominant systematic effects on DTFE MFs, which are caused by finite voxel sizes, halo number densities, halo-weighting schemes, and RSDs, with the goal to comprehend how these effects alter the morphological measurements with DTFE MFs. Note that, starting from this section, we perform the analyses for DTFE MFs at halo field level. 

\begin{figure}
\centering 
\includegraphics[width=0.47\textwidth]{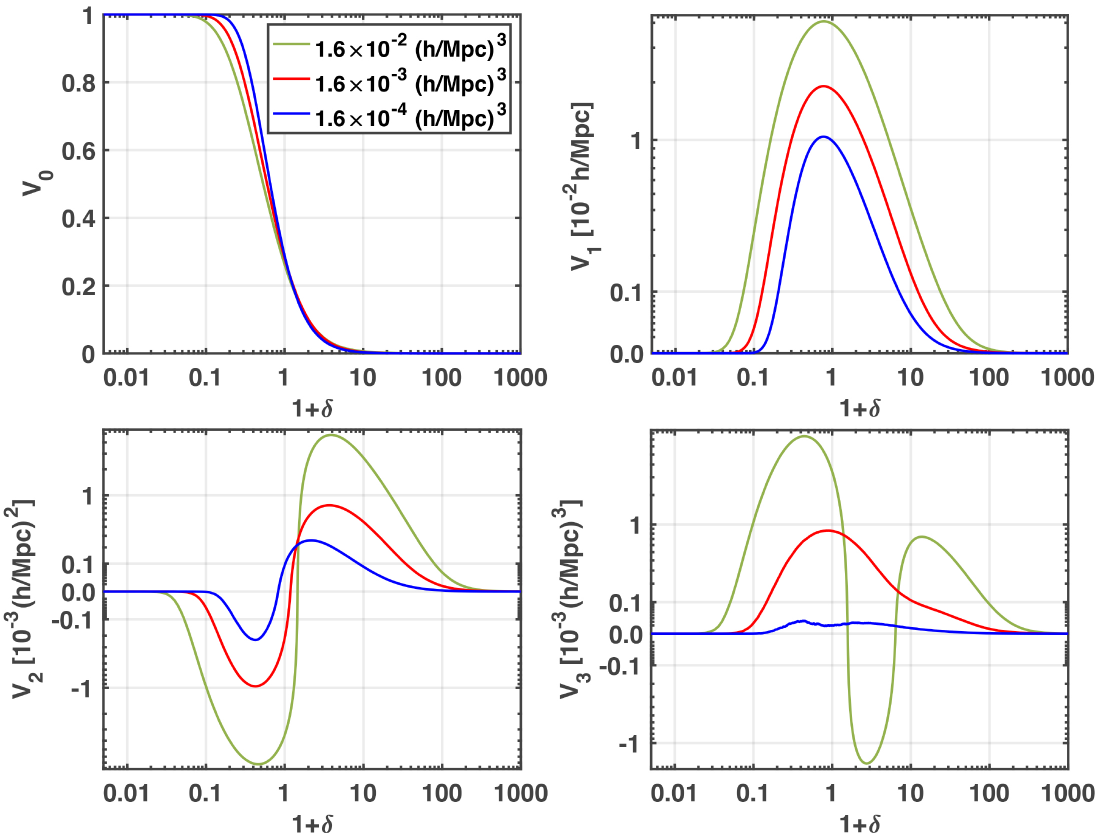}
\caption{The DTFE MFs of uniform-weighted halo fields with various halo number densities in real space. The green, red, and blue lines correspond to halo number densities of $1.6 \times 10^{-2}\,(h^{-1} \mathrm{Mpc})^{-3}$, $1.6 \times 10^{-3}\,(h^{-1} \mathrm{Mpc})^{-3}$, and $1.6 \times 10^{-4}\,(h^{-1} \mathrm{Mpc})^{-3}$, respectively. The amplitudes of MFs decrease with the decrease of halo number density, indicating information losses in lower halo number density cases.}
\label{fig:num_den}
\end{figure}

\begin{figure*}
\centering 
\includegraphics[width=0.8\textwidth]{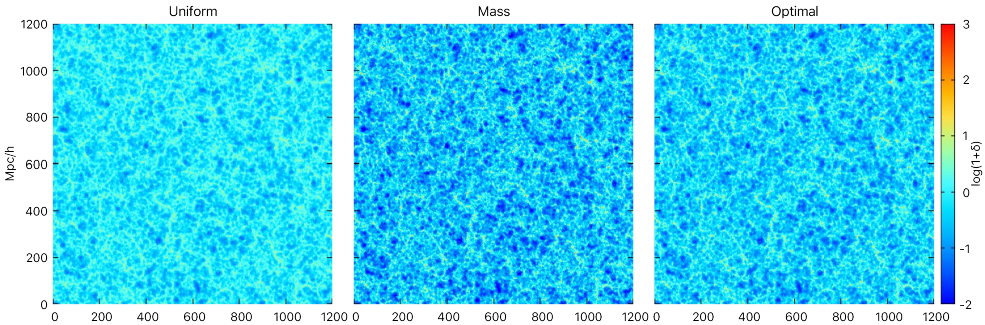}
\caption{The visualization of two-dimensional slices of different weighted DTFE halo fields in real space. The corresponding halo number density for these fields is $1.6 \times 10^{-2}\,(h^{-1} \mathrm{Mpc})^{-3}$. From left to right, each panel corresponds to uniform-, mass-, and optimal-weighted halo field,  respectively. Each slice shows the same $1200 \times 1200\,(h^{-1} \mathrm{Mpc})^2$ region with $10.55\,h^{-1}\rm{Mpc}$ thickness of the simulation box.}
\label{fig:halo_DTFE}
\end{figure*}

\begin{figure*}
\centering 
\includegraphics[width=1.0\textwidth]{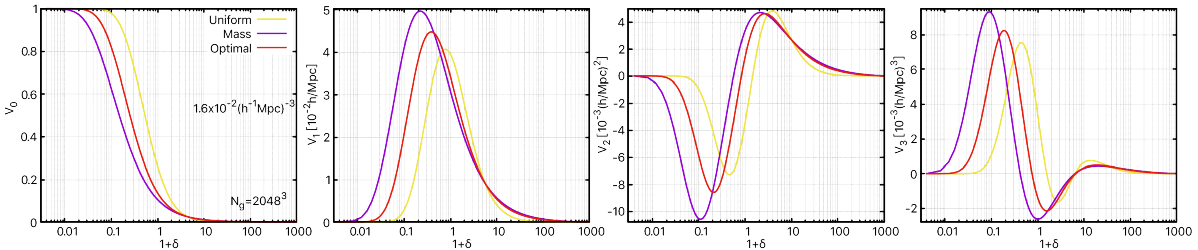}
\caption{The DTFE MFs of various weighted halo fields (cf.\ Fig.\ \ref{fig:halo_DTFE}) in real space. The corresponding halo number density for these fields is $1.6 \times 10^{-2}\,(h^{-1} \mathrm{Mpc})^{-3}$. The yellow, blue, and red lines are for the uniform-, mass-, and optimal-weighted halo fields, respectively.}
\label{fig:weighting}
\end{figure*}
\subsection{Systematic effects from finite voxel sizes} \label{sec:fvs}
In the implementation of DTFE, density field is ultimately sampled on a regular grid (cf.\ Section \ref{sec:DTFE}), which effectively eliminates the adaptive nature of DTFE below the scale of voxel size. As grid becomes coarser, certain fine-scale or faint structures of cosmic web (e.g., substructures of voids, small filamentary features, etc.) tend to be insufficiently resolved. This deficiency will result in the amplitude drops for $V_1$, $V_2$, and $V_3$ (cf.\ \citealt{2014ApJS..212...22K} and \citealt{2017ApJ...836...45A} for the scenario of genus with Gaussian smoothing), irrespective of halo number densities, weighting schemes (cf.\ Section \ref{sec:hws}), etc. We demonstrate it in Fig.\ \ref{fig:grid_res}, where DTFE MFs are measured from halo fields in real space with grid resolutions of $N_\mathrm{g}=4096^3$, $N_\mathrm{g}=2048^3$, and $N_\mathrm{g}=1024^3$ (corresponding to voxel sizes of $0.586\, h^{-1}\mathrm{Mpc}$, $0.293\, h^{-1}\mathrm{Mpc}$, and $1.171\, h^{-1}\mathrm{Mpc}$), respectively. Here, we employ the case of uniform weighting with $\bar{n}_\mathrm{h}=1.6 \times 10^{-2}\,(h^{-1} \mathrm{Mpc})^{-3}$ to quote our results. In essence, the finite voxel size effects (i.e., smoothing effects cause by grid window) are entirely numerical artifacts, which should be minimized by adopting sufficiently large grid sizes while keeping acceptable memory overheads. In the following, we choose $N_\mathrm{g}=2048^3$ to construct halo fields.

\subsection{Systematic effects from halo number densities} \label{sec:hnd}
In observation, the morphological properties of LSS can only be measured from biased tracers. The situation becomes more complicated due to the entanglement between the effects of shot noise and halo/galaxy bias, as compared with the case of matter field (cf.\ Section \ref{sec:DM MFs}). In our study, it is quite non-trivial to separate these two effects, as the effective masses of our halo samples, which essentially determine the biasing effects\footnote{It should be theoretically intriguing to explore `morphological bias' by analyzing MFs of halo samples within different mass bins (cf.\ the `topological bias' found in \citealt{2024MNRAS.529.4325B}). As this falls outside the scope of this paper, systematic studies are left for future research.}, have intrinsic relations with halo number densities. On the other hand, certain degrees of smoothness are also inevitably introduced in DTFE fields due to the limited halo number densities, which determine the effective smoothing lengths of the sophisticated DTFE windows. Therefore, the systematic effects caused by halo number density are actually combined effects jointly determined by halo bias, shot noise, and DTFE smoothing. 

To investigate the combined effects, we measure DTFE MFs of uniform-weighted halo fields in real space constructed from halo samples with different number densities. The results are presented in Fig.\ \ref{fig:num_den}. We note that these effects are the strongest effects within our tested domain. As observed, a reduction in halo number density results in the compression of MFs along the $1+\delta \sim 1$ direction and leads to lower amplitudes for $V_1$, $V_2$, and $V_3$. This occurs because lower halo number densities induce larger smoothing effects, producing smaller root-mean-square (r.m.s) values of densities. Additionally, the decrease in halo number density also makes the halos less effective as tracers of the underlying matter field, naturally leading to the losses of intrinsic structural information of LSS. In particular, when halo number density becomes sufficiently low (e.g., the case of $\bar{n}_\mathrm{h}=1.6 \times 10^{-4}\,(h^{-1} \mathrm{Mpc})^{-3}$), $V_3$ is consistently nonnegative across all density thresholds (i.e., $V_3 \ge 0$ for $1+\delta \in (0, +\infty)$), which indicates that the halo field is dominated by structures of isolated objects (i.e., `meatball' structures).

\subsection{Systematic effects from halo-weighting schemes} \label{sec:hws}
DTFE halo fields can be constructed under different halo weightings (cf.\ Section \ref{sec:DTFE}), which can yield different halo biases (\citealt{2021ApJS..254....4L}). In this work, we consider three weighting schemes: uniform, mass (\citealt{2009PhRvL.103i1303S}), and optimal\footnote{The `optimal weighting' refers to the weighting scheme that can minimize the stochasticity of halos with respect to underlying dark matter (\citealt{2010PhRvD..82d3515H}; \citealt{2011MNRAS.412..995C}; \citealt{2021ApJS..254....4L}).} (\citealt{2010PhRvD..82d3515H}; \citealt{2011MNRAS.412..995C}) weightings, represented by the forms $w_i(M) = 1, M_i, M_i + M_0$, respectively. Here, $M_0$ is a free parameter, and we adopt $M_0=3M_{\mathrm{min}}$\footnote{Note that this is an empirical relation found in \citealt{2010PhRvD..82d3515H}, which may not hold under a different condition, e.g., using a different halo-finder algorithm (cf.\ \citealt{2021ApJS..254....4L}). Here, we just naively employ this relation as a proxy for the optimal weighting in our tests.} (\citealt{2010PhRvD..82d3515H}), where $M_{\mathrm{min}}$ is the mass cutoff (cf.\ Section \ref{sec:N-body}). Since the last two weightings both depend on halo masses, we collectively denote them as mass-dependent weightings. These two weightings have been utilized in various cosmological studies [e.g., primordial non-Gaussianities (\citealt{2011PhRvD..84h3509H}), growth rate of structure formation (\citealt{2012PhRvD..86j3513H}), and initial condition reconstruction (\citealt{2021ApJS..254....4L}), etc.], given that they can significantly improve the correlation between halo field and underlying matter field (cf.\ \citealt{2009PhRvL.103i1303S}; \citealt{2010PhRvD..82d3515H}; \citealt{2021ApJS..254....4L}). 

Compared to uniform weighting, mass-dependent weightings tend to upweight the regions with higher halo abundance, as these regions are more likely to contain massive halos (cf.\ \citealt{2012MNRAS.419.2133H}; \citealt{2015MNRAS.451.4266Z}). As a result, the density contrasts between high- and low-density regions are intensified in mass-dependent weighted halo fields, leading to the stretching of corresponding MFs in two opposite directions (cf.\ Fig.\ \ref{fig:halo_DTFE} and Fig.\ \ref{fig:weighting}). Also, due to halo fields being dominated by low-density regions, the MFs of mass-dependent weighted halo fields are visually shifted towards the direction of low-density thresholds. Moreover, we observe that the amplitudes of $V_1$, $V_2$, and $V_3$ for mass-dependent weightings are enhanced relative to the case of uniform weighting. This implies that mass-dependent weightings presumably aid in resolving more structural elements when estimating halo fields, consistent with previous findings that mass information can enhance the correlation between halo field and underlying matter field (\citealt{2009PhRvL.103i1303S}; \citealt{2010PhRvD..82d3515H}; \citealt{2021ApJS..254....4L}).

\subsection{Systematic effects from RSDs} \label{sec:RSDs}
In galaxy surveys, LSS is actually mapped in redshift space, where the positions of galaxies are misrepresented due to peculiar velocities, induced by gravitational field, along the line of sight (LOS). This phenomenon is referred to as redshift space distortions (RSDs). The RSDs blur density field in redshift space, leading to a striking anisotropic feature in the direction of LOS. On larger scales, coherent infall of galaxies produces squashed pancake-like distortions, known as Kaiser's effect (\citealt{1987MNRAS.227....1K}). Whereas, on smaller scales, peculiar velocities of bound objects tend to generate elongated structures, known as fingers-of-God (FOG) effect (\citealt{1972MNRAS.156P...1J}). In cosmology, these effects provide a generic way to probe peculiar velocity field, and are commonly used to measure the growth rate of structure formation. 

\begin{figure}
\centering 
\includegraphics[width=0.48\textwidth]{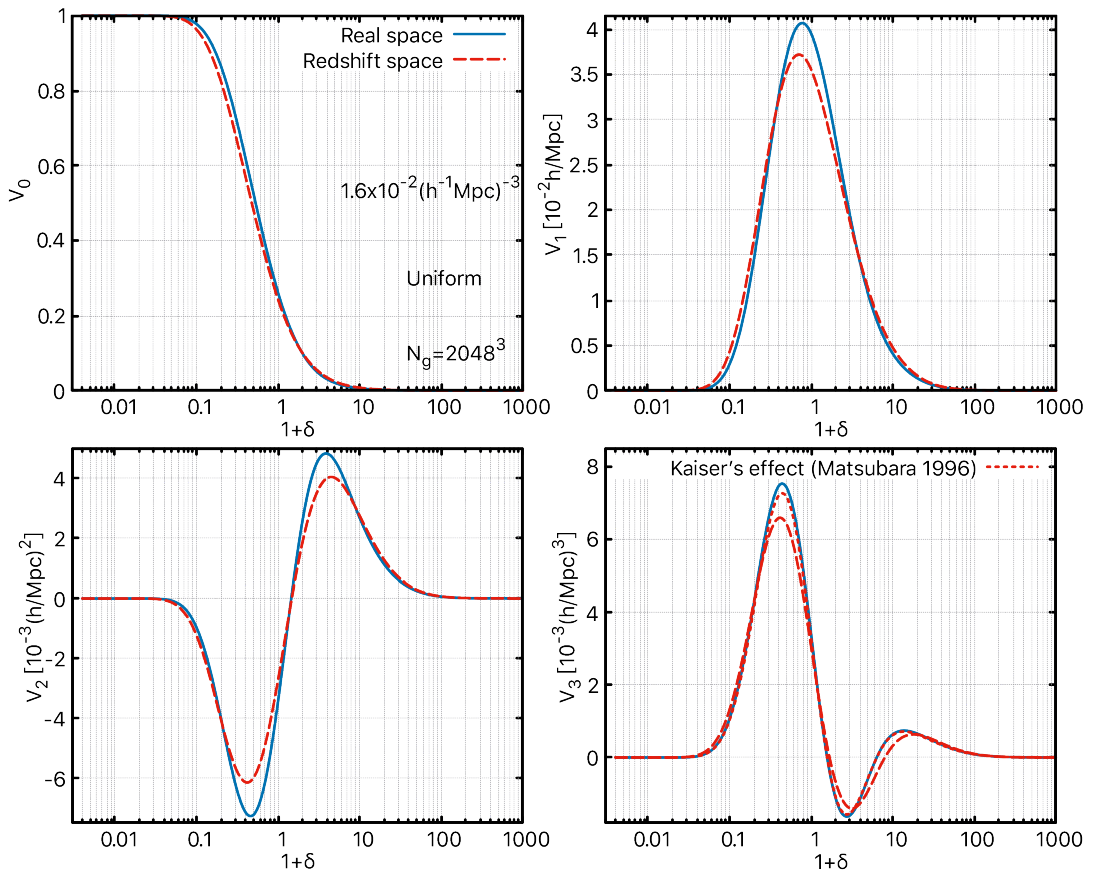}
\caption{The DTFE MFs of uniform-weighted halo fields in real space and redshift space. The corresponding halo number density for these fields is $1.6 \times 10^{-2}\,(h^{-1} \mathrm{Mpc})^{-3}$. The solid and dashed lines represent the MFs in real and redshift space, respectively. As a reference, $V_3$ with linear Kaiser's effect is also plotted (cf.\ the dotted line). As shown, the main RSD effect on MFs is the reduction in amplitudes of $V_1$, $V_2$, and $V_3$.}
\label{fig:RSDs}
\end{figure}

To investigate the effects of RSDs on DTFE MFs, we compute the MFs of uniform-weighted halo fields with $\bar{n}_\mathrm{h}=1.6 \times 10^{-2}\,(h^{-1} \mathrm{Mpc})^{-3}$ separately in real and redshift space (cf.\ Fig.\ \ref{fig:RSDs}). In redshift space, the distant-observer approximation is adopted to obtain halo positions,
\begin{equation}
\mathbf{s}=\mathbf{r}+\frac{(1+z)\mathbf{v}_{\|}} {H(z)},
\end{equation}
where $\mathbf{r}$, $z$, $\mathbf{v}_{\|}$, and $H(z)$ is the halo position in real space, redshift, LOS component of halo peculiar velocity, and Hubble parameter, respectively. We observe that in redshift space, all curves of MFs are stretched in two opposite directions compared to their counterparts in real space, indicating that RSD effects increase the r.m.s values of densities (\citealt{2021arXiv210803851J}). Nevertheless, the main effect of RSDs is the decrease in amplitudes of $V_1$, $V_2$, and $V_3$.

\begin{figure*}
\centering 
\includegraphics[width=\textwidth]{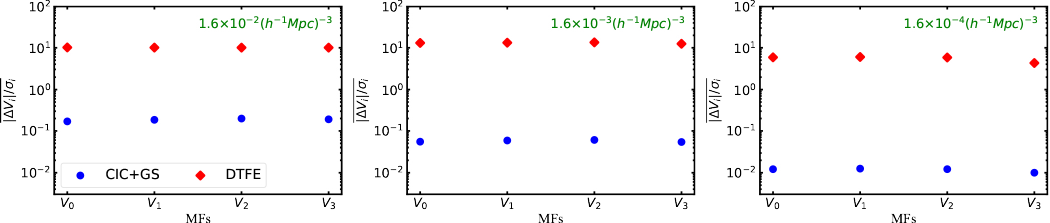}
\caption{The mean RSD S/N ratios. They are obtained by taking the average of $\left|\Delta \mathrm{V}_{i}\right| / \sigma_{i}$ (cf.\ Fig.\ \ref{fig:RSD_SNR}) in density threshold range of $[0.003,1000]$, i.e., $\overline{\left|\Delta V_{i}\right| / \sigma_{i}}$. The results for CIC+GS and DTFE MFs are marked with blue circles and red diamonds, respectively. As shown, the y-axis values of red diamonds are much higher than those of blue circles by around $2$ orders of magnitude. This means that the discriminative power of DTFE MFs for RSD effects is much stronger than that of CIC+GS MFs.}
\label{fig:ave_SNR}
\end{figure*}

Since valuable information about velocity field is encoded in RSD effects, an accurate modelling for these effects on MFs would open up a unique window to extract information on structure growth rate. In linear regime, \citealt{1996ApJ...457...13M} found that Kaiser's effect on genus (equivalent to $V_3$) can be predicted by        
\begin{equation} \label{eq:2}
G^{(\mathrm{z})}(\nu)=\frac{3 \sqrt{3}}{2} \sqrt{C}(1-C) G^{(\mathrm{r})}(\nu),
\end{equation}
where $G^{(\mathrm{z})}(\nu)$ and $G^{(\mathrm{r})}(\nu)$ is the genus in redshift and real space, respectively. The parameter $C$ is expressed as 
\begin{equation}
C=\frac{1}{3} \frac{1+\frac{6}{5} f b^{-1}+\frac{3}{7}\left(f b^{-1}\right)^{2}}{1+\frac{2}{3} f b^{-1}+\frac{1}{5}\left(f b^{-1}\right)^{2}}.
\end{equation}
Here, $b$ is the halo/galaxy bias, and $f$ is the dimensionless linear growth rate, defined as
\begin{equation}
f \equiv \frac{d \ln D}{d \ln a} \approx \Omega_\mathrm{m}^{4/7}+\frac{\Omega_{\Lambda}}{70}\left(1+\frac{\Omega_\mathrm{m}}{2}\right),
\end{equation}
where $D$ is the linear growth factor, and $a$ is the expansion parameter (\citealt{1991MNRAS.251..128L}; \citealt{2001MNRAS.322..419H}). Equation (\ref{eq:2}) suggests that the extent of amplitude decline caused by RSDs depends on growth rate and halo/galaxy bias, such that this effect can be utilized to constrain these cosmological parameters. 

DTFE MFs can naturally capture non-linear structure formation signatures induced by FOG effect. This is because DTFE tends to preserve maximum amount of non-linear structures, which are typically smoothed out in the CIC+GS scheme with a large $R_\mathrm{G}$ (cf.\ \citealt{2014ApJS..212...22K} and \citealt{2022ApJ...928..108A} for (quasi-) linear-scale scenarios). Therefore, our results do not fit well with linear theory predictions (e.g., the Equation (\ref{eq:2}) under Kaiser approximation; also cf.\ Fig.\ \ref{fig:RSDs}). Indeed, using $N$-body simulations, \citealt{1996ApJ...460...51M} found that the amplitude of genus is more suppressed than expected by linear theory (also cf.\ \citealt{2014ApJS..212...22K}), consistent with our results (cf.\ the bottom-right panel of Fig.\ \ref{fig:RSDs}). Nevertheless, in principle, we can still apply emulator-based approaches to parameter estimates (cf.\ \citealt{2019JCAP...06..019M}; \citealt{2015PhRvD..91j3511P}), where significant RSD signatures should be critical to improving constraining power of MFs on cosmological parameters. This subject is beyond the scope of this work, and we defer the investigations to future studies. 

\section{The Statistical Power of DTFE MFs in Extracting Cosmological Information} \label{sec:DTFE power}
In this section, we proceed to illustrate the strong statistical power of DTFE MFs in extracting cosmological information. For practical purposes, we restrict our analyses to extracting the cosmological information encoded in RSDs. Similar to Section \ref{sec:DM MFs}, we also present the results of CIC+GS MFs as references for comparisons, but with a different strategy for determining smoothing lengths. Here, the smoothing lengths are determined by the mean halo spacing, i.e., $\bar{d} \sim \bar{n}^{-1/3}_\mathrm{h}$. For our halo samples with number densities in descending order, they are $R_\mathrm{G}=6.8\,h^{-1}\mathrm{Mpc}$, $R_\mathrm{G}=14.6\,h^{-1}\mathrm{Mpc}$, and $R_\mathrm{G}=31.4\,h^{-1}\mathrm{Mpc}$, respectively. Note that these smoothing lengths are traditionally regarded as the smallest smoothing lengths that can be employed for CIC+GS MFs. Therefore, the comparative results displayed in this section are relatively conservative. 

\subsection{The discriminative power for RSDs} \label{sec:Discrimination} 
\begin{figure*}
\centering 
\includegraphics[width=\textwidth]{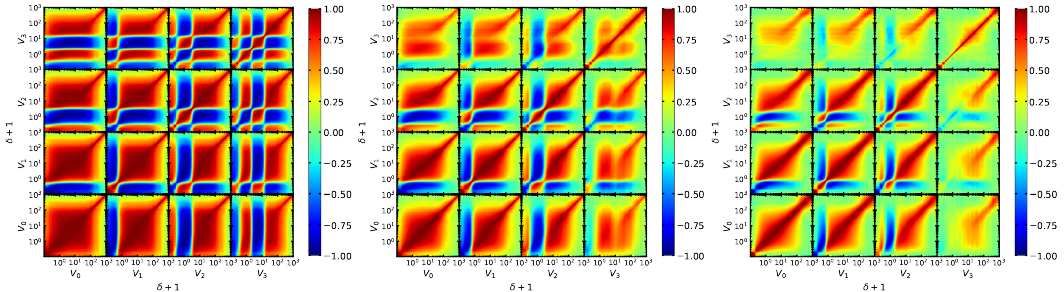}
\caption{The correlation matrices for DTFE MFs of uniform-weighted halo fields with different number densities. From left to right, each panel shows the case of halo number density of $1.6 \times 10^{-2}\,(h^{-1} \mathrm{Mpc})^{-3}$, $1.6 \times 10^{-3}\,(h^{-1} \mathrm{Mpc})^{-3}$, and $1.6 \times 10^{-4}\,(h^{-1} \mathrm{Mpc})^{-3}$, respectively. }
\label{fig:CM_uniform}
\end{figure*}  
    
\begin{figure*}
\centering 
\includegraphics[width=\textwidth]{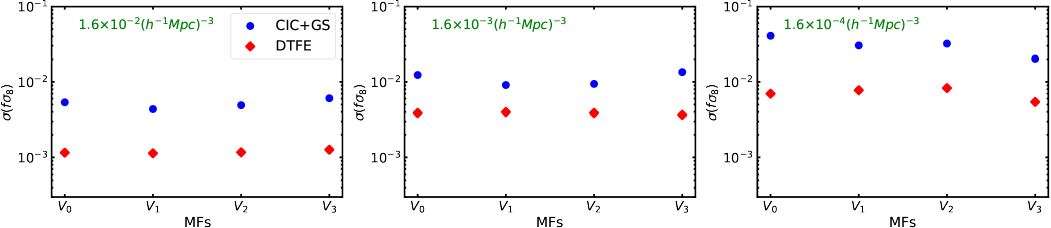}
\caption{The Fisher error forecasts for structure growth rate, with $n_\mathrm{b}=32$ (cf.\ Section \ref{app:Fisher Forecast} for technical details). The results for CIC+GS and DTFE MFs are marked with blue circles and red diamonds, respectively. As shown, the y-axis values of red diamonds are much lower than those of blue circles by a factor of around $3$-$5$. This means that compared with traditional CIC+GS MFs, DTFE MFs are more sensitive to cosmological parameters, because they contain more morphological information of LSS.}
\label{fig:fsigma8_32points}
\end{figure*} 

In this subsection, we quantitatively assess the discriminative power of DTFE MFs for RSD effects. To this end, we calculate the differences in MFs of uniform-weighted halo fields with various number densities between real and redshift space,  
\begin{equation}\label{eq:3}
\Delta V_i\left( \nu \right)=V_i^{(\mathrm{z})}\left(\nu \right)-V_i^{(\mathrm{r})}\left( \nu \right),
\end{equation}
where $V_i^{(\mathrm{r})}\left(\nu \right)$ and $V_i^{(\mathrm{z})}\left( \nu \right)$, with $i=0,1,2,3$, represent the MFs in real and redshift space, respectively. The results are shown in Fig.\ \ref{fig:RSD_SNR}, where the error bars are obtained by
\begin{equation}
\sigma = \sqrt{\sigma_\mathrm{z}^2+\sigma_\mathrm{r}^2}.
\end{equation}
Here, $\sigma_\mathrm{z}$ and $\sigma_\mathrm{r}$ denote the errors of MFs (cf.\ Section \ref{sec:MFs}) in real and redshift space, respectively. 

We see that, for $V_1$, $V_2$, and $V_3$, RSD signals in DTFE MFs are significantly higher than those in CIC+GS MFs, regardless of halo number density.  And, our results are well consistent with those found in previous works (e.g., \citealt{2014ApJS..212...22K}; \citealt{2021arXiv210803851J}; \citealt{2021MNRAS.508.3771L}; \citealt{2022ApJ...928..108A}), but showing more pronounced signals in DTFE case. Moreover, intriguingly, despite our analyses being performed at halo field level, we find that the signs and trends of $\Delta V_i\left( \nu \right)$ in our work are broadly the same as those in \citealt{2021arXiv210803851J}, where CIC+GS DM fields were employed for analyses. For interpretations of $\Delta V_i\left( \nu \right)$, we refer interested readers to that paper to find more details. In Fig.\ \ref{fig:RSD_SNR}, we also present the S/N ratios of RSD effects, defined as $\left|\Delta V_{i}\right| / \sigma_{i}$. As shown, the amplitudes of RSD S/N ratios for DTFE MFs are also larger than those for CIC+GS MFs, except for $\left|\Delta \mathrm{V}_{0}\right| / \sigma_{0}$ with $\bar{n}_\mathrm{h}=1.6 \times 10^{-2}\,(h^{-1} \mathrm{Mpc})^{-3}$ (cf.\ the bottom subplot in left-top panel of Fig.\ \ref{fig:RSD_SNR}), where the amplitudes in the two cases are basically comparable.

In fact, comparing S/N ratio amplitudes is not an effective way for assessing the relative performance of these two methods in extracting RSD signals. A more suitable approach is to compare the areas between the S/N curves and the x-axis. For this reason, we calculate the mean RSD S/N ratios, which are defined as
\begin{equation} \label{eq:4}
\overline{\left| \Delta V_{i} \right| / \sigma_{i}} \equiv \frac{\int_{\nu_\mathrm{min}}^{\nu_\mathrm{max}} \left| \Delta V_{i} \right| / \sigma_{i}} {\nu_{\mathrm{max}}-\nu_{\mathrm{min}}},
\end{equation}
where $\nu_\mathrm{min}=0.003$ and $\nu_\mathrm{max}=1000$. Note that the numerator on the right-hand side of Equation (\ref{eq:4}) is exactly the area between the S/N curve and the x-axis within the range of $[\nu_\mathrm{min}, \nu_\mathrm{max}]$. The results are shown in Fig.\  \ref{fig:ave_SNR}. Expectedly, DTFE leads to remarkable improvements over CIC+GS in terms of discriminative power for RSD effects, by $\sim2$ orders of magnitude (cf.\ Fig.\ \ref{fig:ave_SNR}). Unquestionably, this is because DTFE MFs are more informative and thus more sensible to any modifications in halo fields. 

Additionally, due to information losses, $\overline{\left|\Delta V_{i}\right| / \sigma_{i}}$ becomes lower as halo number density decreases, regardless of the schemes for measuring MFs. In particular, when halo number density becomes too low, halo fields will be dominated by structures of isolated objects, making the intrinsic topologies of these structures less susceptible to the deformations caused by RSDs (cf.\ Fig.\ \ref{fig:RSD_SNR} and Fig.\ \ref{fig:ave_SNR}). On the other hand, we also notice a somewhat counter-intuitive result, i.e., for DTFE MFs with $\bar{n}_\mathrm{h}=1.6 \times 10^{-3}\,(h^{-1} \mathrm{Mpc})^{-3}$, $\overline{\left|\Delta V_{i}\right| / \sigma_{i}}$ is instead higher than that of the case with $\bar{n}_\mathrm{h}=1.6 \times 10^{-2}\,(h^{-1} \mathrm{Mpc})^{-3}$. This may be caused by data truncation effect: density thresholds are limited to $[0.003,1000]$ in our calculation, while there are still RSD signals above $\nu=1000$ (cf.\ Fig.\ \ref{fig:RSD_SNR}); for larger $\bar{n}_\mathrm{h}$, DTFE MFs could have captured signals from higher-density regions, but more signals were abandoned. As for CIC+GS MFs, they do not suffer from such data truncation effect, so their corresponding results appear rational.

\subsection{The constraining power on structure growth rate} \label{sec:Sensitivity} 
In the analyses of last subsection, we actually ignore the correlations between data points at different density thresholds. However, in reality, the correlations do exist and are quite significant. This can be seen from Fig.\ \ref{fig:CM_uniform}, where correlation matrices for DTFE MFs of uniform-weighted halo fields under various number densities are presented. One can observe that the correlation matrices exhibit certain regular patterns. By comparing Fig.\ \ref{fig:num_den} and Fig.\ \ref{fig:CM_uniform}, we notice that the data points on MF-curve segments with same trend are positively correlated, while the data points on rising MF-curve segments are negatively correlated with those on falling MF-curve segments. 

In this subsection, by taking into account these correlations, we go a step further to investigate the sensitivities of DTFE MFs to cosmological parameters, i.e., the constraining power. As a case study, we conduct simple Fisher forecasts for the errors on structure growth rate $f\sigma_8(z=0.01)$, a key cosmological parameter often constrained in RSD studies (cf.\ Appendix \ref{app:Fisher Forecast} for technical details). Here, $\sigma_8$ is the amplitude of density fluctuations within a sphere of comoving radius $R=8\ h^{-1}\mathrm{Mpc}$, which can be obtained by
\begin{equation}
\sigma^2_{8}=\frac{1}{2 \pi^{2}} \int P_{\mathrm{m}}(k) \left|W\left(k R\right)\right|^{2} k^{2} \mathrm{d} k,
\end{equation} 
where 
\begin{equation}
W(kR)=\frac{3[\sin (kR)-kR \cos (kR)]}{k^3R^3}
\end{equation} 
is the top-hat window function in Fourier space and $P_\mathrm{m}(k)$ is the matter power spectrum. In this study, Fisher forecasts are performed for each $V_i(\nu)$, under various halo number densities, and the results are shown in Fig.\ \ref{fig:fsigma8_32points}.

As shown, the predictive errors increase with the decrease in halo number densities, attributed to information losses. In particular, DTFE MFs yield much stronger constraining power, exhibiting remarkable improvement by a factor of $\sim$ $3$-$5$ over CIC+GS MFs within our tested domain. The results unequivocally demonstrate that DTFE MFs outperform traditional CIC+GS MFs in constraining cosmology. Again, this is due to the fact that DTFE MFs can capture more morphological information of LSS from halo distribution. In Fig.\ \ref{fig:fsigma8_32points}, the results are obtained by utilizing $V_i$ with 32 data points. For consistency check, we also employ $V_i$ with 64 data points for Fisher forecasts (cf.\ Fig.\ \ref{fig:fsigma8_64points}). The results from both approaches are consistent, and they are specific to the survey volume of our simulation box, i.e., $1.728\,(h^{-1}\mathrm{Gpc})^3$. If volume increases, the predictive errors will be further reduced. It is noteworthy that various numerical and observational systematic effects (e.g., data selection, irregular survey mask, and those investigated in Section \ref{sec:Systematic Effects}) can affect the predictions, but hard to change our conclusions.  

\section{Summaries and Discussions} \label{sec:Summary}
In cosmology, constructing density fields from point sets is a basic task to perform grid-based analyses of LSS. The choice of a particular method is often a compromise between desired field properties and limitations of the method. In particular, constructing continuous fields is essential to specify iso-density surfaces for estimating MFs. To achieve this, traditional mass assignment methods typically require tracer samples with high number densities to ensure sufficient sampling per grid cell. Otherwise, poorly sampled under-dense regions will be dominated by prominent shot noises, severely limiting their applicabilities to sparse datasets. Therefore, these methods are always utilized in conjunction with smoothing recipes at the cost of erasing substantial structural information. It is obviously not an optimal solution because the ultimate goal should be to preserve the intricate LSS multi-scale patterns as faithfully as possible. A significant advancement toward this objective can be attained by leveraging DTFE technique, which can produce piece-wise continuous fields with unique features of being parameter-free, self-adaptive in scale, and preserving mass conservation.

In this work, we propose to optimize the extractions of morphological information from cosmic web tracers with DTFE MFs. We perform systematic analyses in a step-by-step manner, starting from matter field level and progressing to halo field level:
\begin{itemize}
\item At matter field level, we first investigate shot noise effects on CIC MFs, elucidating the challenges posed by severe discreteness in the density fields,  constructed by traditional mass assignment methods, for proper morphological measurements from sparse tracers with MFs. Then, we measure CIC+GS MFs (i.e., the traditional scheme) and DTFE MFs (i.e., the new scheme) from the same downgraded particle samples and compare them to preliminarily demonstrate the superiorities of DTFE method in measuring MFs from point sets. For CIC+GS MFs, we explore the corresponding shot noise effects and propose a strategy for determining smoothing lengths to sufficiently eliminate shot noises without excessively erasing structural information. 
\item At halo field level, we first numerically study various dominant systematic effects on DTFE MFs, induced by finite voxel sizes, halo number densities, halo-weighting schemes, and redshift space distortions. Then, we showcase the robust statistical power of DTFE MFs for extracting cosmological information encoded in RSDs. We find that DTFE MFs remarkably outperform traditional CIC+GS MFs by $\sim$ $2$ orders of magnitude in discriminative power for RSD effects and by a factor of $\sim$ $3$-$5$ in constraining power on structure growth rate. This is because DTFE scheme can help conserve maximum morphological information of LSS from sparse tracers, rendering DTFE MFs more sensitive to cosmological parameters. 
\end{itemize}
In view of the strong statistical power of DTFE MFs, we will employ this method to extract various critical cosmological signatures [e.g., neutrino masses (\citealt{2020PhRvD.101f3515L}), modified gravities (\citealt{2017PhRvL.118r1301F}), and primordial non-Gaussianities (\citealt{2006ApJ...653...11H}; \citealt{2008MNRAS.385.1613H})] imprinted on halo/galaxy fields, in our ongoing projects.

The implementation scheme for measuring MFs proposed in this paper is conceptually related to other Delaunay-based methods developed in previous works (\citealt{2010arXiv1006.4178A}; \citealt{2021MNRAS.508.3771L}). These works compute MFs from triangulated iso-density surfaces, which are directly specified from Delaunay tessellation of a point set, without interpolating density field onto a regular grid. When density values at tessellation vertexes in these methods are estimated in the same manner as DTFE, their performance in measuring MFs should be highly similar to that of our method, with an added advantage of being free from finite voxel size issue. Alternatively, there might also be a possibility to directly evaluate MFs from alpha shapes of discrete tracers, which are specified via filtrations of Delaunay tessellation (\citealt{2011LNCS.6970...60V}). Despite all that, the method presented in this work stands out for its simplicity and convenience in implementation, as it does not require any additional complex code designs for extracting MFs from triangulated isosurfaces. This is undeniably one of the main merits of our method.

Moreover, It's well-known that Delaunay-based methods are sensitive to point perturbations, leading to substantial rearrangements in Delaunay tessellation. These methods also tend to produce prominent undesired spike-like artifacts due to highly elongated tetrahedra, which hinders the accurate identification of faint structures in point distribution. These issues seem still to be particularly problematic for morphological studies of LSS. One potential solution \footnote{Recently, \citealt{2024arXiv240216234F} proposed a phase-space DTFE (PS-DTFE), a hybrid method trying to combine advantages of phase-space density estimator (\citealt{2012PhRvD..85h3005S}; \citealt{2012MNRAS.427...61A}) and DTFE. However, PS-DTFE field exhibits discontinuities at fold caustic surfaces.  And, it also relies on prior knowledge of tracers before shell crossing, limiting its feasibility to simulated data.} to address these concerns is the use of an ensemble-based DTFE technique (\citealt{2021MNRAS.503..557A}). This method computes the mean DTFE field from an ensemble of point realizations by perturbing the original point set following geometric constraints. It can be regarded as a natural generalization of DTFE and shares the same advantageous characteristics of DTFE. Therefore, the measurements of MFs with ensemble-based DTFE can be easily implemented on top of our method and merit investigations in future studies.

\section*{Acknowledgments}
Y.L.\ would like to thank Cheng Zhao, Charling Tao, Yuting Wang, Shuo Yuan, Zhao Chen, Zhengxiang Li, and Dandan Xu for useful communications. We also thank the anonymous referee for their very detailed and helpful comments, which helped make the paper more informative. This work was supported by National Science Foundation of China (Nos.\ 12303005, 12273015, 12273020, 11621303, 12173030), National Key R\&D Program of China (Nos.\ 2023YFA1605601, 2023YFA1607800, 2023YFA1607802), and the science research grants from China Manned Space Project with No.\ CMS-CSST-2021-A03. Y.L.\ acknowledges support from Shuimu Tsinghua Scholar Program (No.\ 2022SM173). Y.Y. acknowledges the sponsorship from Yangyang Development Fund. This work made use of the Gravity Supercomputer at the Department of Astronomy, Shanghai Jiao Tong University.


\appendix
\section{The Determinations of Smoothing Lengths} \label{app:Smoothing Lengths} 
\begin{figure*}
\centering 
\includegraphics[width=1.0\textwidth]{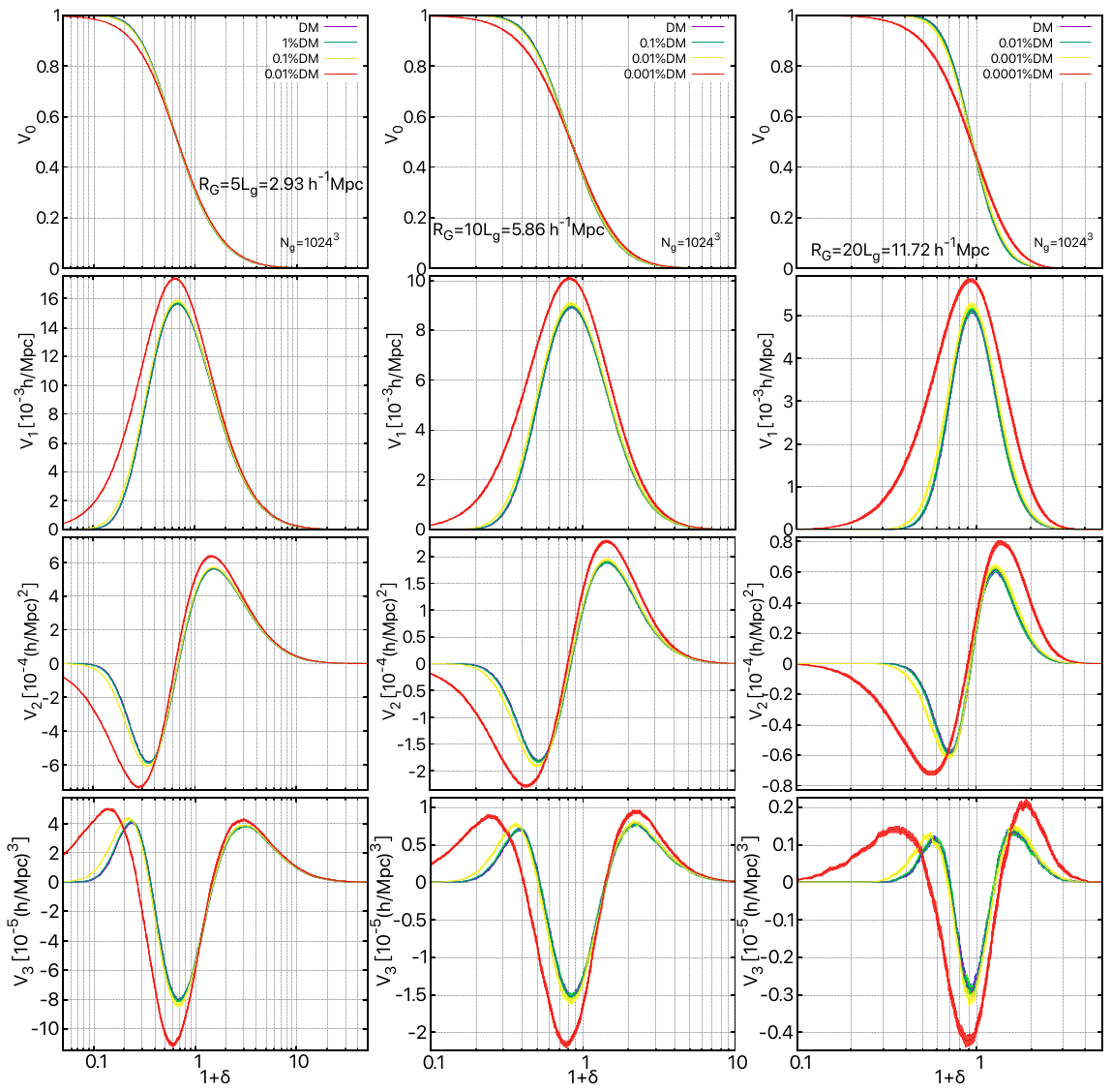}
\caption{The MFs of noise-free field and various downgraded fields under different smoothing lengths. Left panels: MFs of noise-free field (blue lines) and $1\%$ (green lines), $0.1\%$ (yellow lines), $0.01\%$ (red lines) downgraded fields, under $R_G=5L_g=2.93\,h^{-1}\mathrm{Mpc}$. Middle panels: MFs of noise-free field (blue lines) and $0.1\%$ (green lines), $0.01\%$ (yellow lines), $0.001\%$ (red lines) downgraded fields, under $R_G=10L_g=5.86\,h^{-1}\mathrm{Mpc}$. Right panels: MFs of noise-free field (blue lines) and $0.01\%$ (green lines), $0.001\%$ (yellow lines), $0.0001\%$ (red lines) downgraded fields, under $R_G=20L_g=11.72\,h^{-1}\mathrm{Mpc}$. Here, $L_g=L/N_g^{1/3} \simeq 0.59\, h^{-1}\mathrm{Mpc}$ is the grid cell size. Shaded bands show 1-$\sigma$ scatters. As shown, the MFs of $\%1$, $\%0.1$, $\%0.01$ downgraded fields under smoothing lengths of $R_G=2.93\,h^{-1}\mathrm{Mpc}$, $R_G=5.86\,h^{-1}\mathrm{Mpc}$, $R_G=11.72\,h^{-1}\mathrm{Mpc}$ overlap with the MFs of noise-free field under the corresponding smoothing lengths, respectively, within $1$-$\sigma$ accuracy range. That is, in each panel, the green line basically overlaps with the blue line. This means that under these smoothing lengths, the shot noise effects on MFs of corresponding downgraded fields can be sufficiently smoothed out. As downgrading level increases, these smoothing lengths are inadequate to eliminate shot noise effects on MFs. Consequently, in each panel, the yellow and red lines deviate from the blue line, illustrating how shot noises affect the measurements of CIC+GS MFs.}
\label{fig:Gau_len}
\end{figure*}

In this appendix, we illustrate our strategy for determining the smoothing lengths adopted in Section \ref{sec:DM CIC+GS MFs} and display the effects of shot noise on CIC+GS MFs. In previous works (e.g., \citealt{2005ApJ...633....1P}; \citealt{2010ApJS..190..181C}; \citealt{2014ApJ...796...86P}; \citealt{2014ApJS..212...22K}), smoothing lengths were empirically determined with the restriction $R_G \ge \bar{d} \sim \bar{n}_\mathrm{tracer}^{-1/3}$, where $\bar{n}_\mathrm{tracer}$ is tracer number density. Actually, in our tests, setting $R_G = \bar{d}$ cannot provide adequate smoothing for completely eliminating shot noise effects. Our strategy is to strike a balance between removing shot noise effects and retaining sufficient information. To achieve this, $R_G$ is determined through an iterative and stepwise refined process, such that the differences of MFs between downgraded field and noise-free field are controlled within $1$-$\sigma$ deviation. We note that this strategy seems to have never been proposed before. Therefore, it is worth exploring it theoretically.

The results are visualized in Fig.\ \ref{fig:Gau_len}, where MFs of noise-free field and three successive downgraded fields adopt the same $R_G$ for each column. It explicitly illustrates the specified smoothing lengths, i.e., $R_G=2.93\,h^{-1}\mathrm{Mpc}$, $R_G=5.86\,h^{-1}\mathrm{Mpc}$, and $R_G=11.72\,h^{-1}\mathrm{Mpc}$ for the $1\%$, $0.1\%$, and $0.01\%$ downgraded fields, respectively. Meanwhile, we can also see that shot noises characteristically make the amplitudes of $V_1$, $V_2$, and $V_3$ increase (cf.\ \citealt{2014ApJS..212...22K}) and result in MF curves being stretched in two opposite directions. This is because discrete effects generate more pseudo structures on excursion sets, leading to choppier density fields with larger r.m.s of densities. This process for $R_G$ determinations would also be applicable to the cases of halo fields. In consideration that it is not the main concern of our paper, we leave the systematic investigations on this topic to future investigations.

\section{The RSD Signals and S/N Ratios Extracted from DTFE MFs} \label{app:RSD Signals}
\begin{figure*}
\centering 
\includegraphics[width=1.0\textwidth]{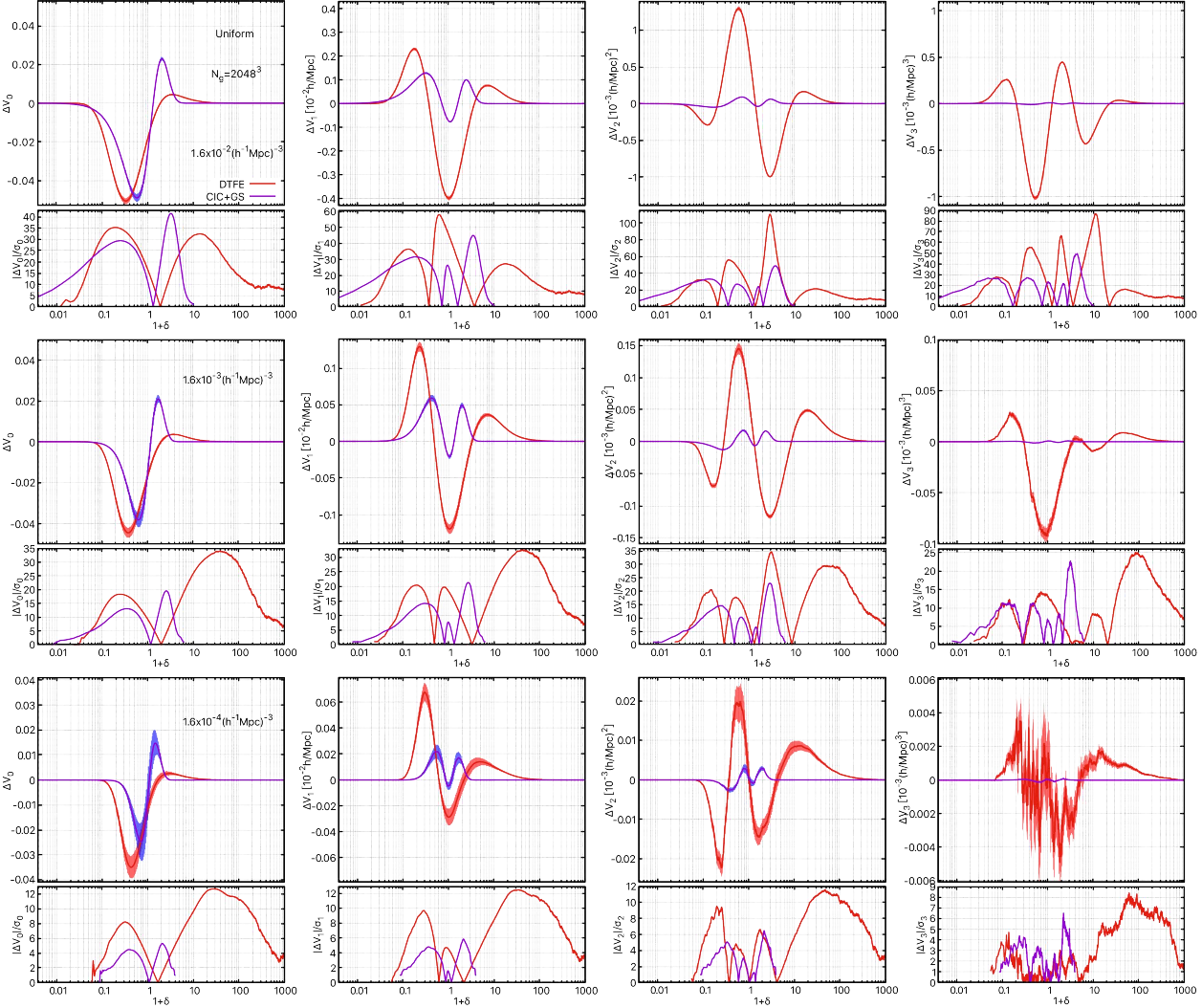}
\caption{The RSD signals and corresponding S/N ratios in MFs. The RSD signals are obtained by measuring the differences between MFs of uniform-weighted halo fields in redshift space and real space (i.e., $\Delta \mathrm{V}_{i}$; cf.\ Equation (\ref{eq:3})). The S/N ratios are calculated through $\left|\Delta \mathrm{V}_{i}\right| / \sigma_{i}$, where $\sigma_{i}$ is the 1-$\sigma$ error on $\Delta V_i$. From left to right columns, the results for $V_0$, $V_1$, $V_2$, and $V_3$ are presented, respectively.  From top to bottom, each row corresponds to the case of halo number density of $1.6 \times 10^{-2}(h^{-1}\, \mathrm{Mpc})^{-3}$, $1.6 \times 10^{-3}(h^{-1}\, \mathrm{Mpc})^{-3}$, and $1.6 \times 10^{-4}\,(h^{-1} \mathrm{Mpc})^{-3}$, respectively.  For each panel, top subplot shows RSD signals and bottom subplot presents RSD S/N ratios, where blue and red curves are for CIC+GS and DTFE cases, respectively. In CIC+GS case, for halo number densities in descending order,  the smoothing lengths $R_G=6.8\,h^{-1}\mathrm{Mpc}$, $R_G=14.6\,h^{-1}\mathrm{Mpc}$, and $R_G=31.4\,h^{-1}\mathrm{Mpc}$ are adopted, respectively. These smoothing lengths are determined by the mean halo spacings of our halo samples. Shaded band shows 1-$\sigma$ scatter. }
\label{fig:RSD_SNR}
\end{figure*}

In this appendix, we provide the RSD signals (cf.\ Equation \ref{eq:3}) and the associated RSD S/N ratios extracted from DTFE MFs (cf.\ Section \ref{sec:Discrimination}). The results are shown in Fig.\ \ref{fig:RSD_SNR}. The figure is decided to be displayed here due to its large size as well as the fact that the main information it conveys is already effectively presented in Fig.\ \ref{fig:ave_SNR}.  
  
\begin{table*}
\resizebox{0.96\linewidth}{!}{
\begin{tabular}{lccc}
\hline \hline \begin{tabular}{l}
\text{$[\nu_\mathrm{lower},\nu_\mathrm{upper}]$,$n_{\mathrm{d}_{32}}$,$n_{\mathrm{d}_{64}}$}
\end{tabular} & \begin{tabular}{c}
\text{$\bar{n}_\mathrm{h}=1.6 \times 10^{-2}\,(h^{-1} \mathrm{Mpc})^{-3}$}
\end{tabular} & \begin{tabular}{c}
\text{$\bar{n}_\mathrm{h}=1.6 \times 10^{-3}\,(h^{-1} \mathrm{Mpc})^{-3}$}
\end{tabular} & \begin{tabular}{c}
\text{$\bar{n}_\mathrm{h}=1.6 \times 10^{-4}\,(h^{-1} \mathrm{Mpc})^{-3}$}
\end{tabular} \\
\hline \begin{tabular}{l}
\textbf{CIC+GS MFs} 
\end{tabular} & \begin{tabular}{c}
$[0.012,6.907],17,36$
\end{tabular} & \begin{tabular}{c}
$[0.048,4.605],12,25$ 
\end{tabular} & \begin{tabular}{c}
$[0.168,3.028],8,15$
\end{tabular} \\ \begin{tabular}{l}
\textbf{DTFE MFs} 
\end{tabular} & \begin{tabular}{c}
$[0.032,1000],26,56$ 
\end{tabular} & \begin{tabular}{c}
$[0.048,1000],25,52$ 
\end{tabular} & \begin{tabular}{c}
$[0.072,1000],24,48$ \\
\end{tabular} \\
\hline \hline
\end{tabular}
}
\caption{The density threshold ranges and numbers of data points used in our Fisher forecasts. For both CIC+GS and DTFE MFs, we show these values in various halo number density cases. Here, $n_{\mathrm{d}_{32}}$ and $n_{\mathrm{d}_{64}}$ indicate the numbers of remaining data points derived from the original $n_\mathrm{b}=32$ and $n_\mathrm{b}=64$ data points, respectively. In lower halo number density cases, the induced larger smoothing effects make the density fields more uniform. As a consequence, the MFs are squeezed towards intermediate density thresholds (cf.\ Section \ref{sec:hnd}). Therefore, the actually used density threshold ranges become narrower, and the corresponding numbers of remaining data points decrease (cf.\ Fig.\ \ref{fig:Ranges_64points}). }
\label{tab:info}
\end{table*}

\begin{figure*}
\centering 
\includegraphics[width=1.0\textwidth]{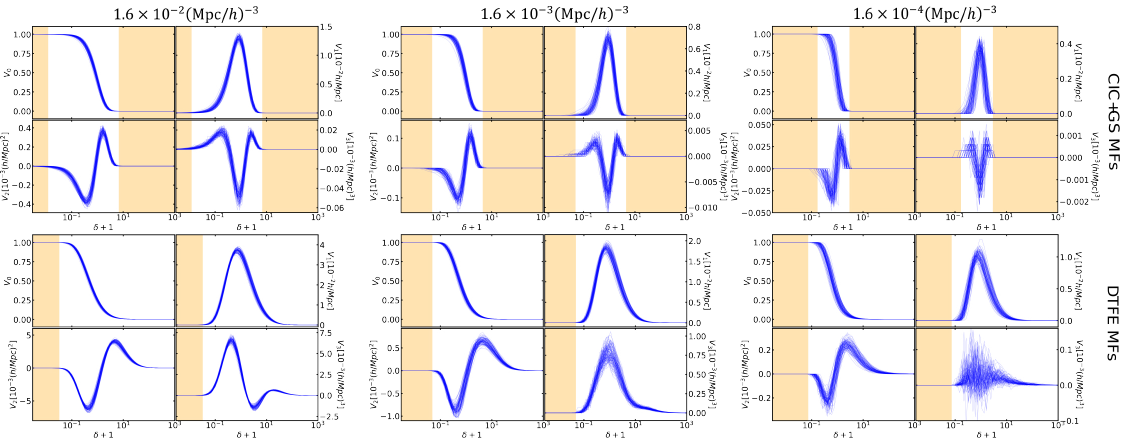}
\caption{The MFs of subfields. The upper panels are for CIC+MFs, and the lower panels are for DTFE MFs. From left to right, each panel shows the case of halo number density of $1.6 \times 10^{-2}\,(h^{-1} \mathrm{Mpc})^{-3}$, $1.6 \times 10^{-3}\,(h^{-1} \mathrm{Mpc})^{-3}$, and $1.6 \times 10^{-4}\,(h^{-1} \mathrm{Mpc})^{-3}$, respectively.  For each panel, we show 100 sets of MFs, which are randomly selected from the MFs of 512 subfields. Here, each $V_i$ has $n_\mathrm{b}=64$ data points, and the unshielded regions indicate the density threshold ranges actually used in our calculations for Fisher forecasts (cf.\ Table \ref{tab:info}). } 
\label{fig:Ranges_64points}
\end{figure*}

\section{The Methodology of Fisher Forecasts for Errors on Structure Growth Rate} \label{app:Fisher Forecast} 
In our work, Gaussian likelihood is assumed, and the dependence of covariance matrix on model parameters is also ignored. Therefore, the elements of Fisher matrix $\boldsymbol{F}$ can be written as 
\begin{equation}\label{eq:5}
F_{\alpha \beta}=\frac{\partial \boldsymbol{\mu}^{T}}{\partial \theta_{\alpha}} \boldsymbol{C}^{-1} \frac{\partial \boldsymbol{\mu}}{\partial \theta_{\beta}},
\end{equation}
where $\alpha$ and $\beta$ label the parameters of interest, $\boldsymbol{\mu}$, $\boldsymbol{\theta}$, and $\boldsymbol{C}$ are the mean of data vector $\boldsymbol{x}$, vector of model parameters, and
data covariance matrix, respectively (cf.\ \citealt{2020A&A...642A.191E}). The elements of $\boldsymbol{C}$ are computed as
\begin{equation}
C_{i j}=\sum_{r=1}^{N} \frac{(x_{i}^{r}-\mu_{i})(x_{j}^{r}-\mu_{j})}{N-1},
\end{equation}
where $\mu_{i(j)}=\sum_{r} x_{i(j)}^{r}/N$,  $i(j)=1, 2, 3 \dots n_\mathrm{b}$, $r$ represents the $r$-th data realization, $N$ and $n_\mathrm{b}$ denote the number of data realizations and number of data vector elements, respectively. Furthermore, we rescale the inverse covariance matrix $\boldsymbol{C}^{-1}$ with a factor $(N-n_\mathrm{b}-2)/(N-1)$ to correct for the bias induced by finite number of data realizations\footnote{The validity of this correction assumes Gaussian errors and uncorrelated data vectors. This is not strictly true for most scenarios. To minimize impacts of this assumption, $N$ needs to be sufficiently larger than $n_\mathrm{b}$. } (\citealt{2007A&A...464..399H}). Finally, the diagonal element of the inverse of Fisher matrix $(\mathcal{F})_{i i}^{-1}$ yields the lower limit to the error of $i$-th model parameter (marginalized over other parameters), 
\begin{equation}
\sigma_{\theta_i}^{2} \geq(\mathcal{F})_{i i}^{-1}.
\end{equation}

In our scenario, data vectors refer to the MFs in redshift space $V_i^{(\mathrm{z})}(\nu)$, and data realizations refer to the $V_i^{(\mathrm{z})}(\nu)$ measured from $n_\mathrm{f}=8^3=512$ subfields (cf.\ Section \ref{sec:MFs}). Specifically, the covariance matrices are first calculated by using the $512$ subfields with RSDs and then are rescaled by $1/n_\mathrm{f}$\footnote{This method is based on an implicit assumption that all subfields are statistically independent. This is not really satisfied in practice. Therefore, it only gives rough estimations (cf.\ \citealt{2017A&A...604A.104L}; \citealt{2023arXiv230504520J}). Since we mainly focus on comparisons of the two schemes for measuring MFs, this issue, in principle, does not affect our conclusions. To solve this issue, one can use a large number of simulation realizations to give accurate but expensive covariance estimates. We leave this for future works.}. To calculate the derivatives in Equation \ref{eq:5}, we construct two additional halo fields with modified RSDs by artificially increasing and decreasing the peculiar velocity of each halo by a factor of $0.03$, respectively. Then, the response functions are given as $\Delta \boldsymbol{\mu}=\boldsymbol{\mu}_{\mathrm{v}_+}-\boldsymbol{\mu}_{\mathrm{v}_-}=\Delta V_i^{(\mathrm{z})}(\nu)$, where subscripts $\mathrm{v}_+$ and $\mathrm{v}_-$ indicate the two cases of increasing and decreasing halo peculiar velocities, respectively. Moreover, in RSD studies, there is an approximate relational expression between structure growth rate $f\sigma_8(z)$ and halo/galaxy velocity bias $b_\mathrm{v}$ (cf.\ \citealt{2018ApJ...861...58C}),
\begin{equation}
\left.\frac{\delta\left(f \sigma_{8}\right)}{f \sigma_{8}}\right|_{k, z} \simeq-\left.\frac{\delta b_\mathrm{v}}{b_\mathrm{v}}\right|_{k, z}.
\end{equation} 
Therefore, in our scenario, the change in $f\sigma_8$ can be expressed as $\Delta\left(f \sigma_{8}\right) \simeq-\frac{\Delta b_\mathrm{v}}{b_\mathrm{v}}f \sigma_{8}=-\frac{b_{\mathrm{v}_+}-b_{\mathrm{v}_-}}{b_\mathrm{v}}f \sigma_{8}=-0.06f \sigma_{8}$. At this point,  we get
\begin{equation}
\frac{\partial \boldsymbol{\mu}}{\partial \theta} \simeq \frac{\boldsymbol{\mu}(\theta+\mathrm{d} \theta)-\boldsymbol{\mu}(\theta-\mathrm{d}\theta)}{2\mathrm{d}\theta}=\frac{\Delta V_i^{(\mathrm{z})}(\nu)}{\Delta\left(f \sigma_{8}\right)}.
\end{equation} 

\begin{figure*}
\centering 
\includegraphics[width=1.0\textwidth]{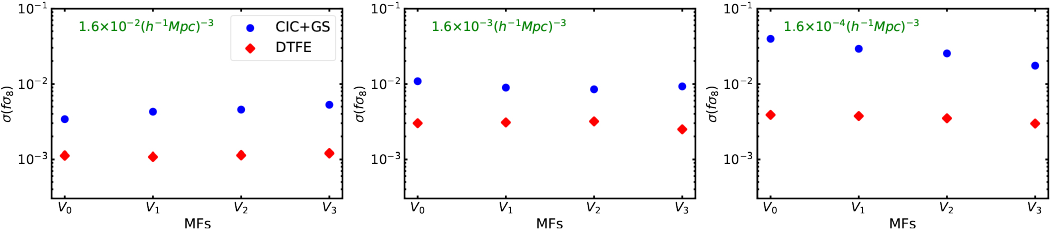}
\caption{The same as Fig.\ \ref{fig:fsigma8_32points}, but for the case of $n_\mathrm{b}=64$. }
\label{fig:fsigma8_64points}
\end{figure*}

In our calculation, we select $n_\mathrm{b}$ data points for each $V_i(\nu)$. These  data points are evenly distributed within $[\nu_\mathrm{min},\nu_\mathrm{max}]$ in logarithmic coordinate. To avoid any problems sourced from the irreversibility of covariance matrices, the used density threshold range $[\nu_\mathrm{lower},\nu_\mathrm{upper}]$ should be narrowed as halo number density decreases. This will result in a decrease in the number of data points actually used in calculations. We illustrate this in Fig.\ \ref{fig:Ranges_64points}\footnote{We only display the results of $n_\mathrm{b}=64$, considering the striking similarity between these results and those of $n_\mathrm{b}=32$.} (cf.\ the unshielded regions) and Table \ref{tab:info}. In order to draw conclusions as reliable as possible, two $n_\mathrm{b}$ cases are tested, namely $n_\mathrm{b}=32$ (cf.\ Fig.\ \ref{fig:fsigma8_32points}) and $n_\mathrm{b}=64$ (cf.\ Fig.\ \ref{fig:fsigma8_64points}). This provides a consistency check for our methodology. In general, in comparison with Fig.\ \ref{fig:fsigma8_32points}, the predictive errors in Fig.\ \ref{fig:fsigma8_64points} are relatively smaller, depending on different cases. Specifically, as the number of data points increases, we find the followings: 
\begin{enumerate}
  \item[(a)] For CIC+GS MFs, the predictive errors exhibit minimal changes, except for a few cases, e.g., the case of $V_0$ with $n_\mathrm{h}=1.6 \times 10^{-2}\,(h^{-1}\mathrm{Mpc})^{-3}$;
  \item[(b)] For DTFE MFs, the predictive errors remain relatively stable in the case of $n_\mathrm{h}=1.6 \times 10^{-2}\,(h^{-1}\mathrm{Mpc})^{-3}$, but show a noticeable decrease in the cases of $n_\mathrm{h}=1.6 \times 10^{-3}\,(h^{-1}\mathrm{Mpc})^{-3}$ and $n_\mathrm{h}=1.6 \times 10^{-4}\,(h^{-1}\mathrm{Mpc})^{-3}$. 
\end{enumerate}
Nevertheless, the performance of DTFE MFs still significantly outperforms that of CIC+GS MFs, thus our conclusions do not change.

\bibliographystyle{yahapj}
\bibliography{mf}

\begin{thebibliography}{}
\providecommand\natexlab[1]{#1}
\providecommand\JournalTitle[1]{#1}

\bibitem[{{Abel} {et~al.}(2012){Abel}, {Hahn}, \&
  {Kaehler}}]{2012MNRAS.427...61A}
{Abel}, T., {Hahn}, O., \& {Kaehler}, R. 2012,
  \href{http://dx.doi.org/10.1111/j.1365-2966.2012.21754.x}{\JournalTitle{\mnras},
  427, 61}

\bibitem[{{Alam} {et~al.}(2017){Alam}, {Ata}, {Bailey}, {Beutler}, {Bizyaev},
  {Blazek}, {Bolton}, {Brownstein}, {Burden}, {Chuang}, {Comparat}, {Cuesta},
  {Dawson}, {Eisenstein}, {Escoffier}, {Gil-Mar{\'\i}n}, {Grieb}, {Hand}, {Ho},
  {Kinemuchi}, {Kirkby}, {Kitaura}, {Malanushenko}, {Malanushenko}, {Maraston},
  {McBride}, {Nichol}, {Olmstead}, {Oravetz}, {Padmanabhan},
  {Palanque-Delabrouille}, {Pan}, {Pellejero-Ibanez}, {Percival}, {Petitjean},
  {Prada}, {Price-Whelan}, {Reid}, {Rodr{\'\i}guez-Torres}, {Roe}, {Ross},
  {Ross}, {Rossi}, {Rubi{\~n}o-Mart{\'\i}n}, {Saito}, {Salazar-Albornoz},
  {Samushia}, {S{\'a}nchez}, {Satpathy}, {Schlegel}, {Schneider},
  {Sc{\'o}ccola}, {Seo}, {Sheldon}, {Simmons}, {Slosar}, {Strauss}, {Swanson},
  {Thomas}, {Tinker}, {Tojeiro}, {Maga{\~n}a}, {Vazquez}, {Verde}, {Wake},
  {Wang}, {Weinberg}, {White}, {Wood-Vasey}, {Y{\`e}che}, {Zehavi}, {Zhai}, \&
  {Zhao}}]{2017MNRAS.470.2617A}
{Alam}, S., {Ata}, M., {Bailey}, S., {et~al.} 2017,
  \href{http://dx.doi.org/10.1093/mnras/stx721}{\JournalTitle{\mnras}, 470,
  2617}

\bibitem[{{Appleby} {et~al.}(2018{\natexlab{a}}){Appleby}, {Chingangbam},
  {Park}, {Hong}, {Kim}, \& {Ganesan}}]{2018ApJ...858...87A}
{Appleby}, S., {Chingangbam}, P., {Park}, C., {et~al.} 2018{\natexlab{a}},
  \href{http://dx.doi.org/10.3847/1538-4357/aabb53}{\JournalTitle{\apj}, 858,
  87}

\bibitem[{{Appleby} {et~al.}(2018{\natexlab{b}}){Appleby}, {Chingangbam},
  {Park}, {Yogendran}, \& {Joby}}]{2018ApJ...863..200A}
{Appleby}, S., {Chingangbam}, P., {Park}, C., {Yogendran}, K.~P., \& {Joby},
  P.~K. 2018{\natexlab{b}},
  \href{http://dx.doi.org/10.3847/1538-4357/aacf8c}{\JournalTitle{\apj}, 863,
  200}

\bibitem[{{Appleby} {et~al.}(2020){Appleby}, {Park}, {Hong}, {Hwang}, \&
  {Kim}}]{2020ApJ...896..145A}
{Appleby}, S., {Park}, C., {Hong}, S.~E., {Hwang}, H.~S., \& {Kim}, J. 2020,
  \href{http://dx.doi.org/10.3847/1538-4357/ab952e}{\JournalTitle{\apj}, 896,
  145}

\bibitem[{{Appleby} {et~al.}(2021){Appleby}, {Park}, {Hong}, {Hwang}, {Kim}, \&
  {Tonegawa}}]{2021ApJ...907...75A}
{Appleby}, S., {Park}, C., {Hong}, S.~E., {et~al.} 2021,
  \href{http://dx.doi.org/10.3847/1538-4357/abcebb}{\JournalTitle{\apj}, 907,
  75}

\bibitem[{{Appleby} {et~al.}(2017){Appleby}, {Park}, {Hong}, \&
  {Kim}}]{2017ApJ...836...45A}
{Appleby}, S., {Park}, C., {Hong}, S.~E., \& {Kim}, J. 2017,
  \href{http://dx.doi.org/10.3847/1538-4357/836/1/45}{\JournalTitle{\apj}, 836,
  45}

\bibitem[{{Appleby} {et~al.}(2018{\natexlab{c}}){Appleby}, {Park}, {Hong}, \&
  {Kim}}]{2018ApJ...853...17A}
{Appleby}, S., {Park}, C., {Hong}, S.~E., \& {Kim}, J. 2018{\natexlab{c}},
  \href{http://dx.doi.org/10.3847/1538-4357/aaa24f}{\JournalTitle{\apj}, 853,
  17}

\bibitem[{{Appleby} {et~al.}(2022){Appleby}, {Park}, {Pranav}, {Hong}, {Hwang},
  {Kim}, \& {Buchert}}]{2022ApJ...928..108A}
{Appleby}, S., {Park}, C., {Pranav}, P., {et~al.} 2022,
  \href{http://dx.doi.org/10.3847/1538-4357/ac562a}{\JournalTitle{\apj}, 928,
  108}

\bibitem[{{Aragon-Calvo}(2021)}]{2021MNRAS.503..557A}
{Aragon-Calvo}, M.~A. 2021,
  \href{http://dx.doi.org/10.1093/mnras/stab403}{\JournalTitle{\mnras}, 503,
  557}

\bibitem[{{Arag{\'o}n-Calvo} {et~al.}(2007){Arag{\'o}n-Calvo}, {Jones}, {van de
  Weygaert}, \& {van der Hulst}}]{2007A&A...474..315A}
{Arag{\'o}n-Calvo}, M.~A., {Jones}, B.~J.~T., {van de Weygaert}, R., \& {van
  der Hulst}, J.~M. 2007,
  \href{http://dx.doi.org/10.1051/0004-6361:20077880}{\JournalTitle{\aap}, 474,
  315}

\bibitem[{{Arag{\'o}n-Calvo} {et~al.}(2010){Arag{\'o}n-Calvo}, {Platen}, {van
  de Weygaert}, \& {Szalay}}]{2010ApJ...723..364A}
{Arag{\'o}n-Calvo}, M.~A., {Platen}, E., {van de Weygaert}, R., \& {Szalay},
  A.~S. 2010,
  \href{http://dx.doi.org/10.1088/0004-637X/723/1/364}{\JournalTitle{\apj},
  723, 364}

\bibitem[{{Aragon-Calvo} {et~al.}(2010){Aragon-Calvo}, {Shandarin}, \&
  {Szalay}}]{2010arXiv1006.4178A}
{Aragon-Calvo}, M.~A., {Shandarin}, S.~F., \& {Szalay}, A. 2010,
  \JournalTitle{arXiv e-prints}, arXiv:1006.4178

\bibitem[{{Arnalte-Mur} {et~al.}(2012){Arnalte-Mur}, {Labatie}, {Clerc},
  {Mart{\'\i}nez}, {Starck}, {Lachi{\`e}ze-Rey}, {Saar}, \&
  {Paredes}}]{2012A&A...542A..34A}
{Arnalte-Mur}, P., {Labatie}, A., {Clerc}, N., {et~al.} 2012,
  \href{http://dx.doi.org/10.1051/0004-6361/201118017}{\JournalTitle{\aap},
  542, A34}

\bibitem[{{Barrow} {et~al.}(1985){Barrow}, {Bhavsar}, \&
  {Sonoda}}]{1985MNRAS.216...17B}
{Barrow}, J.~D., {Bhavsar}, S.~P., \& {Sonoda}, D.~H. 1985,
  \href{http://dx.doi.org/10.1093/mnras/216.1.17}{\JournalTitle{\mnras}, 216,
  17}

\bibitem[{{Basilakos}(2003)}]{2003MNRAS.344..602B}
{Basilakos}, S. 2003,
  \href{http://dx.doi.org/10.1046/j.1365-8711.2003.06845.x}{\JournalTitle{\mnras},
  344, 602}

\bibitem[{{Bautista} {et~al.}(2021){Bautista}, {Paviot}, {Vargas Maga{\~n}a},
  {de la Torre}, {Fromenteau}, {Gil-Mar{\'\i}n}, {Ross}, {Burtin}, {Dawson},
  {Hou}, {Kneib}, {de Mattia}, {Percival}, {Rossi}, {Tojeiro}, {Zhao}, {Zhao},
  {Alam}, {Brownstein}, {Chapman}, {Choi}, {Chuang}, {Escoffier}, {de la
  Macorra}, {du Mas des Bourboux}, {Mohammad}, {Moon}, {M{\"u}ller},
  {Nadathur}, {Newman}, {Schneider}, {Seo}, \& {Wang}}]{2021MNRAS.500..736B}
{Bautista}, J.~E., {Paviot}, R., {Vargas Maga{\~n}a}, M., {et~al.} 2021,
  \href{http://dx.doi.org/10.1093/mnras/staa2800}{\JournalTitle{\mnras}, 500,
  736}

\bibitem[{{Bermejo} {et~al.}(2024){Bermejo}, {Wilding}, {van de Weygaert},
  {Jones}, {Vegter}, \& {Efstathiou}}]{2024MNRAS.529.4325B}
{Bermejo}, R., {Wilding}, G., {van de Weygaert}, R., {et~al.} 2024,
  \href{http://dx.doi.org/10.1093/mnras/stae543}{\JournalTitle{\mnras}, 529,
  4325}

\bibitem[{{Bernardeau} \& {van de Weygaert}(1996)}]{1996MNRAS.279..693B}
{Bernardeau}, F., \& {van de Weygaert}, R. 1996,
  \href{http://dx.doi.org/10.1093/mnras/279.2.693}{\JournalTitle{\mnras}, 279,
  693}

\bibitem[{{Blake} {et~al.}(2014){Blake}, {James}, \&
  {Poole}}]{2014MNRAS.437.2488B}
{Blake}, C., {James}, J.~B., \& {Poole}, G.~B. 2014,
  \href{http://dx.doi.org/10.1093/mnras/stt2062}{\JournalTitle{\mnras}, 437,
  2488}

\bibitem[{{Cai} {et~al.}(2011){Cai}, {Bernstein}, \&
  {Sheth}}]{2011MNRAS.412..995C}
{Cai}, Y.-C., {Bernstein}, G., \& {Sheth}, R.~K. 2011,
  \href{http://dx.doi.org/10.1111/j.1365-2966.2010.17969.x}{\JournalTitle{\mnras},
  412, 995}

\bibitem[{{Cautun} {et~al.}(2013){Cautun}, {van de Weygaert}, \&
  {Jones}}]{2013MNRAS.429.1286C}
{Cautun}, M., {van de Weygaert}, R., \& {Jones}, B. J.~T. 2013,
  \href{http://dx.doi.org/10.1093/mnras/sts416}{\JournalTitle{\mnras}, 429,
  1286}

\bibitem[{{Cautun} \& {van de
  Weygaert}(2011{\natexlab{a}})}]{2011arXiv1105.0370C}
{Cautun}, M.~C., \& {van de Weygaert}, R. 2011{\natexlab{a}},
  \JournalTitle{arXiv e-prints}, arXiv:1105.0370

\bibitem[{{Cautun} \& {van de
  Weygaert}(2011{\natexlab{b}})}]{2011ascl.Soft05003c}
{Cautun}, M.~C., \& {van de Weygaert}, R. 2011{\natexlab{b}}, {The DTFE public
  software: The Delaunay Tessellation Field Estimator code}

\bibitem[{{Chan} {et~al.}(2014){Chan}, {Hamaus}, \&
  {Desjacques}}]{2014PhRvD..90j3521C}
{Chan}, K.~C., {Hamaus}, N., \& {Desjacques}, V. 2014,
  \href{http://dx.doi.org/10.1103/PhysRevD.90.103521}{\JournalTitle{\prd}, 90,
  103521}

\bibitem[{{Chen} {et~al.}(2018){Chen}, {Zhang}, {Zheng}, {Yu}, \&
  {Jing}}]{2018ApJ...861...58C}
{Chen}, J., {Zhang}, P., {Zheng}, Y., {Yu}, Y., \& {Jing}, Y. 2018,
  \href{http://dx.doi.org/10.3847/1538-4357/aaca2f}{\JournalTitle{\apj}, 861,
  58}

\bibitem[{{Cheng} \& {M{\'e}nard}(2021)}]{2021MNRAS.507.1012C}
{Cheng}, S., \& {M{\'e}nard}, B. 2021,
  \href{http://dx.doi.org/10.1093/mnras/stab2102}{\JournalTitle{\mnras}, 507,
  1012}

\bibitem[{{Cheng} {et~al.}(2020){Cheng}, {Ting}, {M{\'e}nard}, \&
  {Bruna}}]{2020MNRAS.499.5902C}
{Cheng}, S., {Ting}, Y.-S., {M{\'e}nard}, B., \& {Bruna}, J. 2020,
  \href{http://dx.doi.org/10.1093/mnras/staa3165}{\JournalTitle{\mnras}, 499,
  5902}

\bibitem[{{Choi} {et~al.}(2010){Choi}, {Park}, {Kim}, {Gott}, {Weinberg},
  {Vogeley}, {Kim}, \& {SDSS Collaboration}}]{2010ApJS..190..181C}
{Choi}, Y.-Y., {Park}, C., {Kim}, J., {et~al.} 2010,
  \href{http://dx.doi.org/10.1088/0067-0049/190/1/181}{\JournalTitle{\apjs},
  190, 181}

\bibitem[{{Codis} {et~al.}(2013){Codis}, {Pichon}, {Pogosyan}, {Bernardeau}, \&
  {Matsubara}}]{2013MNRAS.435..531C}
{Codis}, S., {Pichon}, C., {Pogosyan}, D., {Bernardeau}, F., \& {Matsubara}, T.
  2013, \href{http://dx.doi.org/10.1093/mnras/stt1316}{\JournalTitle{\mnras},
  435, 531}

\bibitem[{{de Jong} {et~al.}(2012){de Jong}, {Bellido-Tirado}, {Chiappini},
  {Depagne}, {Haynes}, {Johl}, {Schnurr}, {Schwope}, {Walcher}, {Dionies},
  {Haynes}, {Kelz}, {Kitaura}, {Lamer}, {Minchev}, {M{\"u}ller}, {Nuza},
  {Olaya}, {Piffl}, {Popow}, {Steinmetz}, {Ural}, {Williams}, {Winkler},
  {Wisotzki}, {Ansorge}, {Banerji}, {Gonzalez Solares}, {Irwin}, {Kennicutt},
  {King}, {McMahon}, {Koposov}, {Parry}, {Sun}, {Walton}, {Finger}, {Iwert},
  {Krumpe}, {Lizon}, {Vincenzo}, {Amans}, {Bonifacio}, {Cohen}, {Francois},
  {Jagourel}, {Mignot}, {Royer}, {Sartoretti}, {Bender}, {Grupp}, {Hess},
  {Lang-Bardl}, {Muschielok}, {B{\"o}hringer}, {Boller}, {Bongiorno}, {Brusa},
  {Dwelly}, {Merloni}, {Nandra}, {Salvato}, {Pragt}, {Navarro}, {Gerlofsma},
  {Roelfsema}, {Dalton}, {Middleton}, {Tosh}, {Boeche}, {Caffau}, {Christlieb},
  {Grebel}, {Hansen}, {Koch}, {Ludwig}, {Quirrenbach}, {Sbordone}, {Seifert},
  {Thimm}, {Trifonov}, {Helmi}, {Trager}, {Feltzing}, {Korn}, \&
  {Boland}}]{2012SPIE.8446E..0TD}
{de Jong}, R.~S., {Bellido-Tirado}, O., {Chiappini}, C., {et~al.} 2012,
  \href{http://dx.doi.org/10.1117/12.926239}{in Society of Photo-Optical
  Instrumentation Engineers (SPIE) Conference Series, Vol. 8446, Ground-based
  and Airborne Instrumentation for Astronomy IV, ed. I.~S. {McLean}, S.~K.
  {Ramsay}, \& H.~{Takami}}, 84460T

\bibitem[{{de Lapparent} {et~al.}(1991){de Lapparent}, {Geller}, \&
  {Huchra}}]{1991ApJ...369..273D}
{de Lapparent}, V., {Geller}, M.~J., \& {Huchra}, J.~P. 1991,
  \href{http://dx.doi.org/10.1086/169759}{\JournalTitle{\apj}, 369, 273}

\bibitem[{{DESI Collaboration} {et~al.}(2016){DESI Collaboration}, {Aghamousa},
  {Aguilar}, {Ahlen}, {Alam}, {Allen}, {Allende Prieto}, {Annis}, {Bailey},
  {Balland}, {Ballester}, {Baltay}, {Beaufore}, {Bebek}, {Beers}, {Bell},
  {Bernal}, {Besuner}, {Beutler}, {Blake}, {Bleuler}, {Blomqvist}, {Blum},
  {Bolton}, {Briceno}, {Brooks}, {Brownstein}, {Buckley-Geer}, {Burden},
  {Burtin}, {Busca}, {Cahn}, {Cai}, {Cardiel-Sas}, {Carlberg}, {Carton},
  {Casas}, {Castander}, {Cervantes-Cota}, {Claybaugh}, {Close}, {Coker},
  {Cole}, {Comparat}, {Cooper}, {Cousinou}, {Crocce}, {Cuby}, {Cunningham},
  {Davis}, {Dawson}, {de la Macorra}, {De Vicente}, {Delubac}, {Derwent},
  {Dey}, {Dhungana}, {Ding}, {Doel}, {Duan}, {Ealet}, {Edelstein},
  {Eftekharzadeh}, {Eisenstein}, {Elliott}, {Escoffier}, {Evatt}, {Fagrelius},
  {Fan}, {Fanning}, {Farahi}, {Farihi}, {Favole}, {Feng}, {Fernandez},
  {Findlay}, {Finkbeiner}, {Fitzpatrick}, {Flaugher}, {Flender}, {Font-Ribera},
  {Forero-Romero}, {Fosalba}, {Frenk}, {Fumagalli}, {Gaensicke}, {Gallo},
  {Garcia-Bellido}, {Gaztanaga}, {Pietro Gentile Fusillo}, {Gerard},
  {Gershkovich}, {Giannantonio}, {Gillet}, {Gonzalez-de-Rivera},
  {Gonzalez-Perez}, {Gott}, {Graur}, {Gutierrez}, {Guy}, {Habib}, {Heetderks},
  {Heetderks}, {Heitmann}, {Hellwing}, {Herrera}, {Ho}, {Holland}, {Honscheid},
  {Huff}, {Hutchinson}, {Huterer}, {Hwang}, {Illa Laguna}, {Ishikawa},
  {Jacobs}, {Jeffrey}, {Jelinsky}, {Jennings}, {Jiang}, {Jimenez}, {Johnson},
  {Joyce}, {Jullo}, {Juneau}, {Kama}, {Karcher}, {Karkar}, {Kehoe}, {Kennamer},
  {Kent}, {Kilbinger}, {Kim}, {Kirkby}, {Kisner}, {Kitanidis}, {Kneib},
  {Koposov}, {Kovacs}, {Koyama}, {Kremin}, {Kron}, {Kronig}, {Kueter-Young},
  {Lacey}, {Lafever}, {Lahav}, {Lambert}, {Lampton}, {Landriau}, {Lang},
  {Lauer}, {Le Goff}, {Le Guillou}, {Le Van Suu}, {Lee}, {Lee}, {Leitner},
  {Lesser}, {Levi}, {L'Huillier}, {Li}, {Liang}, {Lin}, {Linder}, {Loebman},
  {Luki{\'c}}, {Ma}, {MacCrann}, {Magneville}, {Makarem}, {Manera}, {Manser},
  {Marshall}, {Martini}, {Massey}, {Matheson}, {McCauley}, {McDonald},
  {McGreer}, {Meisner}, {Metcalfe}, {Miller}, {Miquel}, {Moustakas}, {Myers},
  {Naik}, {Newman}, {Nichol}, {Nicola}, {Nicolati da Costa}, {Nie}, {Niz},
  {Norberg}, {Nord}, {Norman}, {Nugent}, {O'Brien}, {Oh}, {Olsen}, {Padilla},
  {Padmanabhan}, {Padmanabhan}, {Palanque-Delabrouille}, {Palmese},
  {Pappalardo}, {P{\^a}ris}, {Park}, {Patej}, {Peacock}, {Peiris}, {Peng},
  {Percival}, {Perruchot}, {Pieri}, {Pogge}, {Pollack}, {Poppett}, {Prada},
  {Prakash}, {Probst}, {Rabinowitz}, {Raichoor}, {Ree}, {Refregier}, {Regal},
  {Reid}, {Reil}, {Rezaie}, {Rockosi}, {Roe}, {Ronayette}, {Roodman}, {Ross},
  {Ross}, {Rossi}, {Rozo}, {Ruhlmann-Kleider}, {Rykoff}, {Sabiu}, {Samushia},
  {Sanchez}, {Sanchez}, {Schlegel}, {Schneider}, {Schubnell}, {Secroun},
  {Seljak}, {Seo}, {Serrano}, {Shafieloo}, {Shan}, {Sharples}, {Sholl},
  {Shourt}, {Silber}, {Silva}, {Sirk}, {Slosar}, {Smith}, {Smoot}, {Som},
  {Song}, {Sprayberry}, {Staten}, {Stefanik}, {Tarle}, {Sien Tie}, {Tinker},
  {Tojeiro}, {Valdes}, {Valenzuela}, {Valluri}, {Vargas-Magana}, {Verde},
  {Walker}, {Wang}, {Wang}, {Weaver}, {Weaverdyck}, {Wechsler}, {Weinberg},
  {White}, {Yang}, {Yeche}, {Zhang}, {Zhao}, {Zheng}, {Zhou}, {Zhou}, {Zhu},
  {Zou}, \& {Zu}}]{2016arXiv161100036D}
{DESI Collaboration}, {Aghamousa}, A., {Aguilar}, J., {et~al.} 2016,
  \JournalTitle{arXiv e-prints}, arXiv:1611.00036

\bibitem[{{Dor{\'e}} {et~al.}(2014){Dor{\'e}}, {Bock}, {Ashby}, {Capak},
  {Cooray}, {de Putter}, {Eifler}, {Flagey}, {Gong}, {Habib}, {Heitmann},
  {Hirata}, {Jeong}, {Katti}, {Korngut}, {Krause}, {Lee}, {Masters},
  {Mauskopf}, {Melnick}, {Mennesson}, {Nguyen}, {{\"O}berg}, {Pullen},
  {Raccanelli}, {Smith}, {Song}, {Tolls}, {Unwin}, {Venumadhav}, {Viero},
  {Werner}, \& {Zemcov}}]{2014arXiv1412.4872D}
{Dor{\'e}}, O., {Bock}, J., {Ashby}, M., {et~al.} 2014, \JournalTitle{arXiv
  e-prints}, arXiv:1412.4872

\bibitem[{{Dor{\'e}} {et~al.}(2018){Dor{\'e}}, {Hirata}, {Wang}, {Weinberg},
  {Baronchelli}, {Benson}, {Capak}, {Choi}, {Eifler}, {Hemmati}, {Ho}, {Izard},
  {Jain}, {Jarvis}, {Kiessling}, {Krause}, {Massara}, {Masters}, {Merson},
  {Miyatake}, {Plazas Malagon}, {Mandelbaum}, {Samushia}, {Shapiro}, {Simet},
  {Spergel}, {Teplitz}, {Troxel}, {Bean}, {Colbert}, {Heinrich}, {Heitmann},
  {Helou}, {Hudson}, {Huff}, {Leauthaud}, {MacCrann}, {Padmanabhan}, {Pisani},
  {Rhodes}, {Rozo}, {Seiffert}, {Smith}, {Takada}, {von der Linden}, {Lupton},
  {Yoshida}, {Wu}, \& {Zu}}]{2018arXiv180403628D}
{Dor{\'e}}, O., {Hirata}, C., {Wang}, Y., {et~al.} 2018, \JournalTitle{arXiv
  e-prints}, arXiv:1804.03628

\bibitem[{{Douglass} {et~al.}(2022){Douglass}, {Veyrat}, {O'Neill}, {BenZvi},
  {Zaidouni}, \& {Guzzetti}}]{2022JOSS....7.4033D}
{Douglass}, K., {Veyrat}, D., {O'Neill}, S., {et~al.} 2022,
  \href{http://dx.doi.org/10.21105/joss.04033}{\JournalTitle{The Journal of
  Open Source Software}, 7, 4033}

\bibitem[{{Douglass} {et~al.}(2023){Douglass}, {Veyrat}, \&
  {BenZvi}}]{2023ApJS..265....7D}
{Douglass}, K.~A., {Veyrat}, D., \& {BenZvi}, S. 2023,
  \href{http://dx.doi.org/10.3847/1538-4365/acabcf}{\JournalTitle{\apjs}, 265,
  7}

\bibitem[{{Ducout} {et~al.}(2013){Ducout}, {Bouchet}, {Colombi}, {Pogosyan}, \&
  {Prunet}}]{2013MNRAS.429.2104D}
{Ducout}, A., {Bouchet}, F.~R., {Colombi}, S., {Pogosyan}, D., \& {Prunet}, S.
  2013, \href{http://dx.doi.org/10.1093/mnras/sts483}{\JournalTitle{\mnras},
  429, 2104}

\bibitem[{{Eisenstein} {et~al.}(2005){Eisenstein}, {Zehavi}, {Hogg},
  {Scoccimarro}, {Blanton}, {Nichol}, {Scranton}, {Seo}, {Tegmark}, {Zheng},
  {Anderson}, {Annis}, {Bahcall}, {Brinkmann}, {Burles}, {Castander},
  {Connolly}, {Csabai}, {Doi}, {Fukugita}, {Frieman}, {Glazebrook}, {Gunn},
  {Hendry}, {Hennessy}, {Ivezi{\'c}}, {Kent}, {Knapp}, {Lin}, {Loh}, {Lupton},
  {Margon}, {McKay}, {Meiksin}, {Munn}, {Pope}, {Richmond}, {Schlegel},
  {Schneider}, {Shimasaku}, {Stoughton}, {Strauss}, {SubbaRao}, {Szalay},
  {Szapudi}, {Tucker}, {Yanny}, \& {York}}]{2005ApJ...633..560E}
{Eisenstein}, D.~J., {Zehavi}, I., {Hogg}, D.~W., {et~al.} 2005,
  \href{http://dx.doi.org/10.1086/466512}{\JournalTitle{\apj}, 633, 560}

\bibitem[{{Elbers} \& {van de Weygaert}(2019)}]{2019MNRAS.486.1523E}
{Elbers}, W., \& {van de Weygaert}, R. 2019,
  \href{http://dx.doi.org/10.1093/mnras/stz908}{\JournalTitle{\mnras}, 486,
  1523}

\bibitem[{{Elbers} \& {van de Weygaert}(2023)}]{2023MNRAS.520.2709E}
{Elbers}, W., \& {van de Weygaert}, R. 2023,
  \href{http://dx.doi.org/10.1093/mnras/stad120}{\JournalTitle{\mnras}, 520,
  2709}

\bibitem[{{Ellis} \& {Dawson}(2019)}]{2019BAAS...51g..45E}
{Ellis}, R., \& {Dawson}, K. 2019,
  \href{http://dx.doi.org/10.48550/arXiv.1907.06797}{in Bulletin of the
  American Astronomical Society, Vol.~51}, 45

\bibitem[{{Emberson} {et~al.}(2017){Emberson}, {Yu}, {Inman}, {Zhang}, {Pen},
  {Harnois-D{\'e}raps}, {Yuan}, {Teng}, {Zhu}, {Chen}, \&
  {Xing}}]{2017RAA....17...85E}
{Emberson}, J.~D., {Yu}, H.-R., {Inman}, D., {et~al.} 2017,
  \href{http://dx.doi.org/10.1088/1674-4527/17/8/85}{\JournalTitle{Research in
  Astronomy and Astrophysics}, 17, 085}

\bibitem[{{Euclid Collaboration} {et~al.}(2020){Euclid Collaboration},
  {Blanchard}, {Camera}, {Carbone}, {Cardone}, {Casas}, {Clesse}, {Ili{\'c}},
  {Kilbinger}, {Kitching}, {Kunz}, {Lacasa}, {Linder}, {Majerotto},
  {Markovi{\v{c}}}, {Martinelli}, {Pettorino}, {Pourtsidou}, {Sakr},
  {S{\'a}nchez}, {Sapone}, {Tutusaus}, {Yahia-Cherif}, {Yankelevich},
  {Andreon}, {Aussel}, {Balaguera-Antol{\'\i}nez}, {Baldi}, {Bardelli},
  {Bender}, {Biviano}, {Bonino}, {Boucaud}, {Bozzo}, {Branchini}, {Brau-Nogue},
  {Brescia}, {Brinchmann}, {Burigana}, {Cabanac}, {Capobianco}, {Cappi},
  {Carretero}, {Carvalho}, {Casas}, {Castander}, {Castellano}, {Cavuoti},
  {Cimatti}, {Cledassou}, {Colodro-Conde}, {Congedo}, {Conselice}, {Conversi},
  {Copin}, {Corcione}, {Coupon}, {Courtois}, {Cropper}, {Da Silva}, {de la
  Torre}, {Di Ferdinando}, {Dubath}, {Ducret}, {Duncan}, {Dupac}, {Dusini},
  {Fabbian}, {Fabricius}, {Farrens}, {Fosalba}, {Fotopoulou}, {Fourmanoit},
  {Frailis}, {Franceschi}, {Franzetti}, {Fumana}, {Galeotta}, {Gillard},
  {Gillis}, {Giocoli}, {G{\'o}mez-Alvarez}, {Graci{\'a}-Carpio}, {Grupp},
  {Guzzo}, {Hoekstra}, {Hormuth}, {Israel}, {Jahnke}, {Keihanen}, {Kermiche},
  {Kirkpatrick}, {Kohley}, {Kubik}, {Kurki-Suonio}, {Ligori}, {Lilje}, {Lloro},
  {Maino}, {Maiorano}, {Marggraf}, {Martinet}, {Marulli}, {Massey},
  {Medinaceli}, {Mei}, {Mellier}, {Metcalf}, {Metge}, {Meylan}, {Moresco},
  {Moscardini}, {Munari}, {Nichol}, {Niemi}, {Nucita}, {Padilla}, {Paltani},
  {Pasian}, {Percival}, {Pires}, {Polenta}, {Poncet}, {Pozzetti}, {Racca},
  {Raison}, {Renzi}, {Rhodes}, {Romelli}, {Roncarelli}, {Rossetti}, {Saglia},
  {Schneider}, {Scottez}, {Secroun}, {Sirri}, {Stanco}, {Starck}, {Sureau},
  {Tallada-Cresp{\'\i}}, {Tavagnacco}, {Taylor}, {Tenti}, {Tereno},
  {Toledo-Moreo}, {Torradeflot}, {Valenziano}, {Vassallo}, {Verdoes Kleijn},
  {Viel}, {Wang}, {Zacchei}, {Zoubian}, \& {Zucca}}]{2020A&A...642A.191E}
{Euclid Collaboration}, {Blanchard}, A., {Camera}, S., {et~al.} 2020,
  \href{http://dx.doi.org/10.1051/0004-6361/202038071}{\JournalTitle{\aap},
  642, A191}

\bibitem[{{Fang} {et~al.}(2019){Fang}, {Forero-Romero}, {Rossi}, {Li}, \&
  {Feng}}]{2019MNRAS.485.5276F}
{Fang}, F., {Forero-Romero}, J., {Rossi}, G., {Li}, X.-D., \& {Feng}, L.-L.
  2019, \href{http://dx.doi.org/10.1093/mnras/stz773}{\JournalTitle{\mnras},
  485, 5276}

\bibitem[{{Fang} {et~al.}(2017){Fang}, {Li}, \& {Zhao}}]{2017PhRvL.118r1301F}
{Fang}, W., {Li}, B., \& {Zhao}, G.-B. 2017,
  \href{http://dx.doi.org/10.1103/PhysRevLett.118.181301}{\JournalTitle{\prl},
  118, 181301}

\bibitem[{{Feldbrugge}(2024)}]{2024arXiv240216234F}
{Feldbrugge}, J. 2024,
  \href{http://dx.doi.org/10.48550/arXiv.2402.16234}{\JournalTitle{arXiv
  e-prints}, arXiv:2402.16234}

\bibitem[{{Gay} {et~al.}(2012){Gay}, {Pichon}, \&
  {Pogosyan}}]{2012PhRvD..85b3011G}
{Gay}, C., {Pichon}, C., \& {Pogosyan}, D. 2012,
  \href{http://dx.doi.org/10.1103/PhysRevD.85.023011}{\JournalTitle{\prd}, 85,
  023011}

\bibitem[{{Giri} \& {Mellema}(2021)}]{2021MNRAS.505.1863G}
{Giri}, S.~K., \& {Mellema}, G. 2021,
  \href{http://dx.doi.org/10.1093/mnras/stab1320}{\JournalTitle{\mnras}, 505,
  1863}

\bibitem[{{Gong} {et~al.}(2019){Gong}, {Liu}, {Cao}, {Chen}, {Fan}, {Li}, {Li},
  {Li}, {Zhang}, \& {Zhan}}]{2019ApJ...883..203G}
{Gong}, Y., {Liu}, X., {Cao}, Y., {et~al.} 2019,
  \href{http://dx.doi.org/10.3847/1538-4357/ab391e}{\JournalTitle{\apj}, 883,
  203}

\bibitem[{{Gott} {et~al.}(1986){Gott}, {Melott}, \&
  {Dickinson}}]{1986ApJ...306..341G}
{Gott}, J.~Richard, I., {Melott}, A.~L., \& {Dickinson}, M. 1986,
  \href{http://dx.doi.org/10.1086/164347}{\JournalTitle{\apj}, 306, 341}

\bibitem[{{Gott} {et~al.}(1987){Gott}, {Weinberg}, \&
  {Melott}}]{1987ApJ...319....1G}
{Gott}, J.~Richard, I., {Weinberg}, D.~H., \& {Melott}, A.~L. 1987,
  \href{http://dx.doi.org/10.1086/165427}{\JournalTitle{\apj}, 319, 1}

\bibitem[{{Gott} {et~al.}(1989){Gott}, {Miller}, {Thuan}, {Schneider},
  {Weinberg}, {Gammie}, {Polk}, {Vogeley}, {Jeffrey}, {Bhavsar}, {Melott},
  {Giovanelli}, {Hayes}, {Tully}, \& {Hamilton}}]{1989ApJ...340..625G}
{Gott}, J.~Richard, I., {Miller}, J., {Thuan}, T.~X., {et~al.} 1989,
  \href{http://dx.doi.org/10.1086/167425}{\JournalTitle{\apj}, 340, 625}

\bibitem[{{Haas} {et~al.}(2012){Haas}, {Schaye}, \&
  {Jeeson-Daniel}}]{2012MNRAS.419.2133H}
{Haas}, M.~R., {Schaye}, J., \& {Jeeson-Daniel}, A. 2012,
  \href{http://dx.doi.org/10.1111/j.1365-2966.2011.19863.x}{\JournalTitle{\mnras},
  419, 2133}

\bibitem[{Hadwiger(Springer, Berlin, 1957)}]{Hadwiger1957Vorlesungen}
Hadwiger, H. Springer, Berlin, 1957, Vorlesungen \"{u}ber Inhalt,
  Oberfl\"{a}che und Isoperimetrie

\bibitem[{{Hamaus} {et~al.}(2011){Hamaus}, {Seljak}, \&
  {Desjacques}}]{2011PhRvD..84h3509H}
{Hamaus}, N., {Seljak}, U., \& {Desjacques}, V. 2011,
  \href{http://dx.doi.org/10.1103/PhysRevD.84.083509}{\JournalTitle{\prd}, 84,
  083509}

\bibitem[{{Hamaus} {et~al.}(2012){Hamaus}, {Seljak}, \&
  {Desjacques}}]{2012PhRvD..86j3513H}
{Hamaus}, N., {Seljak}, U., \& {Desjacques}, V. 2012,
  \href{http://dx.doi.org/10.1103/PhysRevD.86.103513}{\JournalTitle{\prd}, 86,
  103513}

\bibitem[{{Hamaus} {et~al.}(2010){Hamaus}, {Seljak}, {Desjacques}, {Smith}, \&
  {Baldauf}}]{2010PhRvD..82d3515H}
{Hamaus}, N., {Seljak}, U., {Desjacques}, V., {Smith}, R.~E., \& {Baldauf}, T.
  2010,
  \href{http://dx.doi.org/10.1103/PhysRevD.82.043515}{\JournalTitle{\prd}, 82,
  043515}

\bibitem[{{Hamilton}(2001)}]{2001MNRAS.322..419H}
{Hamilton}, A.~J.~S. 2001,
  \href{http://dx.doi.org/10.1046/j.1365-8711.2001.04137.x}{\JournalTitle{\mnras},
  322, 419}

\bibitem[{{Hamilton} {et~al.}(1986){Hamilton}, {Gott}, \&
  {Weinberg}}]{1986ApJ...309....1H}
{Hamilton}, A.~J.~S., {Gott}, J.~Richard, I., \& {Weinberg}, D. 1986,
  \href{http://dx.doi.org/10.1086/164571}{\JournalTitle{\apj}, 309, 1}

\bibitem[{{Harnois-D{\'e}raps} {et~al.}(2013){Harnois-D{\'e}raps}, {Pen},
  {Iliev}, {Merz}, {Emberson}, \& {Desjacques}}]{2013MNRAS.436..540H}
{Harnois-D{\'e}raps}, J., {Pen}, U.-L., {Iliev}, I.~T., {et~al.} 2013,
  \href{http://dx.doi.org/10.1093/mnras/stt1591}{\JournalTitle{\mnras}, 436,
  540}

\bibitem[{{Hartlap} {et~al.}(2007){Hartlap}, {Simon}, \&
  {Schneider}}]{2007A&A...464..399H}
{Hartlap}, J., {Simon}, P., \& {Schneider}, P. 2007,
  \href{http://dx.doi.org/10.1051/0004-6361:20066170}{\JournalTitle{\aap}, 464,
  399}

\bibitem[{{Hikage} {et~al.}(2008){Hikage}, {Coles}, {Grossi}, {Moscardini},
  {Dolag}, {Branchini}, \& {Matarrese}}]{2008MNRAS.385.1613H}
{Hikage}, C., {Coles}, P., {Grossi}, M., {et~al.} 2008,
  \href{http://dx.doi.org/10.1111/j.1365-2966.2008.12944.x}{\JournalTitle{\mnras},
  385, 1613}

\bibitem[{{Hikage} {et~al.}(2006){Hikage}, {Komatsu}, \&
  {Matsubara}}]{2006ApJ...653...11H}
{Hikage}, C., {Komatsu}, E., \& {Matsubara}, T. 2006,
  \href{http://dx.doi.org/10.1086/508653}{\JournalTitle{\apj}, 653, 11}

\bibitem[{{Hikage} {et~al.}(2003){Hikage}, {Schmalzing}, {Buchert}, {Suto},
  {Kayo}, {Taruya}, {Vogeley}, {Hoyle}, {Gott}, \&
  {Brinkmann}}]{2003PASJ...55..911H}
{Hikage}, C., {Schmalzing}, J., {Buchert}, T., {et~al.} 2003,
  \href{http://dx.doi.org/10.1093/pasj/55.5.911}{\JournalTitle{\pasj}, 55, 911}

\bibitem[{{Hill} {et~al.}(2008){Hill}, {Gebhardt}, {Komatsu}, {Drory},
  {MacQueen}, {Adams}, {Blanc}, {Koehler}, {Rafal}, {Roth}, {Kelz}, {Gronwall},
  {Ciardullo}, \& {Schneider}}]{2008ASPC..399..115H}
{Hill}, G.~J., {Gebhardt}, K., {Komatsu}, E., {et~al.} 2008, in Astronomical
  Society of the Pacific Conference Series, Vol. 399, Panoramic Views of Galaxy
  Formation and Evolution, ed. T.~{Kodama}, T.~{Yamada}, \& K.~{Aoki}, 115

\bibitem[{{Hou} {et~al.}(2021){Hou}, {Cahn}, {Philcox}, \&
  {Slepian}}]{2021arXiv210801714H}
{Hou}, J., {Cahn}, R.~N., {Philcox}, O. H.~E., \& {Slepian}, Z. 2021,
  \JournalTitle{arXiv e-prints}, arXiv:2108.01714

\bibitem[{{Jackson}(1972)}]{1972MNRAS.156P...1J}
{Jackson}, J.~C. 1972,
  \href{http://dx.doi.org/10.1093/mnras/156.1.1P}{\JournalTitle{\mnras}, 156,
  1P}

\bibitem[{{Jalali Kanafi} {et~al.}(2023){Jalali Kanafi}, {Ansarifard}, \&
  {Movahed}}]{2023arXiv231113520J}
{Jalali Kanafi}, M.~H., {Ansarifard}, S., \& {Movahed}, S.~M.~S. 2023,
  \href{http://dx.doi.org/10.48550/arXiv.2311.13520}{\JournalTitle{arXiv
  e-prints}, arXiv:2311.13520}

\bibitem[{{James}(2012)}]{2012ApJ...751...40J}
{James}, J.~B. 2012,
  \href{http://dx.doi.org/10.1088/0004-637X/751/1/40}{\JournalTitle{\apj}, 751,
  40}

\bibitem[{{Jamieson} \& {Loverde}(2021)}]{2021PhRvD.103j3522J}
{Jamieson}, D., \& {Loverde}, M. 2021,
  \href{http://dx.doi.org/10.1103/PhysRevD.103.103522}{\JournalTitle{\prd},
  103, 103522}

\bibitem[{{Jiang} {et~al.}(2021){Jiang}, {Liu}, {Fang}, \&
  {Zhao}}]{2021arXiv210803851J}
{Jiang}, A., {Liu}, W., {Fang}, W., \& {Zhao}, W. 2021, \JournalTitle{arXiv
  e-prints}, arXiv:2108.03851

\bibitem[{{Jiang} {et~al.}(2023){Jiang}, {Liu}, {Li}, {Barrera-Hinojosa},
  {Zhang}, \& {Fang}}]{2023arXiv230504520J}
{Jiang}, A., {Liu}, W., {Li}, B., {et~al.} 2023,
  \href{http://dx.doi.org/10.48550/arXiv.2305.04520}{\JournalTitle{arXiv
  e-prints}, arXiv:2305.04520}

\bibitem[{{Jing}(2019)}]{2019SCPMA..6219511J}
{Jing}, Y. 2019,
  \href{http://dx.doi.org/10.1007/s11433-018-9286-x}{\JournalTitle{Science
  China Physics, Mechanics, and Astronomy}, 62, 19511}

\bibitem[{{Jing} \& {Suto}(2002)}]{2002ApJ...574..538J}
{Jing}, Y.~P., \& {Suto}, Y. 2002,
  \href{http://dx.doi.org/10.1086/341065}{\JournalTitle{\apj}, 574, 538}

\bibitem[{{Jing} {et~al.}(2007){Jing}, {Suto}, \& {Mo}}]{2007ApJ...657..664J}
{Jing}, Y.~P., {Suto}, Y., \& {Mo}, H.~J. 2007,
  \href{http://dx.doi.org/10.1086/511130}{\JournalTitle{\apj}, 657, 664}

\bibitem[{Jost \& Jost(2008)}]{jost2008riemannian}
Jost, J., \& Jost, J. 2008, Riemannian geometry and geometric analysis, Vol.
  42005 (Springer)

\bibitem[{{Kaiser}(1987)}]{1987MNRAS.227....1K}
{Kaiser}, N. 1987,
  \href{http://dx.doi.org/10.1093/mnras/227.1.1}{\JournalTitle{\mnras}, 227, 1}

\bibitem[{{Kim} {et~al.}(2014){Kim}, {Choi}, {Kim}, {Kim}, {Lee}, {Shin}, \&
  {Kim}}]{2014ApJS..212...22K}
{Kim}, Y.-R., {Choi}, Y.-Y., {Kim}, S.~S., {et~al.} 2014,
  \href{http://dx.doi.org/10.1088/0067-0049/212/2/22}{\JournalTitle{\apjs},
  212, 22}

\bibitem[{{Kreisch} {et~al.}(2019){Kreisch}, {Pisani}, {Carbone}, {Liu},
  {Hawken}, {Massara}, {Spergel}, \& {Wandelt}}]{2019MNRAS.488.4413K}
{Kreisch}, C.~D., {Pisani}, A., {Carbone}, C., {et~al.} 2019,
  \href{http://dx.doi.org/10.1093/mnras/stz1944}{\JournalTitle{\mnras}, 488,
  4413}

\bibitem[{{Lacasa} \& {Kunz}(2017)}]{2017A&A...604A.104L}
{Lacasa}, F., \& {Kunz}, M. 2017,
  \href{http://dx.doi.org/10.1051/0004-6361/201730784}{\JournalTitle{\aap},
  604, A104}

\bibitem[{{Lahav} {et~al.}(1991){Lahav}, {Lilje}, {Primack}, \&
  {Rees}}]{1991MNRAS.251..128L}
{Lahav}, O., {Lilje}, P.~B., {Primack}, J.~R., \& {Rees}, M.~J. 1991,
  \href{http://dx.doi.org/10.1093/mnras/251.1.128}{\JournalTitle{\mnras}, 251,
  128}

\bibitem[{{Laureijs} {et~al.}(2011){Laureijs}, {Amiaux}, {Arduini},
  {Augu{\`e}res}, {Brinchmann}, {Cole}, {Cropper}, {Dabin}, {Duvet}, {Ealet},
  {Garilli}, {Gondoin}, {Guzzo}, {Hoar}, {Hoekstra}, {Holmes}, {Kitching},
  {Maciaszek}, {Mellier}, {Pasian}, {Percival}, {Rhodes}, {Saavedra Criado},
  {Sauvage}, {Scaramella}, {Valenziano}, {Warren}, {Bender}, {Castander},
  {Cimatti}, {Le F{\`e}vre}, {Kurki-Suonio}, {Levi}, {Lilje}, {Meylan},
  {Nichol}, {Pedersen}, {Popa}, {Rebolo Lopez}, {Rix}, {Rottgering},
  {Zeilinger}, {Grupp}, {Hudelot}, {Massey}, {Meneghetti}, {Miller}, {Paltani},
  {Paulin-Henriksson}, {Pires}, {Saxton}, {Schrabback}, {Seidel}, {Walsh},
  {Aghanim}, {Amendola}, {Bartlett}, {Baccigalupi}, {Beaulieu}, {Benabed},
  {Cuby}, {Elbaz}, {Fosalba}, {Gavazzi}, {Helmi}, {Hook}, {Irwin}, {Kneib},
  {Kunz}, {Mannucci}, {Moscardini}, {Tao}, {Teyssier}, {Weller}, {Zamorani},
  {Zapatero Osorio}, {Boulade}, {Foumond}, {Di Giorgio}, {Guttridge}, {James},
  {Kemp}, {Martignac}, {Spencer}, {Walton}, {Bl{\"u}mchen}, {Bonoli},
  {Bortoletto}, {Cerna}, {Corcione}, {Fabron}, {Jahnke}, {Ligori}, {Madrid},
  {Martin}, {Morgante}, {Pamplona}, {Prieto}, {Riva}, {Toledo}, {Trifoglio},
  {Zerbi}, {Abdalla}, {Douspis}, {Grenet}, {Borgani}, {Bouwens}, {Courbin},
  {Delouis}, {Dubath}, {Fontana}, {Frailis}, {Grazian}, {Koppenh{\"o}fer},
  {Mansutti}, {Melchior}, {Mignoli}, {Mohr}, {Neissner}, {Noddle}, {Poncet},
  {Scodeggio}, {Serrano}, {Shane}, {Starck}, {Surace}, {Taylor},
  {Verdoes-Kleijn}, {Vuerli}, {Williams}, {Zacchei}, {Altieri}, {Escudero
  Sanz}, {Kohley}, {Oosterbroek}, {Astier}, {Bacon}, {Bardelli}, {Baugh},
  {Bellagamba}, {Benoist}, {Bianchi}, {Biviano}, {Branchini}, {Carbone},
  {Cardone}, {Clements}, {Colombi}, {Conselice}, {Cresci}, {Deacon}, {Dunlop},
  {Fedeli}, {Fontanot}, {Franzetti}, {Giocoli}, {Garcia-Bellido}, {Gow},
  {Heavens}, {Hewett}, {Heymans}, {Holland}, {Huang}, {Ilbert}, {Joachimi},
  {Jennins}, {Kerins}, {Kiessling}, {Kirk}, {Kotak}, {Krause}, {Lahav}, {van
  Leeuwen}, {Lesgourgues}, {Lombardi}, {Magliocchetti}, {Maguire}, {Majerotto},
  {Maoli}, {Marulli}, {Maurogordato}, {McCracken}, {McLure}, {Melchiorri},
  {Merson}, {Moresco}, {Nonino}, {Norberg}, {Peacock}, {Pello}, {Penny},
  {Pettorino}, {Di Porto}, {Pozzetti}, {Quercellini}, {Radovich}, {Rassat},
  {Roche}, {Ronayette}, {Rossetti}, {Sartoris}, {Schneider}, {Semboloni},
  {Serjeant}, {Simpson}, {Skordis}, {Smadja}, {Smartt}, {Spano}, {Spiro},
  {Sullivan}, {Tilquin}, {Trotta}, {Verde}, {Wang}, {Williger}, {Zhao},
  {Zoubian}, \& {Zucca}}]{2011arXiv1110.3193L}
{Laureijs}, R., {Amiaux}, J., {Arduini}, S., {et~al.} 2011, \JournalTitle{arXiv
  e-prints}, arXiv:1110.3193

\bibitem[{{Lippich} \& {S{\'a}nchez}(2021)}]{2021MNRAS.508.3771L}
{Lippich}, M., \& {S{\'a}nchez}, A.~G. 2021,
  \href{http://dx.doi.org/10.1093/mnras/stab2820}{\JournalTitle{\mnras}, 508,
  3771}

\bibitem[{{Liu} {et~al.}(2015{\natexlab{a}}){Liu}, {Petri}, {Haiman}, {Hui},
  {Kratochvil}, \& {May}}]{2015PhRvD..91f3507L}
{Liu}, J., {Petri}, A., {Haiman}, Z., {et~al.} 2015{\natexlab{a}},
  \href{http://dx.doi.org/10.1103/PhysRevD.91.063507}{\JournalTitle{\prd}, 91,
  063507}

\bibitem[{{Liu} {et~al.}(2022){Liu}, {Jiang}, \& {Fang}}]{2022arXiv220402945L}
{Liu}, W., {Jiang}, A., \& {Fang}, W. 2022, \JournalTitle{arXiv e-prints},
  arXiv:2204.02945

\bibitem[{{Liu} {et~al.}(2015{\natexlab{b}}){Liu}, {Pan}, {Li}, {Shan}, {Wang},
  {Fu}, {Fan}, {Kneib}, {Leauthaud}, {Van Waerbeke}, {Makler}, {Moraes},
  {Erben}, \& {Charbonnier}}]{2015MNRAS.450.2888L}
{Liu}, X., {Pan}, C., {Li}, R., {et~al.} 2015{\natexlab{b}},
  \href{http://dx.doi.org/10.1093/mnras/stv784}{\JournalTitle{\mnras}, 450,
  2888}

\bibitem[{{Liu} {et~al.}(2021){Liu}, {Yu}, \& {Li}}]{2021ApJS..254....4L}
{Liu}, Y., {Yu}, Y., \& {Li}, B. 2021,
  \href{http://dx.doi.org/10.3847/1538-4365/abe868}{\JournalTitle{\apjs}, 254,
  4}

\bibitem[{{Liu} {et~al.}(2020){Liu}, {Yu}, {Yu}, \&
  {Zhang}}]{2020PhRvD.101f3515L}
{Liu}, Y., {Yu}, Y., {Yu}, H.-R., \& {Zhang}, P. 2020,
  \href{http://dx.doi.org/10.1103/PhysRevD.101.063515}{\JournalTitle{\prd},
  101, 063515}

\bibitem[{{LSST Science Collaboration} {et~al.}(2009){LSST Science
  Collaboration}, {Abell}, {Allison}, {Anderson}, {Andrew}, {Angel}, {Armus},
  {Arnett}, {Asztalos}, {Axelrod}, {Bailey}, {Ballantyne}, {Bankert},
  {Barkhouse}, {Barr}, {Barrientos}, {Barth}, {Bartlett}, {Becker}, {Becla},
  {Beers}, {Bernstein}, {Biswas}, {Blanton}, {Bloom}, {Bochanski}, {Boeshaar},
  {Borne}, {Bradac}, {Brandt}, {Bridge}, {Brown}, {Brunner}, {Bullock},
  {Burgasser}, {Burge}, {Burke}, {Cargile}, {Chandrasekharan}, {Chartas},
  {Chesley}, {Chu}, {Cinabro}, {Claire}, {Claver}, {Clowe}, {Connolly}, {Cook},
  {Cooke}, {Cooray}, {Covey}, {Culliton}, {de Jong}, {de Vries}, {Debattista},
  {Delgado}, {Dell'Antonio}, {Dhital}, {Di Stefano}, {Dickinson}, {Dilday},
  {Djorgovski}, {Dobler}, {Donalek}, {Dubois-Felsmann}, {Durech},
  {Eliasdottir}, {Eracleous}, {Eyer}, {Falco}, {Fan}, {Fassnacht}, {Ferguson},
  {Fernandez}, {Fields}, {Finkbeiner}, {Figueroa}, {Fox}, {Francke}, {Frank},
  {Frieman}, {Fromenteau}, {Furqan}, {Galaz}, {Gal-Yam}, {Garnavich},
  {Gawiser}, {Geary}, {Gee}, {Gibson}, {Gilmore}, {Grace}, {Green}, {Gressler},
  {Grillmair}, {Habib}, {Haggerty}, {Hamuy}, {Harris}, {Hawley}, {Heavens},
  {Hebb}, {Henry}, {Hileman}, {Hilton}, {Hoadley}, {Holberg}, {Holman},
  {Howell}, {Infante}, {Ivezic}, {Jacoby}, {Jain}, {R}, {Jedicke}, {Jee},
  {Garrett Jernigan}, {Jha}, {Johnston}, {Jones}, {Juric}, {Kaasalainen},
  {Styliani}, {Kafka}, {Kahn}, {Kaib}, {Kalirai}, {Kantor}, {Kasliwal},
  {Keeton}, {Kessler}, {Knezevic}, {Kowalski}, {Krabbendam}, {Krughoff},
  {Kulkarni}, {Kuhlman}, {Lacy}, {Lepine}, {Liang}, {Lien}, {Lira}, {Long},
  {Lorenz}, {Lotz}, {Lupton}, {Lutz}, {Macri}, {Mahabal}, {Mandelbaum},
  {Marshall}, {May}, {McGehee}, {Meadows}, {Meert}, {Milani}, {Miller},
  {Miller}, {Mills}, {Minniti}, {Monet}, {Mukadam}, {Nakar}, {Neill}, {Newman},
  {Nikolaev}, {Nordby}, {O'Connor}, {Oguri}, {Oliver}, {Olivier}, {Olsen},
  {Olsen}, {Olszewski}, {Oluseyi}, {Padilla}, {Parker}, {Pepper}, {Peterson},
  {Petry}, {Pinto}, {Pizagno}, {Popescu}, {Prsa}, {Radcka}, {Raddick},
  {Rasmussen}, {Rau}, {Rho}, {Rhoads}, {Richards}, {Ridgway}, {Robertson},
  {Roskar}, {Saha}, {Sarajedini}, {Scannapieco}, {Schalk}, {Schindler},
  {Schmidt}, {Schmidt}, {Schneider}, {Schumacher}, {Scranton}, {Sebag},
  {Seppala}, {Shemmer}, {Simon}, {Sivertz}, {Smith}, {Allyn Smith}, {Smith},
  {Spitz}, {Stanford}, {Stassun}, {Strader}, {Strauss}, {Stubbs}, {Sweeney},
  {Szalay}, {Szkody}, {Takada}, {Thorman}, {Trilling}, {Trimble}, {Tyson}, {Van
  Berg}, {Vanden Berk}, {VanderPlas}, {Verde}, {Vrsnak}, {Walkowicz},
  {Wandelt}, {Wang}, {Wang}, {Warner}, {Wechsler}, {West}, {Wiecha},
  {Williams}, {Willman}, {Wittman}, {Wolff}, {Wood-Vasey}, {Wozniak}, {Young},
  {Zentner}, \& {Zhan}}]{2009arXiv0912.0201L}
{LSST Science Collaboration}, {Abell}, P.~A., {Allison}, J., {et~al.} 2009,
  \JournalTitle{arXiv e-prints}, arXiv:0912.0201

\bibitem[{{Marques} {et~al.}(2019){Marques}, {Liu}, {Zorrilla Matilla},
  {Haiman}, {Bernui}, \& {Novaes}}]{2019JCAP...06..019M}
{Marques}, G.~A., {Liu}, J., {Zorrilla Matilla}, J.~M., {et~al.} 2019,
  \href{http://dx.doi.org/10.1088/1475-7516/2019/06/019}{\JournalTitle{\jcap},
  2019, 019}

\bibitem[{{Martinez} {et~al.}(1993){Martinez}, {Paredes}, \&
  {Saar}}]{1993MNRAS.260..365M}
{Martinez}, V.~J., {Paredes}, S., \& {Saar}, E. 1993,
  \href{http://dx.doi.org/10.1093/mnras/260.2.365}{\JournalTitle{\mnras}, 260,
  365}

\bibitem[{{Mart{\'\i}nez} {et~al.}(2005){Mart{\'\i}nez}, {Starck}, {Saar},
  {Donoho}, {Reynolds}, {de la Cruz}, \& {Paredes}}]{2005ApJ...634..744M}
{Mart{\'\i}nez}, V.~J., {Starck}, J.-L., {Saar}, E., {et~al.} 2005,
  \href{http://dx.doi.org/10.1086/497125}{\JournalTitle{\apj}, 634, 744}

\bibitem[{{Matsubara}(1994)}]{1994ApJ...434L..43M}
{Matsubara}, T. 1994,
  \href{http://dx.doi.org/10.1086/187570}{\JournalTitle{\apjl}, 434, L43}

\bibitem[{{Matsubara}(1996)}]{1996ApJ...457...13M}
{Matsubara}, T. 1996,
  \href{http://dx.doi.org/10.1086/176708}{\JournalTitle{\apj}, 457, 13}

\bibitem[{{Matsubara}(2003)}]{2003ApJ...584....1M}
{Matsubara}, T. 2003,
  \href{http://dx.doi.org/10.1086/345521}{\JournalTitle{\apj}, 584, 1}

\bibitem[{{Matsubara}(2010)}]{2010PhRvD..81h3505M}
{Matsubara}, T. 2010,
  \href{http://dx.doi.org/10.1103/PhysRevD.81.083505}{\JournalTitle{\prd}, 81,
  083505}

\bibitem[{{Matsubara} {et~al.}(2022){Matsubara}, {Hikage}, \&
  {Kuriki}}]{2022PhRvD.105b3527M}
{Matsubara}, T., {Hikage}, C., \& {Kuriki}, S. 2022,
  \href{http://dx.doi.org/10.1103/PhysRevD.105.023527}{\JournalTitle{\prd},
  105, 023527}

\bibitem[{{Matsubara} \& {Kuriki}(2021)}]{2021PhRvD.104j3522M}
{Matsubara}, T., \& {Kuriki}, S. 2021,
  \href{http://dx.doi.org/10.1103/PhysRevD.104.103522}{\JournalTitle{\prd},
  104, 103522}

\bibitem[{{Matsubara} \& {Suto}(1996)}]{1996ApJ...460...51M}
{Matsubara}, T., \& {Suto}, Y. 1996,
  \href{http://dx.doi.org/10.1086/176951}{\JournalTitle{\apj}, 460, 51}

\bibitem[{Mecke(1979)}]{https://doi.org/10.1002/zamm.19790590633}
Mecke, J. 1979,
  \href{http://dx.doi.org/https://doi.org/10.1002/zamm.19790590633}{\JournalTitle{ZAMM
  - Journal of Applied Mathematics and Mechanics / Zeitschrift für Angewandte
  Mathematik und Mechanik}, 59, 286}

\bibitem[{{Mecke} {et~al.}(1994){Mecke}, {Buchert}, \&
  {Wagner}}]{1994A&A...288..697M}
{Mecke}, K.~R., {Buchert}, T., \& {Wagner}, H. 1994, \JournalTitle{\aap}, 288,
  697

\bibitem[{{Melott} {et~al.}(1989){Melott}, {Cohen}, {Hamilton}, {Gott}, \&
  {Weinberg}}]{1989ApJ...345..618M}
{Melott}, A.~L., {Cohen}, A.~P., {Hamilton}, A. J.~S., {Gott}, J.~Richard, I.,
  \& {Weinberg}, D.~H. 1989,
  \href{http://dx.doi.org/10.1086/167935}{\JournalTitle{\apj}, 345, 618}

\bibitem[{{Melott} \& {Dominik}(1993)}]{1993ApJS...86....1M}
{Melott}, A.~L., \& {Dominik}, K.~G. 1993,
  \href{http://dx.doi.org/10.1086/191770}{\JournalTitle{\apjs}, 86, 1}

\bibitem[{{Melott} {et~al.}(1988){Melott}, {Weinberg}, \&
  {Gott}}]{1988ApJ...328...50M}
{Melott}, A.~L., {Weinberg}, D.~H., \& {Gott}, J.~Richard, I. 1988,
  \href{http://dx.doi.org/10.1086/166267}{\JournalTitle{\apj}, 328, 50}

\bibitem[{Milnor(1963)}]{milnor1963morse}
Milnor, J.~W. 1963, Morse theory No.~51 (Princeton university press)

\bibitem[{{Moon} {et~al.}(2023){Moon}, {Valcin}, {Rashkovetskyi}, {Saulder},
  {Aguilar}, {Ahlen}, {Alam}, {Bailey}, {Baltay}, {Blum}, {Brooks}, {Burtin},
  {Chaussidon}, {Dawson}, {de la Macorra}, {de M attia}, {Dhungana},
  {Eisenstein}, {Flaugher}, {Font-Ribera}, {Forero-Romero}, {Garcia-Quintero},
  {Gontcho A Gontcho}, {Guy}, {Hanif}, {Honscheid}, {Ishak}, {Kehoe}, {Kim},
  {Kisner}, {Kremin}, {Landriau}, {Le Guillou}, {Levi}, {Manera}, {Martini},
  {McDonald}, {Meisner}, {Miquel}, {Moustakas}, {Myers}, {Nadathur}, {Neveux},
  {Newman}, {Nie}, {Padmanabhan}, {Palanque-Delabrouille}, {Percival},
  {P{\'e}rez Fern{\'a}ndez}, {Poppett}, {Prada}, {Raichoor}, {Ross}, {Rossi},
  {Samushia}, {Schlegel}, {Seo}, {Tarl{\'e}}, {Vargas Magana}, {Variu},
  {Weaver}, {White}, {Y{\`e}che}, {Yuan}, {Zhao}, {Zhou}, {Zhou}, \&
  {Zou}}]{2023MNRAS.525.5406M}
{Moon}, J., {Valcin}, D., {Rashkovetskyi}, M., {et~al.} 2023,
  \href{http://dx.doi.org/10.1093/mnras/stad2618}{\JournalTitle{\mnras}, 525,
  5406}

\bibitem[{Morse(1934)}]{morse1934calculus}
Morse, M. 1934, The calculus of variations in the large, Vol.~18 (American
  Mathematical Soc.)

\bibitem[{{Nadathur} {et~al.}(2019){Nadathur}, {Carter}, {Percival}, {Winther},
  \& {Bautista}}]{2019PhRvD.100b3504N}
{Nadathur}, S., {Carter}, P.~M., {Percival}, W.~J., {Winther}, H.~A., \&
  {Bautista}, J.~E. 2019,
  \href{http://dx.doi.org/10.1103/PhysRevD.100.023504}{\JournalTitle{\prd},
  100, 023504}

\bibitem[{{Naidoo} {et~al.}(2021){Naidoo}, {Massara}, \&
  {Lahav}}]{2021arXiv211112088N}
{Naidoo}, K., {Massara}, E., \& {Lahav}, O. 2021, \JournalTitle{arXiv
  e-prints}, arXiv:2111.12088

\bibitem[{{Neyrinck}(2008)}]{2008MNRAS.386.2101N}
{Neyrinck}, M.~C. 2008,
  \href{http://dx.doi.org/10.1111/j.1365-2966.2008.13180.x}{\JournalTitle{\mnras},
  386, 2101}

\bibitem[{{Novikov} {et~al.}(2006){Novikov}, {Colombi}, \&
  {Dor{\'e}}}]{2006MNRAS.366.1201N}
{Novikov}, D., {Colombi}, S., \& {Dor{\'e}}, O. 2006,
  \href{http://dx.doi.org/10.1111/j.1365-2966.2005.09925.x}{\JournalTitle{\mnras},
  366, 1201}

\bibitem[{{Paranjape} \& {Alam}(2020)}]{2020MNRAS.495.3233P}
{Paranjape}, A., \& {Alam}, S. 2020,
  \href{http://dx.doi.org/10.1093/mnras/staa1379}{\JournalTitle{\mnras}, 495,
  3233}

\bibitem[{{Parihar} {et~al.}(2014){Parihar}, {Vogeley}, {Gott}, {Choi}, {Kim},
  {Kim}, {Speare}, {Brownstein}, \& {Brinkmann}}]{2014ApJ...796...86P}
{Parihar}, P., {Vogeley}, M.~S., {Gott}, J.~Richard, I., {et~al.} 2014,
  \href{http://dx.doi.org/10.1088/0004-637X/796/2/86}{\JournalTitle{\apj}, 796,
  86}

\bibitem[{{Park} \& {Gott}(1991)}]{1991ApJ...378..457P}
{Park}, C., \& {Gott}, J.~R., I. 1991,
  \href{http://dx.doi.org/10.1086/170445}{\JournalTitle{\apj}, 378, 457}

\bibitem[{{Park} {et~al.}(2005){Park}, {Kim}, \& {Gott}}]{2005ApJ...633....1P}
{Park}, C., {Kim}, J., \& {Gott}, J.~Richard, I. 2005,
  \href{http://dx.doi.org/10.1086/452621}{\JournalTitle{\apj}, 633, 1}

\bibitem[{{Park} \& {Kim}(2010)}]{2010ApJ...715L.185P}
{Park}, C., \& {Kim}, Y.-R. 2010,
  \href{http://dx.doi.org/10.1088/2041-8205/715/2/L185}{\JournalTitle{\apjl},
  715, L185}

\bibitem[{{Percival} {et~al.}(2019){Percival}, {Y{\`e}che}, {Bilicki},
  {Font-Ribera}, {Hathi}, {Howlett}, {Hudson}, {McConnachie}, {Gohar Mohammad},
  {Newman}, {Palanque-Delabrouille}, {Variu}, {Wang}, \&
  {Wilson}}]{2019arXiv190303158P}
{Percival}, W.~J., {Y{\`e}che}, C., {Bilicki}, M., {et~al.} 2019,
  \href{http://dx.doi.org/10.48550/arXiv.1903.03158}{\JournalTitle{arXiv
  e-prints}, arXiv:1903.03158}

\bibitem[{{Petri} {et~al.}(2015){Petri}, {Liu}, {Haiman}, {May}, {Hui}, \&
  {Kratochvil}}]{2015PhRvD..91j3511P}
{Petri}, A., {Liu}, J., {Haiman}, Z., {et~al.} 2015,
  \href{http://dx.doi.org/10.1103/PhysRevD.91.103511}{\JournalTitle{\prd}, 91,
  103511}

\bibitem[{{Philcox} \& {Slepian}(2021)}]{2021arXiv210610278P}
{Philcox}, O. H.~E., \& {Slepian}, Z. 2021, \JournalTitle{arXiv e-prints},
  arXiv:2106.10278

\bibitem[{{Philcox} {et~al.}(2022){Philcox}, {Slepian}, {Hou}, {Warner},
  {Cahn}, \& {Eisenstein}}]{2022MNRAS.509.2457P}
{Philcox}, O. H.~E., {Slepian}, Z., {Hou}, J., {et~al.} 2022,
  \href{http://dx.doi.org/10.1093/mnras/stab3025}{\JournalTitle{\mnras}, 509,
  2457}

\bibitem[{{Planck Collaboration} {et~al.}(2016){Planck Collaboration}, {Ade},
  {Aghanim}, {Arnaud}, {Arroja}, {Ashdown}, {Aumont}, {Baccigalupi},
  {Ballardini}, {Banday}, {Barreiro}, {Bartolo}, {Basak}, {Battaner},
  {Benabed}, {Beno{\^\i}t}, {Benoit-L{\'e}vy}, {Bernard}, {Bersanelli},
  {Bielewicz}, {Bock}, {Bonaldi}, {Bonavera}, {Bond}, {Borrill}, {Bouchet},
  {Boulanger}, {Bucher}, {Burigana}, {Butler}, {Calabrese}, {Cardoso},
  {Catalano}, {Challinor}, {Chamballu}, {Chiang}, {Christensen}, {Church},
  {Clements}, {Colombi}, {Colombo}, {Combet}, {Couchot}, {Coulais}, {Crill},
  {Curto}, {Cuttaia}, {Danese}, {Davies}, {Davis}, {de Bernardis}, {de Rosa},
  {de Zotti}, {Delabrouille}, {D{\'e}sert}, {Diego}, {Dole}, {Donzelli},
  {Dor{\'e}}, {Douspis}, {Ducout}, {Dupac}, {Efstathiou}, {Elsner},
  {En{\ss}lin}, {Eriksen}, {Fergusson}, {Finelli}, {Forni}, {Frailis},
  {Fraisse}, {Franceschi}, {Frejsel}, {Galeotta}, {Galli}, {Ganga}, {Gauthier},
  {Ghosh}, {Giard}, {Giraud-H{\'e}raud}, {Gjerl{\o}w}, {Gonz{\'a}lez-Nuevo},
  {G{\'o}rski}, {Gratton}, {Gregorio}, {Gruppuso}, {Gudmundsson}, {Hamann},
  {Hansen}, {Hanson}, {Harrison}, {Heavens}, {Helou}, {Henrot-Versill{\'e}},
  {Hern{\'a}ndez-Monteagudo}, {Herranz}, {Hildebrandt}, {Hivon}, {Hobson},
  {Holmes}, {Hornstrup}, {Hovest}, {Huang}, {Huffenberger}, {Hurier}, {Jaffe},
  {Jaffe}, {Jones}, {Juvela}, {Keih{\"a}nen}, {Keskitalo}, {Kim}, {Kisner},
  {Knoche}, {Kunz}, {Kurki-Suonio}, {Lacasa}, {Lagache}, {L{\"a}hteenm{\"a}ki},
  {Lamarre}, {Lasenby}, {Lattanzi}, {Lawrence}, {Leonardi}, {Lesgourgues},
  {Levrier}, {Lewis}, {Liguori}, {Lilje}, {Linden-V{\o}rnle},
  {L{\'o}pez-Caniego}, {Lubin}, {Mac{\'\i}as-P{\'e}rez}, {Maggio}, {Maino},
  {Mandolesi}, {Mangilli}, {Marinucci}, {Maris}, {Martin},
  {Mart{\'\i}nez-Gonz{\'a}lez}, {Masi}, {Matarrese}, {McGehee}, {Meinhold},
  {Melchiorri}, {Mendes}, {Mennella}, {Migliaccio}, {Mitra},
  {Miville-Desch{\^e}nes}, {Moneti}, {Montier}, {Morgante}, {Mortlock}, {Moss},
  {M{\"u}nchmeyer}, {Munshi}, {Murphy}, {Naselsky}, {Nati}, {Natoli},
  {Netterfield}, {N{\o}rgaard-Nielsen}, {Noviello}, {Novikov}, {Novikov},
  {Oxborrow}, {Paci}, {Pagano}, {Pajot}, {Paoletti}, {Pasian}, {Patanchon},
  {Peiris}, {Perdereau}, {Perotto}, {Perrotta}, {Pettorino}, {Piacentini},
  {Piat}, {Pierpaoli}, {Pietrobon}, {Plaszczynski}, {Pointecouteau}, {Polenta},
  {Popa}, {Pratt}, {Pr{\'e}zeau}, {Prunet}, {Puget}, {Rachen}, {Racine},
  {Rebolo}, {Reinecke}, {Remazeilles}, {Renault}, {Renzi}, {Ristorcelli},
  {Rocha}, {Rosset}, {Rossetti}, {Roudier}, {Rubi{\~n}o-Mart{\'\i}n},
  {Rusholme}, {Sandri}, {Santos}, {Savelainen}, {Savini}, {Scott}, {Seiffert},
  {Shellard}, {Shiraishi}, {Smith}, {Spencer}, {Stolyarov}, {Stompor},
  {Sudiwala}, {Sunyaev}, {Sutter}, {Sutton}, {Suur-Uski}, {Sygnet}, {Tauber},
  {Terenzi}, {Toffolatti}, {Tomasi}, {Tristram}, {Troja}, {Tucci}, {Tuovinen},
  {Valenziano}, {Valiviita}, {Van Tent}, {Vielva}, {Villa}, {Wade}, {Wandelt},
  {Wehus}, {Yvon}, {Zacchei}, \& {Zonca}}]{2016A&A...594A..17P}
{Planck Collaboration}, {Ade}, P.~A.~R., {Aghanim}, N., {et~al.} 2016,
  \href{http://dx.doi.org/10.1051/0004-6361/201525836}{\JournalTitle{\aap},
  594, A17}

\bibitem[{{Platen} {et~al.}(2007){Platen}, {van de Weygaert}, \&
  {Jones}}]{2007MNRAS.380..551P}
{Platen}, E., {van de Weygaert}, R., \& {Jones}, B. J.~T. 2007,
  \href{http://dx.doi.org/10.1111/j.1365-2966.2007.12125.x}{\JournalTitle{\mnras},
  380, 551}

\bibitem[{{Pogosyan} {et~al.}(2009){Pogosyan}, {Gay}, \&
  {Pichon}}]{2009PhRvD..80h1301P}
{Pogosyan}, D., {Gay}, C., \& {Pichon}, C. 2009,
  \href{http://dx.doi.org/10.1103/PhysRevD.80.081301}{\JournalTitle{\prd}, 80,
  081301}

\bibitem[{{Pranav} {et~al.}(2017){Pranav}, {Edelsbrunner}, {van de Weygaert},
  {Vegter}, {Kerber}, {Jones}, \& {Wintraecken}}]{2017MNRAS.465.4281P}
{Pranav}, P., {Edelsbrunner}, H., {van de Weygaert}, R., {et~al.} 2017,
  \href{http://dx.doi.org/10.1093/mnras/stw2862}{\JournalTitle{\mnras}, 465,
  4281}

\bibitem[{{Pranav} {et~al.}(2019){Pranav}, {van de Weygaert}, {Vegter},
  {Jones}, {Adler}, {Feldbrugge}, {Park}, {Buchert}, \&
  {Kerber}}]{2019MNRAS.485.4167P}
{Pranav}, P., {van de Weygaert}, R., {Vegter}, G., {et~al.} 2019,
  \href{http://dx.doi.org/10.1093/mnras/stz541}{\JournalTitle{\mnras}, 485,
  4167}

\bibitem[{{Repp} \& {Szapudi}(2020)}]{2020MNRAS.498L.125R}
{Repp}, A., \& {Szapudi}, I. 2020,
  \href{http://dx.doi.org/10.1093/mnrasl/slaa139}{\JournalTitle{\mnras}, 498,
  L125}

\bibitem[{{Romano-D{\'\i}az} \& {van de Weygaert}(2007)}]{2007MNRAS.382....2R}
{Romano-D{\'\i}az}, E., \& {van de Weygaert}, R. 2007,
  \href{http://dx.doi.org/10.1111/j.1365-2966.2007.12190.x}{\JournalTitle{\mnras},
  382, 2}

\bibitem[{{Sahni} {et~al.}(1998){Sahni}, {Sathyaprakash}, \&
  {Shandarin}}]{1998ApJ...495L...5S}
{Sahni}, V., {Sathyaprakash}, B.~S., \& {Shandarin}, S.~F. 1998,
  \href{http://dx.doi.org/10.1086/311214}{\JournalTitle{\apjl}, 495, L5}

\bibitem[{{Schaap}(2007)}]{2007PhDT.......486S}
{Schaap}, W.~E. 2007, PhD thesis, University of Groningen, Netherlands

\bibitem[{{Schaap} \& {van de Weygaert}(2000)}]{2000A&A...363L..29S}
{Schaap}, W.~E., \& {van de Weygaert}, R. 2000, \JournalTitle{\aap}, 363, L29

\bibitem[{{Schlegel} {et~al.}(2019){Schlegel}, {Kollmeier}, \&
  {Ferraro}}]{2019BAAS...51g.229S}
{Schlegel}, D., {Kollmeier}, J.~A., \& {Ferraro}, S. 2019,
  \href{http://dx.doi.org/10.48550/arXiv.1907.11171}{in Bulletin of the
  American Astronomical Society, Vol.~51}, 229

\bibitem[{{Schmalzing} \& {Buchert}(1997)}]{1997ApJ...482L...1S}
{Schmalzing}, J., \& {Buchert}, T. 1997,
  \href{http://dx.doi.org/10.1086/310680}{\JournalTitle{\apjl}, 482, L1}

\bibitem[{Schneider(1993)}]{schneider_1993}
Schneider, R. 1993, Convex Bodies: The Brunn-Minkowski Theory, Encyclopedia of
  Mathematics and its Applications (Cambridge University Press)

\bibitem[{{Seljak} {et~al.}(2009){Seljak}, {Hamaus}, \&
  {Desjacques}}]{2009PhRvL.103i1303S}
{Seljak}, U., {Hamaus}, N., \& {Desjacques}, V. 2009,
  \href{http://dx.doi.org/10.1103/PhysRevLett.103.091303}{\JournalTitle{\prl},
  103, 091303}

\bibitem[{{Shan} {et~al.}(2014){Shan}, {Kneib}, {Comparat}, {Jullo},
  {Charbonnier}, {Erben}, {Makler}, {Moraes}, {Van Waerbeke}, {Courbin},
  {Meylan}, {Tao}, \& {Taylor}}]{2014MNRAS.442.2534S}
{Shan}, H.~Y., {Kneib}, J.-P., {Comparat}, J., {et~al.} 2014,
  \href{http://dx.doi.org/10.1093/mnras/stu1040}{\JournalTitle{\mnras}, 442,
  2534}

\bibitem[{{Shandarin} {et~al.}(2012){Shandarin}, {Habib}, \&
  {Heitmann}}]{2012PhRvD..85h3005S}
{Shandarin}, S., {Habib}, S., \& {Heitmann}, K. 2012,
  \href{http://dx.doi.org/10.1103/PhysRevD.85.083005}{\JournalTitle{\prd}, 85,
  083005}

\bibitem[{{Sheth} {et~al.}(2003){Sheth}, {Sahni}, {Shandarin}, \&
  {Sathyaprakash}}]{2003MNRAS.343...22S}
{Sheth}, J.~V., {Sahni}, V., {Shandarin}, S.~F., \& {Sathyaprakash}, B.~S.
  2003,
  \href{http://dx.doi.org/10.1046/j.1365-8711.2003.06642.x}{\JournalTitle{\mnras},
  343, 22}

\bibitem[{{Sousbie}(2011)}]{2011MNRAS.414..350S}
{Sousbie}, T. 2011,
  \href{http://dx.doi.org/10.1111/j.1365-2966.2011.18394.x}{\JournalTitle{\mnras},
  414, 350}

\bibitem[{{Sousbie} {et~al.}(2008){Sousbie}, {Pichon}, {Colombi}, {Novikov}, \&
  {Pogosyan}}]{2008MNRAS.383.1655S}
{Sousbie}, T., {Pichon}, C., {Colombi}, S., {Novikov}, D., \& {Pogosyan}, D.
  2008,
  \href{http://dx.doi.org/10.1111/j.1365-2966.2007.12685.x}{\JournalTitle{\mnras},
  383, 1655}

\bibitem[{{Sousbie} {et~al.}(2011){Sousbie}, {Pichon}, \&
  {Kawahara}}]{2011MNRAS.414..384S}
{Sousbie}, T., {Pichon}, C., \& {Kawahara}, H. 2011,
  \href{http://dx.doi.org/10.1111/j.1365-2966.2011.18395.x}{\JournalTitle{\mnras},
  414, 384}

\bibitem[{{Su{\'a}rez-P{\'e}rez} {et~al.}(2021){Su{\'a}rez-P{\'e}rez},
  {Camargo}, {Li}, \& {Forero-Romero}}]{2021ApJ...922..204S}
{Su{\'a}rez-P{\'e}rez}, J.~F., {Camargo}, Y., {Li}, X.-D., \& {Forero-Romero},
  J.~E. 2021,
  \href{http://dx.doi.org/10.3847/1538-4357/ac1fed}{\JournalTitle{\apj}, 922,
  204}

\bibitem[{{Sutter} {et~al.}(2015){Sutter}, {Lavaux}, {Hamaus}, {Pisani},
  {Wandelt}, {Warren}, {Villaescusa-Navarro}, {Zivick}, {Mao}, \&
  {Thompson}}]{2015A&C.....9....1S}
{Sutter}, P.~M., {Lavaux}, G., {Hamaus}, N., {et~al.} 2015,
  \href{http://dx.doi.org/10.1016/j.ascom.2014.10.002}{\JournalTitle{Astronomy
  and Computing}, 9, 1}

\bibitem[{{Takada} {et~al.}(2014){Takada}, {Ellis}, {Chiba}, {Greene},
  {Aihara}, {Arimoto}, {Bundy}, {Cohen}, {Dor{\'e}}, {Graves}, {Gunn},
  {Heckman}, {Hirata}, {Ho}, {Kneib}, {Le F{\`e}vre}, {Lin}, {More},
  {Murayama}, {Nagao}, {Ouchi}, {Seiffert}, {Silverman}, {Sodr{\'e}},
  {Spergel}, {Strauss}, {Sugai}, {Suto}, {Takami}, \&
  {Wyse}}]{2014PASJ...66R...1T}
{Takada}, M., {Ellis}, R.~S., {Chiba}, M., {et~al.} 2014,
  \href{http://dx.doi.org/10.1093/pasj/pst019}{\JournalTitle{\pasj}, 66, R1}

\bibitem[{Tomita(1990)}]{9789814368223_0003}
Tomita, H. 1990, STATISTICS AND GEOMETRY OF RANDOM INTERFACE SYSTEMS, 113

\bibitem[{{Uhlemann} {et~al.}(2020){Uhlemann}, {Friedrich},
  {Villaescusa-Navarro}, {Banerjee}, \& {Codis}}]{2020MNRAS.495.4006U}
{Uhlemann}, C., {Friedrich}, O., {Villaescusa-Navarro}, F., {Banerjee}, A., \&
  {Codis}, S. 2020,
  \href{http://dx.doi.org/10.1093/mnras/staa1155}{\JournalTitle{\mnras}, 495,
  4006}

\bibitem[{{van de Weygaert} {et~al.}(2009){van de Weygaert}, {Aragon-Calvo},
  {Jones}, \& {Platen}}]{2009arXiv0912.3448V}
{van de Weygaert}, R., {Aragon-Calvo}, M.~A., {Jones}, B. J.~T., \& {Platen},
  E. 2009, \JournalTitle{arXiv e-prints}, arXiv:0912.3448

\bibitem[{{van de Weygaert} \& {Schaap}(2009)}]{2009LNP...665..291V}
{van de Weygaert}, R., \& {Schaap}, W. 2009,
  \href{http://dx.doi.org/10.1007/978-3-540-44767-2\_11}{in Data Analysis in
  Cosmology, ed. V.~J. {Mart{\'\i}nez}, E.~{Saar},
  E.~{Mart{\'\i}nez-Gonz{\'a}lez}, \& M.~J. {Pons-Border{\'\i}a}, Vol. 665},
  291

\bibitem[{{van de Weygaert} {et~al.}(2011){van de Weygaert}, {Vegter},
  {Edelsbrunner}, {Jones}, {Pranav}, {Park}, {Hellwing}, {Eldering},
  {Kruithof}, {Bos}, {Hidding}, {Feldbrugge}, {ten Have}, {van Engelen},
  {Caroli}, \& {Teillaud}}]{2011LNCS.6970...60V}
{van de Weygaert}, R., {Vegter}, G., {Edelsbrunner}, H., {et~al.} 2011,
  \href{http://dx.doi.org/10.1007/978-3-642-25249-5_3}{in Lecture Notes in
  Computer Science, Vol. 6970}, 60

\bibitem[{{Wang} {et~al.}(2012){Wang}, {Chen}, \& {Park}}]{2012ApJ...747...48W}
{Wang}, X., {Chen}, X., \& {Park}, C. 2012,
  \href{http://dx.doi.org/10.1088/0004-637X/747/1/48}{\JournalTitle{\apj}, 747,
  48}

\bibitem[{{Wiegand} {et~al.}(2014){Wiegand}, {Buchert}, \&
  {Ostermann}}]{2014MNRAS.443..241W}
{Wiegand}, A., {Buchert}, T., \& {Ostermann}, M. 2014,
  \href{http://dx.doi.org/10.1093/mnras/stu1118}{\JournalTitle{\mnras}, 443,
  241}

\bibitem[{{Wilding} {et~al.}(2021){Wilding}, {Nevenzeel}, {van de Weygaert},
  {Vegter}, {Pranav}, {Jones}, {Efstathiou}, \&
  {Feldbrugge}}]{2021MNRAS.507.2968W}
{Wilding}, G., {Nevenzeel}, K., {van de Weygaert}, R., {et~al.} 2021,
  \href{http://dx.doi.org/10.1093/mnras/stab2326}{\JournalTitle{\mnras}, 507,
  2968}

\bibitem[{{Yip} {et~al.}(2024){Yip}, {Biagetti}, {Cole}, {Viswanathan}, \&
  {Shiu}}]{2024arXiv240313985Y}
{Yip}, J. H.~T., {Biagetti}, M., {Cole}, A., {Viswanathan}, K., \& {Shiu}, G.
  2024, \href{http://dx.doi.org/10.48550/arXiv.2403.13985}{\JournalTitle{arXiv
  e-prints}, arXiv:2403.13985}

\bibitem[{{Yu} {et~al.}(2017){Yu}, {Emberson}, {Inman}, {Zhang}, {Pen},
  {Harnois-D{\'e}raps}, {Yuan}, {Teng}, {Zhu}, {Chen}, {Xing}, {Du}, {Zhang},
  {Lu}, \& {Liao}}]{2017NatAs...1E.143Y}
{Yu}, H.-R., {Emberson}, J.~D., {Inman}, D., {et~al.} 2017,
  \href{http://dx.doi.org/10.1038/s41550-017-0143}{\JournalTitle{Nature
  Astronomy}, 1, 0143}

\bibitem[{{Zhang} {et~al.}(2010){Zhang}, {Springel}, \&
  {Yang}}]{2010ApJ...722..812Z}
{Zhang}, Y., {Springel}, V., \& {Yang}, X. 2010,
  \href{http://dx.doi.org/10.1088/0004-637X/722/1/812}{\JournalTitle{\apj},
  722, 812}

\bibitem[{{Zhao} {et~al.}(2015){Zhao}, {Kitaura}, {Chuang}, {Prada}, {Yepes},
  \& {Tao}}]{2015MNRAS.451.4266Z}
{Zhao}, C., {Kitaura}, F.-S., {Chuang}, C.-H., {et~al.} 2015,
  \href{http://dx.doi.org/10.1093/mnras/stv1262}{\JournalTitle{\mnras}, 451,
  4266}

\bibitem[{{Zhao} {et~al.}(2016){Zhao}, {Tao}, {Liang}, {Kitaura}, \&
  {Chuang}}]{2016MNRAS.459.2670Z}
{Zhao}, C., {Tao}, C., {Liang}, Y., {Kitaura}, F.-S., \& {Chuang}, C.-H. 2016,
  \href{http://dx.doi.org/10.1093/mnras/stw660}{\JournalTitle{\mnras}, 459,
  2670}

\bibitem[{{Zunckel} {et~al.}(2011){Zunckel}, {Gott}, \&
  {Lunnan}}]{2011MNRAS.412.1401Z}
{Zunckel}, C., {Gott}, J.~R., \& {Lunnan}, R. 2011,
  \href{http://dx.doi.org/10.1111/j.1365-2966.2010.18015.x}{\JournalTitle{\mnras},
  412, 1401}

\end{thebibliography}

\end{document}